\documentclass[aps,prl,twocolumn,superscriptaddress,floatfix,nofootinbib]{revtex4-2}
\usepackage{epsfig,amsmath,amssymb,color,comment,physics}
\usepackage[makeroom]{cancel}
\usepackage[caption=false]{subfig}
\usepackage{mathrsfs}
\usepackage[countmax]{subfloat}
\usepackage[normalem]{ulem}
\usepackage[english]{babel}
\usepackage{dsfont}
\usepackage{float}
\usepackage[bookmarks=true,colorlinks,linkcolor=black,urlcolor=NavyBlue,citecolor=RoyalBlue]{hyperref}
\usepackage[dvipsnames]{xcolor}
\usepackage{soul} 
\usepackage{color, xcolor}

\hypersetup{
    unicode=false,     
    pdftoolbar=false,  
    pdfmenubar=true,   
    pdffitwindow=false, 
    pdfstartview={FitH},
    pdftitle={},    
    pdfauthor={Authors},     
    pdfsubject={},   
    pdfcreator={},   
    pdfproducer={}, 
    pdfnewwindow=true,
    colorlinks=true,
    linkcolor=black,
    citecolor=blue, 
    filecolor=magenta,
    urlcolor=blue
}

\newcommand{\beginsupplement}{%
    \setcounter{table}{0}
    \renewcommand{\thetable}{S\arabic{table}}%
    \setcounter{figure}{0}
    \renewcommand{\thefigure}{S\arabic{figure}}%
    \setcounter{equation}{0}
    \renewcommand{\theequation}{S\arabic{equation}}%
    \setcounter{section}{0}
    \renewcommand{\thesection}{S\arabic{section}}%
   }

\graphicspath{{./Figures/}} 

\begin{document}

\title{
Liouvillian Spectral Transition in Noisy Quantum Many-Body Scars
}

\author{Jin-Lou Ma}
\thanks{These authors contributed equally}
\affiliation{School of Physics and Zhejiang Key Laboratory of Micro-nano Quantum Chips and Quantum Control, Zhejiang University, Hangzhou $310027$, China}
\author{Zexian Guo}
\thanks{These authors contributed equally}
\affiliation{School of Physics and Zhejiang Key Laboratory of Micro-nano Quantum Chips and Quantum Control, Zhejiang University, Hangzhou $310027$, China}
\author{Yu Gao}
\affiliation{School of Physics and Zhejiang Key Laboratory of Micro-nano Quantum Chips and Quantum Control, Zhejiang University, Hangzhou $310027$, China}

\author{Zlatko Papi\'c}
\email{z.papic@leeds.ac.uk}
\affiliation{School of Physics and Astronomy, University of Leeds, Leeds LS2 9JT, United Kingdom}

\author{Lei Ying}
\email{leiying@zju.edu.cn}
\affiliation{School of Physics and Zhejiang Key Laboratory of Micro-nano Quantum Chips and Quantum Control, Zhejiang University, Hangzhou $310027$, China}

\date{\today}

\begin{abstract}
Understanding the behavior of quantum many-body systems under decoherence is essential for developing robust quantum technologies. Here, we examine the fate of weak ergodicity breaking in systems hosting quantum many-body scars when subject to local pure dephasing---an experimentally relevant form of environmental noise. Focusing on a large class of models with an approximate su(2)-structured scar subspace, we show that scarred eigenmodes of the Liouvillian exhibit a transition reminiscent of spontaneous $\mathbb{PT}$-symmetry breaking as the dephasing strength increases. Unlike previously studied non-Hermitian mechanisms, this transition arises from a distinct quantum jump effect. Remarkably, in platforms such as the XY spin ladder and PXP model of Rydberg atom arrays, the critical dephasing rate shows only weak dependence on the system size, revealing an unexpected robustness of scarred dynamics in noisy environments.
\end{abstract}

\maketitle

\emph{Introduction---}Recent advances in quantum simulations~\cite{Bloch2012,Georgescu2014,Kjaergaard2020,Browaeys2020,MonroeRMP} have opened a window to studying thermalization in isolated quantum many-body systems~\cite{DeutschETH, SrednickiETH, RigolNature, dAlessio2016,Ueda2020}. Many such systems are now understood to host atypical non-thermalizing eigenstates known as quantum many-body scars (QMBSs)~\cite{serbyn2021quantum,Moudgalya_2022,ChadranReview}. QMBSs are ubiquitous in many physical systems, including Rydberg atom arrays~\cite{bernien2017probing,turner2018weak,Ho2019, Choi2019, Surace2020, Khemani2019, lin2019exact,  lin2020quantum, bluvstein2021controlling, mondragon2021fate, Omiya2022,zhao2024observationquantumthermalizationrestricted,kerschbaumer2024quantummanybodyscarspxp,Ding2024RydbergClusters,ivanov2025exactarealawscareigenstates}, the Heisenberg-type spin models~\cite{Moudgalya2018_2,Schecter2019,yang2024phantom,Guo2023Origin}, ultracold atoms in tilted optical lattices~\cite{Hudomal2020bosons,Zhao2020Optical,Desaules2021Tilted,scherg2021observing,Su2023,Adler2024}, superconducting circuits~\cite{zhang2023many,larsen2024experimentalprotocolobservingsingle}, and many others~\cite{ShiraishiMori, NeupertScars, Shibata202Onsager, McClarty2020, Lee2020Colored, pakrouski2020many, Mark2020eta, ren2021quasisymmetry, Mohapatra2023, kolb2023stability, Srivatsa2023, Desaules2023Schwinger, Halimeh2023robustquantummany, Buca2023, Gotta2023, Moudgalya2024, Evrard2024, Dooley2024Dual, pizzi2024quantumscarsmanybodysystems}. 
The  unique properties of QMBSs are also of interest in quantum information processing~\cite{Dooley2021,Desaules2022QFI,Dooley2023}, e.g., they have been used to prepare Greenberger-Horne-Zeilinger states~\cite{omran2019GHZ}.

The existing platforms for quantum simulation, however, are prone to decoherence caused by the surrounding environment. The decoherence -- particularly pure dephasing -- can significantly impact individual qubits~\cite{albash2015decoherence}, limiting their dephasing time to around $20$~$\mu$s in current state-of-the-art superconducting platforms~\cite{WOS:001022795400001}. While dissipation was studied in kinetically-constrained models~\cite{Olmos2012,Macieszczak2016} and time crystals~\cite{Buca2019,Booker2020,Gambetta2019,Kessler2021}, its impact on the robustness of QMBS states remains poorly understood~\cite{Wang2024DecoherenceFree,jiang2025robustnessquantummanybodyscars,garcíagarcía2025lindbladmanybodyscars}. 
Previous studies of QMBSs to perturbations in closed systems~\cite{Lin2020Pert,Surace2021Pert} suggest that they might quickly thermalize in the presence of noise. Moreover, while Ref.~\cite{Wang2024DecoherenceFree} recently demonstrated a construction for embedding QMBS into decoherence-free subspaces, a comprehensive understanding of realistic decoherence effects, such as pure dephasing, remains elusive.

In this paper, we consider a large class of QMBS models coupled to the environment via local pure dephasing $\gamma$ on each site.  We focus on models with a so-called restricted su(2) spectrum generating algebra~\cite{moudgalya2020eta,Mark2020,ODea2020Tunnels,Bull2020}, which includes experimentally-realized QMBSs in several platforms~\cite{bernien2017probing,Su2023,zhang2023many,Dong2023}. In the absence of dephasing, these QMBS eigenstates distribute with nearly equal energy spacings and the system exhibits long-lived coherent dynamics when prepared in special initial states.  As the dephasing increases beyond a critical value $\gamma_\star$, the thermal bulk of the spectrum undergoes a spontaneous Liouvillian $\mathbb{PT}$-symmetry (LPTS) breaking transition~\cite{PhysRevLett.109.090404}. This occurs rapidly, as the critical dephasing is inversely proportional to the square of the Liouvillian eigenvalue density~\cite{PhysRevLett.109.090404} and hence exponentially small in system size, $\gamma_\star\sim\exp(-L)$. By contrast, we find that QMBS eigenvalues undergo a distinct spectral transition at $\gamma_\star^\mathrm{S}$, which scales polynomially with system size, $\gamma_\star^\mathrm{S}\sim L^{-1}$. This difference in scaling has a striking manifestation for the robustness of QMBS signatures under dissipative dynamics. We identify the origin of this phenomenon in  the quantum jumper effect beyond previously considered non-Hermitian mechanisms~\cite{Qianqian2023nonHerm, Shen2024}. We illustrate our conclusions using an experimentally relevant model of superconducing qubits from Refs.~\cite{zhang2023many,Dong2023},  while in End Matter we provide further evidence for the PXP model of Rydberg atom arrays~\cite{bernien2017probing}.

\emph{Creutz Ladder (CL) model---}We consider a paradigmatic QMBS model describing superconducting qubits in a Creutz ladder configuration~\cite{Iadecola2019Ladders,zhang2023many}. 
A variant of this model was used to experimentally demonstrate QMBS-enabled tunability of entanglement in the presence of disorder~\cite{Dong2023}. The total Hamiltonian, $\hat{H}=\hat{H}_{\mathrm{CL}}\!=\!\hat{H}_\mathrm{h,1}+\hat{H}_\mathrm{h,2}+\hat{H}_\mathrm{v}+\hat{H}_\mathrm{x}$, is a sum of
Hamiltonians $\hat H_\mathrm{h,\alpha}$ for each of the two horizontal legs $\alpha\in\{1,2\}$ of the ladder, the vertical coupling between them, $\hat H_\mathrm{v}$, and the cross-coupling $\hat H_\mathrm{x}$ present in experiment~\cite{zhang2023many,Dong2023}: 
\begin{eqnarray}\label{eq:CLHam}
\nonumber \hat{H}_\mathrm{h,\alpha} &=& J_\mathrm{h} \sum_{j}\big[(-1)^{j+\alpha}\hat{\sigma}^{+}_{j,{\alpha}}\hat{\sigma}^{-}_{j+1,\alpha}{+}\mathrm{h.c.}\big],\\
 \hat{H}_\mathrm{v} &=& J\sum_{j}\left(\hat{\sigma}^{+}_{j,1}\hat{\sigma}^{-}_{j,2}+\mathrm{h.c.}\right),   \\
\nonumber \hat{H}_\mathrm{x} &=& \sum_{j}J_{\mathrm{x},j}\left(\hat{\sigma}^{+}_{j,1}\hat{\sigma}^{-}_{j+1,2}+\hat{\sigma}^{+}_{j,2}\hat{\sigma}^{-}_{j+1,1}+\mathrm{h.c.}\right),
\end{eqnarray}
where $\hat{\sigma}^{\pm}_{j,\alpha}$ is the standard raising (lowering) spin-$1/2$ operator for $j$th spin in layer $\alpha$. 
The ladder contains $L=2N$ spins in total. We use `$\uparrow$, $\downarrow$' to denote the standard spin basis states of a single site, assume open boundary conditions, and set $J=1$ as the overall energy scale. Note that the cross-coupling $J_{\mathrm{x},j}$ can be site-dependent or even random. 

When $J_{\mathrm{x},j}\!=\!0$, the CL model hosts exact QMBS eigenstates which are simply the Dicke states of a collective spin of magnitude $N$:
\begin{equation}\label{eq:S0}
 \hat{S}_0 = \sum_{j=1}^N(|\!\!\Uparrow \rangle\langle\Downarrow\!\!|_j+|\!\!\Downarrow\rangle\langle\Uparrow \!\!|_j).
\end{equation}
This operator is defined in the basis $|\!\!\Uparrow\rangle \equiv \big|\substack {\uparrow\\ \downarrow} \big\rangle$, $|\!\!\Downarrow\rangle \equiv \big|\substack {\downarrow\\ \uparrow} \big\rangle$, see Supplementary Material (SM) for details~\cite{SM}. In the limit $J_{\mathrm{x},j}\!=\!0$, the CL Hamiltonian splits into a direct sum of scar and thermal subspaces, $\hat H = \hat S_0 \oplus \hat T$, and the product state $\left|\Pi\right\rangle \!=\! \left| \Uparrow\Uparrow\Uparrow\cdots \right\rangle$ has no component outside $\hat S_0$. Thus, $|\Pi\rangle$ undergoes perfect revivals when $J_{\mathrm{x},j}\!=\!0$, corresponding to the free precession of the collective spin. On the other hand, when $J_{\mathrm{x},j}\!\neq\! 0$, the QMBS eigenstates 
are only approximately given by the Dicke states, and a small coupling $\hat V$ then connects the scar and thermal subspaces, $\hat H = \hat S_0 \oplus \hat T + \hat V$. This causes the revivals to acquire a decaying envelope, as observed in experiment~\cite{Dong2023}.

\begin{figure}
\includegraphics[width=\linewidth]{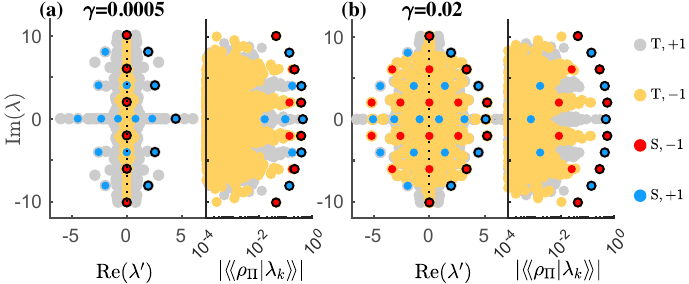}
\caption{
(a)-(b) The Liouvillian spectra (left panel) and the overlap with $|\rho_\Pi\rangle\!\rangle$ (right panel) of the CL model, Eq.~(\ref{eq:CLHam}), for dephasing strengths $\gamma \!=\!$ $0.0005$ (a) and $0.02$ (b). We plot the imaginary part of Liouvillian eigenvalues $\lambda_k$ against the real part of shifted eigenvalues $\lambda_k^\prime$. 
Red and blue dots represent scar (S) eigenvalues, while yellow and gray ones represent thermal (T) eigenvalues, labeled according to their inversion symmetry quantum number, $p\!=\!\pm1$.  Open black circles represent scar states corresponding to local maxima in the overlaps (right panels).
The scar eigenvalues in `$\mathrm{S},{-}1$' sector undergo a spectral transition as $\gamma$ increases, while those in `$\mathrm{S},{+}1$' do not [compare red dots in panel (a) with (b)]. 
The data is obtained by exact diagonalization for couplings $J_\mathrm{h}\!=\!0.66$, $J_{\mathrm{x},j}\!=\!0.1$, system size $L\!=\!10$ with open boundary conditions.
}\label{fig:schematic}
\end{figure}

\emph{Liouvillian spectral transition---}
We now consider each spin to have local decoherence to a bath. Under the Markov approximation, such a system is generally described by a Liouville superoperator~\cite{breuer2002theory}. Here, we focus on pure dephasing, with the corresponding superoperator  
\begin{equation}\label{eq:liouv_supero}
\mathcal{L}=-i\left(
\hat{H}\otimes\hat{I}-\hat{I}\otimes\hat{H}^*
\right)+\gamma\mathcal{D},
\end{equation}
where $\gamma$ is the pure dephasing rate, the dissipator is $\mathcal{D}=\mathcal{D}^\prime-\sum_j \hat{I}_{j}\otimes\hat{I}_j$ with $\mathcal{D}^\prime=\sum_j \hat{\sigma}^z_{j}\otimes\hat{\sigma}^z_j$. 
The operator $|E_n\rangle\langle E_m|$ in $\mathcal{L}$ has a vector representation given by $|E_n\rangle\!\otimes\!|E_m\rangle$~\cite{breuer2002theory}, where $|E_{n}\rangle$ and $|E_m\rangle$ are eigenstates of the entire Hamiltonian with $n,m=1,2,\dots,D$ and $D$ is the Hilbert space dimension.
The $k$th eigenvalue and eigenmode of the Liouville superoperator in Eq.~(\ref{eq:liouv_supero}) are denoted by $\lambda_k$ and $|\lambda_k\rangle\!\rangle$, respectively, with $k=1,2,\cdots,D^2$. 
To understand the spectral properties of $\mathcal{L}$, we perform a transformation $\mathcal{L}^{\prime}\!=\!\mathcal{L}+\gamma\big(\mathrm{Tr}(\mathcal{D})/\mathrm{Tr}(\mathcal{I})\big)\mathcal{I}$,  with $\mathcal{I}\!=\!\hat{I}\!\otimes\!\hat{I}$, which makes the Liouvillian spectrum traceless, i.e., sets the average of $\mathrm{Re}(\lambda^{\prime}_k)$ to zero~\cite{PhysRevLett.109.090404}. 

For an isolated system with cross-couplings turned off, $\gamma=J_{\mathrm{x},j}=0$, the Liouvillian of scar subspace is $\mathcal{S}_0=-i(\hat{S}_0\otimes\hat{I}-\hat{I}\otimes\hat{S}_0)$. The spectrum of  $\mathcal{S}_0$ is purely imaginary and the eigenmodes are tensor products of previously mentioned Dicke states for Eq.~(\ref{eq:S0}): 
\begin{equation}\label{eq:scarS0}
\big|\lambda_{(l,s)}^{(0)}\big\rangle\!\big\rangle \!=\! \big|E_{l+s}^\mathrm{S}\big\rangle\! \otimes \!\big|E_s^\mathrm{S}\big\rangle, \quad 
\lambda_{(l,s)}^{(0)} 
\!=\! i\Big(E_{l+s}^\mathrm{S}\!-\!E_{s}^\mathrm{S}\Big) \!=\! 2il\,,
\end{equation}
where $E^\mathrm{S}$ and $|E^\mathrm{S}\rangle$ represent the eigenenergy and eigenstate of $\hat{S}_0$, respectively.
For later convenience, we use indices  $l \!=\! -N, -\!N{+}1,\cdots,N$ and $s \!=\! 0, 1,\cdots, N\!\!-\!|l|$ to label the scar eigenmodes. For small $\gamma$, we will use the overlap with states in Eq.~(\ref{eq:scarS0}) to identify QMBS eigenmodes of the full Liouvillian. For stronger dephasings, we identify QMBS states by following smooth changes of the spectrum as $\gamma$ is slowly varied.

The Liouvillian spectra of the CL model in Eq.~(\ref{eq:CLHam}) with $J_\mathrm{h}\!=\!0.66$, $J_{\mathrm{x},j}\!=\!0.1$ are presented in Fig.~\ref{fig:schematic}. We show illustrative examples of the spectra before (a) and after (b) the spectral transition. The QMBS eigenmodes (red and blue circles) are distinguished by their enhanced overlap with the state $|\rho_\Pi\rangle\!\rangle=|\Pi\rangle\!\otimes\!|\Pi\rangle$, also shown in Fig.~\ref{fig:schematic}. These eigenvalues are furthermore distinguished by their symmetry quantum number $p=\pm 1$ under inversion, generated by $\prod_j (\hat \sigma_{j,1}^x \hat \sigma_{j,2}^x)\otimes (\hat \sigma_{j,1}^x \hat \sigma_{j,2}^x)$. Before the spectral transition, Fig.~\ref{fig:schematic}(a), the identified QMBS eigenvalues cluster on the imaginary axis, unlike the thermal eigenvalues (yellow and gray circles). After the transition, Fig.~\ref{fig:schematic}(b), the same QMBS eigenvalues exhibit a qualitatively different distribution pattern in the form of a sparse diamond grid across the complex plane. For a system of $N$ qubits per leg, the QMBS eigenvalues undergoing this transition are layers $l=-N,2-N,\dots,N-2,N$ in Eq.~(\ref{eq:scarS0}). Thus, they alternate between inversion symmetry sectors given by $p=(-1)^l$. 
 
The change in the distribution of eigenvalues in the complex plane in Fig.~\ref{fig:schematic} is reminiscent of a LPTS breaking transition~\cite{PhysRevLett.109.090404}. However, most of the thermal
eigenvalues  have already undergone LPTS breaking at $\gamma=0.0005$ in Fig.~\ref{fig:schematic}(a), suggesting that QMBS eigenvalues undergo a distinct transition. Moreover, it is only a subset of those eigenvalues belonging to `$\mathrm{S},-1$' sector in Fig.~\ref{fig:schematic} that undergo such a transition.  Below we show that the spectral transition of these QMBS eigenvalues leaves an imprint on the dissipative dynamics, and we develop their perturbative description to reveal the distinction from LPTS breaking transition of thermal eigenvalues.

\begin{figure}[tb]
\centering
\includegraphics[width=\linewidth]{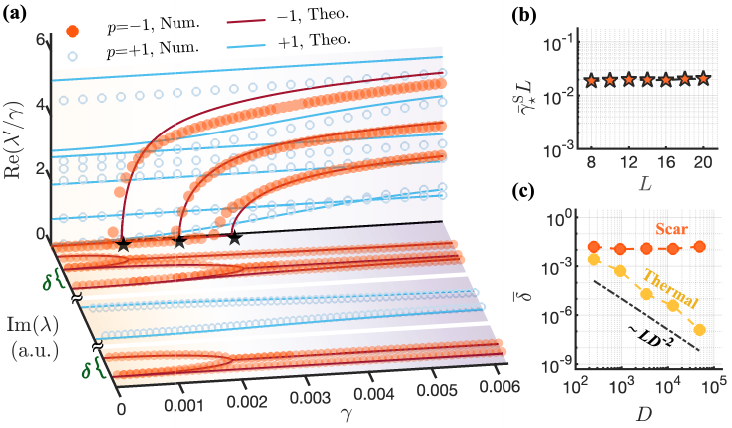}
\caption{ 
(a) The evolution of the Liouvillian spectrum in the CL model, Eq.~(\ref{eq:CLHam}), with  couplings $J_\mathrm{h}\!=\!0.66$, $J_{\mathrm{x},j}\!=\!0.1$ and system size $L\!=\!10$ with open boundary conditions. We plot $\mathrm{Im}(\lambda)$ and $\mathrm{Re}(\lambda^{\prime}/\gamma)$, as the dephasing rate $\gamma$ is varied. For improved visualization, the imaginary eigenvalues have been offset within a small window. Analytical prediction of perturbation theory (lines) are in good agreement with the numerical data, in particular for the spectral transition points $\gamma_{\star}^\mathrm{S(1,2,3)}$ (black stars). 
(b) The average spectral transition point, $\bar{\gamma}_{\star}^S\, L$,  as a function of the system size $L$ obtained in perturbation theory.
(c) The average spacing $\bar{\delta}$ between the scarred and thermal eigenenergies in the dephasing-free case, plotted as a function of the Hilbert space dimension $D$.
}
\label{fig:CL_lpts}
\end{figure}

\emph{Liouvillian perturbation theory.---}To examine the spectral transition in Fig.~\ref{fig:schematic} more quantitatively, we study the spectra as we sweep the dephasing rate $\gamma$ in Fig.~\ref{fig:CL_lpts}(a).  At weak dephasing, several eigenvalues with inversion symmetry $p\!=\!-1$ remain stationary along the real axis, while moving along the imaginary one. As $\gamma$ increases, pairs of these eigenvalues collide on the imaginary axis and abruptly bifurcate into two branches at specific dephasing rates $\gamma_{\star}^\mathrm{S(1,2,3)}$ along the real axis 
[for clarity, only the positive branch is shown in Fig.~\ref{fig:CL_lpts}(a)].

To understand the spectral evolution in Fig.~\ref{fig:CL_lpts}(a), we have developed a Liouvillian degenerate perturbation theory valid for small $\gamma$. If we neglect the coupling between the bath and the thermal subsystem, the effective Liouvillian for the scar subspace is
$
\mathcal{L}_\mathcal{S} \approx \mathcal{S}_0 + \mathcal{E}+ \gamma_{\mathrm{eff}}\mathcal{D}^\prime
$,
where 
$\mathcal{E}={-i\big(\hat{\Sigma} \otimes \hat{I}-\hat{I} \otimes \hat{\Sigma}\big)}$ is the self-energy superoperator that includes the effect of thermal bulk~\cite{SM}. The effective dephasing $\gamma_\mathrm{eff}=2\gamma$ accounts for the magnitude of collective spin being half the system size, $N=L/2$.

In the regime $\Vert \mathcal{E}+\gamma_{\mathrm{eff}}\mathcal{D}^\prime\Vert\ll \Vert\mathcal{S}_0\Vert$, the traceless Liouvillian eigenvalues for the $l$th layer are modified with respect to Eq.~(\ref{eq:scarS0}) into $\lambda^\prime_{(l,s)}\approx \lambda^{(0)}_{(l,s)} + \lambda^{(1)}_{(l,s)}$, where the first-order corrections $\lambda^{(1)}_{(l,s)}$ are obtained by diagonalizing the projection of $\left(\mathcal{E} \!+\! \gamma_{\mathrm{eff}} \mathcal{D}^\prime\right)$ to the degenerate subspace of the $l$th layer. The projected matrix consists of an imaginary diagonal matrix $\mathcal{E}_l$ and a real shift matrix $\mathcal{D}_l^\prime$:
\begin{equation}\label{eq:D_l_main}
\begin{aligned}
   \mathcal{E}_l &= i \times \mathrm{Diag}\Big[\Sigma_{(l,0)},\ \Sigma_{(l,1)},\ \cdots, \Sigma_{(l,N-|l|)}\Big],\\
   \mathcal{D}_l^\prime &=  \sum_{s=0}^{N-|l|-1} d_{(l,s)} \Big[\ \big|\lambda_{(l,s)}^{(0)}\big\rangle\!\big\rangle  \big\langle\!\big\langle\lambda_{(l,s+1)}^{(0)}\big|+\mathrm{h.c.}  \Big], 
\end{aligned}
\end{equation}
with $\Sigma_{(l,s)}=\big\langle E_{s}^\mathrm{S}\big|\hat{\Sigma}\big|E_{s}^\mathrm{S}\big\rangle-\big\langle E_{s+l}^\mathrm{S}\big|\hat{\Sigma}\big|E_{s+l}^\mathrm{S}\big\rangle$ and $d_{(l,s)} =\sqrt{\left(s\!+\!1\right)\left(N\!-\!s\right)\left(|l|\!+\!s\!+\!1\right)\left(N\!-\!|l|\!-\!s\right)}/N$~\cite{SM}.
By diagonalizing $\left(\mathcal{E}_l \!+\! \gamma_{\mathrm{eff}} \mathcal{D}_l^\prime\right)$, we obtain the perturbed eigenvalues for the $l$th layer and plot them against the numerical results for the same parameters in Fig.~\ref{fig:CL_lpts}(a).

The perturbation theory results in Fig.~\ref{fig:CL_lpts}(a) compare well with the numerics for both the imaginary and the real parts of the Liouvillian spectrum. The analytical prediction of $\gamma_\star^{\mathrm{S}(1,2,3)}$ implies that the spectral transition in QMBS Liouvillian spectra originates from the competition between dissipation $\gamma_{\mathrm{eff}}\mathcal{D}^\prime$ and self-energy $\mathcal{E}$. While different QMBS eigenvalues undergo bifurcations at generally different values of $\gamma_\star^{\mathrm{S}(1,2\ldots)}$, the mean value $\bar{\gamma}_\star^\mathrm{S}$ has a robust universal scaling with $L$ and determines the dynamical signatures studied below. Hence, we estimate the spectral transition point for QMBS Liouvillian eigenvalues:
\begin{equation}\label{eq:gammastarpert}
    \bar{\gamma}_\star^\mathrm{S} \sim {\big\Vert\mathcal{E}\big\Vert}\times{\big\Vert\mathcal{D}^\prime\big\Vert}^{-1}\sim L^{-1},
\end{equation}
which follows from the operator norm of $\mathcal{D}^\prime$ being extensive in system size, while $\Vert\mathcal{E}\Vert$ is independent of $L$~\cite{SM}. This estimate, shown in Fig.~\ref{fig:CL_lpts}(b),
is consistent with Fig.~\ref{fig:CL_lpts}(c), where we numerically extract the average spacing $\bar{\delta}$ for scar and thermal eigenvalues in the dephasing-free case ($\gamma\!=\!0$) 
as a function of the Hilbert space dimension $D$. 
The typical spacing for thermal eigenvalues scales as  
$\bar{\delta}\!\sim\! LD^{-2}$, while the scar $\bar{\delta}^\mathrm{S}$ is nearly independent of $D$.
Then, the heuristic estimate of the typical spectral transition point, $\bar{\gamma}_\star^\mathrm{S} \!\sim \!\bar{\delta}^\mathrm{S}\Vert\mathcal{D}^\prime\Vert^{-1}$, 
is in agreement with Eq.~(\ref{eq:gammastarpert}).
This is in stark contrast with the exponential scaling $\bar{\gamma}_\star\!\sim\! D^{-2}$ of thermal eigenvalues~\cite{PhysRevLett.109.090404}.  Thus, 
the presence of QMBSs gives rise to spectral transition points that scale polynomially with system size and have robust dynamical signatures, as shown next.

\begin{figure}[tb]
\centering
\includegraphics[width=\linewidth]{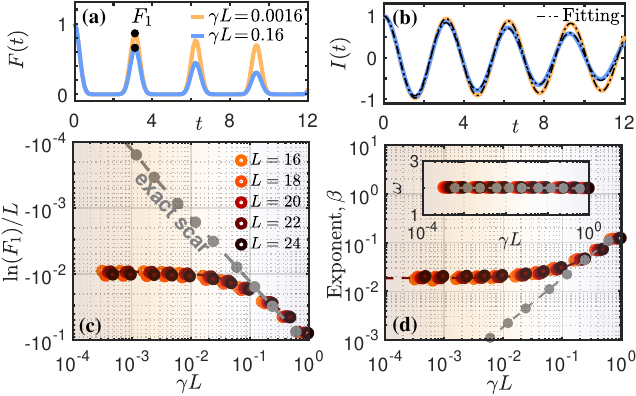}
\caption{The time evolution of the fidelity $F(t)$ (a) and the density imbalance $I(t)$ (b) for two dephasing rates at
system size $L = 16$. The black circles in Fig.~\ref{fig:CL_dyn} (a) represent the first peak of $F(t)$. The black dash-dotted lines in Fig.~\ref{fig:CL_dyn}(b) represent the fit of $\bar{I}(t)$ (see text). The fidelity density at the first revival, $\ln(F_1)/L$ (c) and the decay coefficient $\beta$ of density imbalance (d) for the CL model as a function of dephasing $\gamma$.  
For $\gamma\!<\!\gamma_\star^\mathrm{S}$, both the fidelity density and imbalance decay are well-converged in system size, suggesting the robustness of scar signatures over a finite range of $\gamma L$. By contrast, in the exact scar case with $J_{\mathrm{x},j}\!=\!0$, the imbalance decay is highly sensitive to $\gamma$ across the whole range. All data is for $J_\mathrm{h}\!=\!0.66$, $J_{\mathrm{x},j}\!\in\![0,0.2]$ drawn from a uniform distribution. Parameters $(\beta,\omega)$ are obtained by fitting the imbalance dynamics over the time period $[0,3.2]$. These results are obtained by TEBD algorithm based on ITensor library~\cite{ITensor}. 
}
\label{fig:CL_dyn}
\end{figure}

\emph{Dynamical signatures---}We demonstrate that weak dependence of $\bar{\gamma}_\star^\mathrm{S}$ on system size allows for QMBS signatures to persist in open-system dynamics defined by the evolution equation $\mathrm{d}|\rho(t)\rangle\!\rangle/\mathrm{d} t  \!=\! \mathcal{L} |\rho(t)\rangle\!\rangle$~\cite{breuer2002theory}. 
We integrate this equation starting from the initial state $|\rho_0\rangle\!\rangle\!=\!|\rho_{\Pi}\rangle\!\rangle$ and characterize subsequent time evolution by computing the fidelity and the density imbalance,
\begin{equation}\label{eq:fidelity_imbalance}
F(t) = \Big|\sum_k  \langle\!\langle \rho_0| \lambda_k\rangle\!\rangle^2 e^{\lambda_k t}\Big|, \quad
I(t) =\frac{1}{L}\sum_j\langle\!\langle 
\mathcal{Z}_j \rangle\!\rangle_{0}\langle\!\langle\mathcal{Z}_j\rangle\!\rangle_t,
\end{equation}
where $\langle\!\langle \mathcal{Z}_j\rangle\!\rangle_t=\! \langle\!\langle\rho(t)|\hat{\sigma}^{z}_{j}\!\otimes\!\hat{\sigma}^{z}_{j}|\rho(t)\rangle\!\rangle$.
As the fidelity dynamics directly corresponds to the Liouvillian spectrum, the variance of fidelity density at the first revival point $\ln(F_1)/L$, illustrated in Fig.~\ref{fig:CL_dyn}(a), can reveal the Liouvillian spectrum transition.
Furthermore, the imbalance dynamics takes the form of a damped oscillation [Fig.~\ref{fig:CL_dyn}(b)], which we fit using $\bar{I}(t) \!=\! \exp(-\beta t)\cos(\omega t)$  to extract the decay rate $\beta$ and revival frequency $\omega$. In practice, we found it sufficient to fit the data up to the first revival period~\cite{SM}.

In Fig.~\ref{fig:CL_dyn}(c), we plot $\ln(F_1)/L$ as a function of $\gamma L$, while the extracted imbalance parameters ($\omega$, $\beta$) are plotted in Fig.~\ref{fig:CL_dyn}(d).  Below the spectral transition point, both quantities are well-converged and weakly depend on $\gamma L$, which is indeed consistent with the spectral transition results in Fig.~\ref{fig:schematic} and perturbation theory estimate in Eq.~(\ref{eq:gammastarpert}).

The dissipative dynamics in Fig.~\ref{fig:CL_dyn} highlight the difference between the QMBS systems where the su(2) spectrum generating algebra is exact ($J_{\mathrm{x},j}\!=\!0$) versus only approximately obeyed ($J_{\mathrm{x},j}\!\neq\!0$) within the scar subspace. In Fig.~\ref{fig:CL_dyn}(c)-(d), we notice that the fidelity density and decay exponent $\beta$ in the exact scar case both display power-law dependence on $\gamma L$, in stark contrast with the plateau observed for $J_{\mathrm{x},j}\!\neq\!0$. Thus, only approximate QMBSs undergo the spectral transition at a non-zero $\gamma_\star^\mathrm{S}$. 
This difference in behavior is due to the exact energy spacing at $J_{\mathrm{x},j}\!=\!0$, which leads to degeneracies on the imaginary axis in Eq.~(\ref{eq:scarS0}). Hence, the Liouvillian eigenvalues of exact scars jump into the complex plane for any $\gamma\!\neq\!0$, as confirmed in the SM~\cite{SM}.
However, as the dephasing rate in Fig.~\ref{fig:CL_dyn}(c)-(d) increases beyond $\gamma_\star^\mathrm{S}$, the behaviors of approximate and exact scars become similar.

{\em The nature of the spectral transition.---}In generic chaotic models (without QMBSs), the LPTS breaking can be understood from the symmetry classification developed in Ref.~\cite{PhysRevX.13.031019}. 
In particular, the existence of a superoperator $\mathcal{T}_{-}$ that obeys $\mathcal{T}_{-} \mathcal{L} \mathcal{T}_{-}^{-1}=-\mathcal{L}$ is responsible for the dihedral symmetry of the Liouvillian spectrum and $\mathbb{P}\mathbb{T}$ breaking. At $\gamma \!=\! 0$, $|-(\lambda^{\prime})^\ast\rangle\!\rangle$ and $|\lambda^{\prime}\rangle\!\rangle$ represent the same state which lies on the imaginary axis.  Beyond a certain $\gamma$, 
$\mathbb{P}\mathbb{T} |\lambda^{\prime} \rangle\! \rangle \!=\! |\!-\!(\lambda^{\prime})^\ast\rangle\!\rangle$ continues to hold, but  $\lambda^\prime$ and $-(\lambda^\prime)^\ast$ are no longer the same. Thus, after this critical point, pairs of degenerate eigenvalues move symmetrically away
from the imaginary axis along the real axis, resulting in 
LPTS breaking~\cite{PhysRevLett.109.090404}.  

One may wonder if a similar, symmetry-based argument can be applied to the QMBS spectral transition observed in Fig.~\ref{fig:schematic} above. For the CL model, there indeed exists an analogous operator $\mathcal{T}_{-}$ (see~\cite{SM} for its explicit form). However, the spectral transition in the CL model only occurs in inversion symmetry sector $p\!=\!-1(+1)$, corresponding to odd (even) $N$. See $N\!=\!5$ case in Fig.~\ref{fig:schematic}. In contrast, the symmetry is already broken for infinitesimal $\gamma$ in $p\!=\!+1(-1)$ sector, also correlating with even (odd) $N$. Thus, $\mathcal{T}_{-}$ by itself does not fully account for the observations in Fig.~\ref{fig:schematic}. As a more transparent example, in End Matter, we perform the same analysis for the PXP model of Rydberg atom arrays, finding a similar transition in its Liouvillian spectrum as a function of dephasing. The PXP model manifestly lacks the $\mathcal{T}_{-}$ symmetry~\cite{SM}, while it displays a spectral transition similar to Fig.~\ref{fig:schematic} and robustness of QMBS dynamical signatures similar to Fig.~\ref{fig:CL_dyn}. Thus, we conclude the QMBS spectral transition is not due to symmetry breaking, but due to the smallness of matrix elements connecting the QMBS subspace with thermal bulk of the spectrum.

\emph{Conclusions---}By analyzing the Liouvillian spectra and dynamics in the presence of dephasing, we demonstrated that QMBSs can undergo a dissipative transition that resembles LPTS breaking. This behavior distinguishes approximate scar states from thermal states; the former exhibit a certain degree of robustness to a dephasing dissipation, reminiscent of their behavior under closed system dynamics when prepared in a thermal Gibbs state~\cite{Desaules2024FiniteTemperature}. We note that quantum noise is crucial for observing the QMBS Liouvillian spectral transition phenomena as the latter do not persist under a non-Hermitian Hamiltonian approximation~\cite{SM}. 

We supported our conclusions using the CL and PXP models realized in superconducting and Rydberg platforms. Notably, these two platforms allow for controlled dephasing noise~\cite{jurcevic2022effective,PhysRevLett.121.123603} and can be used test our predictions for the imbalance dynamics by tuning the dephasing rate. Further numerical evidence for the Liouvillian spectral transition in the 1D tilted Fermi-Hubbard model~\cite{Desaules2021Tilted}, and ladder models with Hilbert space fragmentation~\cite{10.21468/SciPostPhys.11.4.074} is presented in the SM~\cite{SM}. 
All of these models are described by the similar effective Hamiltonian $\hat H = \hat S_0 \oplus \hat T + \hat V$, suggesting broader applicability of our conclusions to models with this structure. Furthermore, the example of Hilbert space fragmentation suggests that the Liouvillian spectral transition may extend beyond QMBSs to other types of weak ergodicity breaking, warranting further study. 
Another open question concerns other types of dissipation beyond pure dephasing and their impact on the spectral transition.
Finally,  in closed systems, QMBSs have been proposed for quantum sensing~\cite{Dooley2021}, and it would be interesting to explore the sensitivity of such applications to dephasing noise.

\emph{Acknowledgments---}We thank Profs. Jiasen Jin, Zheng-Wei Zhou, Wen-Ge Wang, Xiong-Jun Liu, Dong-Lin Deng for helpful discussions. We also appreciate Dr. Jie Ren discussion on TEBD simulation for quantum open systems. 
This work was supported by the National Natural Science Foundation of China (Nos. 12375021 and 12247101), the Zhejiang Provincial Natural Science Foundation of China (No. LD25A050002), the National Key Research and Development Program of China (No. 2022YFA1404203) and the Fundamental Research Funds for the Central Universities (Grant No. lzujbky-2024-jdzx06).
Z.P. acknowledges support by the Leverhulme Trust Research Leadership Award RL-2019-015 and EPSRC Grant EP/Z533634/1. Statement of compliance with EPSRC policy framework on research data: This publication is theoretical work that does not require supporting research data. This research was supported in part by grant NSF PHY-2309135 to the Kavli Institute for Theoretical Physics (KITP).

\newpage 
\cleardoublepage 
\onecolumngrid
\vspace{+1cm}
\begin{center}
{\Large {\bf End Matter}} 
\end{center}
\vspace{+0.5cm}

\twocolumngrid

\emph{Transition in the PXP model---}Another paradigmatic model of QMBSs is the PXP model describing Rydberg atom arrays~\cite{bernien2017probing}.
The PXP Hamiltonian is~\cite{FendleySachdev,Lesanovsky2012} 
\begin{equation}\label{eq:PXP}
 \hat{H}=\hat{H}_{\mathrm{PXP}}= \Omega \sum^L_j\hat{P}_{j-1}\hat{\sigma}^x_{j}\hat{P}_{j+1}, \quad \hat{P}_{j}\!=\!(1-\hat{\sigma}^z_j)/2,  
\end{equation}
where $\hat{\sigma}_j^{x,y,z}$ are the Pauli matrices and we set the Rabi frequency to $\Omega\!=\!1$. We impose periodic boundary conditions, with $L+1 \equiv 1$. The PXP model features revival dynamics when the system is quenched from the initial state $|\mathbb{Z}_2\rangle\!\equiv\! |{\uparrow}{\downarrow}{\uparrow}{\downarrow}{\cdots}\rangle$~\cite{turner2018weak,Turner2018PRB}. Below we demonstrate that the PXP model with pure dephasing exhibits similar behavior to the CL model.

\begin{figure}[htb]
\centering
\includegraphics[width=1\linewidth]{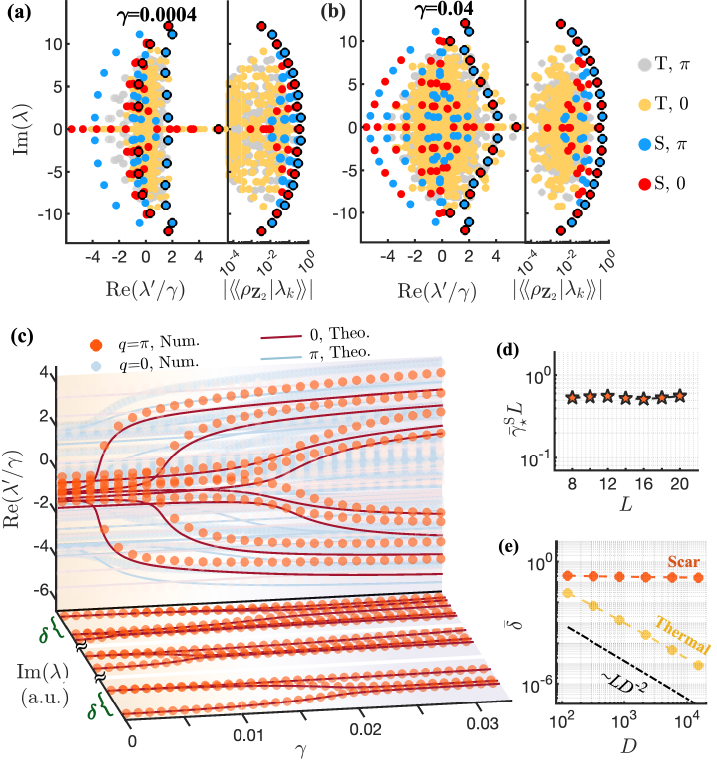}
\caption{ 
(a)-(b) The Liouvillian eigenspectra of the PXP model for dephasing strengths $\gamma\!=\!0.0004$ and $\gamma\!=\!0.04$. Red and blue dots represent scar (S) eigenvalues, while yellow and gray ones represent thermal (T) eigenvalues. These are further classified by momenta $q\!=\!0$ and $\pi$ under translations. 
The right panels in (a)-(b) show the overlaps of Liouvillian eigenmodes with $|\rho_\mathrm{Z_2}\rangle\!\rangle$, with black open circles highlighting local maxima.
(c) The evolution of the Liouvillian spectrum, $\mathrm{Im}(\lambda)$ and $\mathrm{Re}(\lambda^{\prime}/\gamma)$, as the dephasing rate $\gamma$ is varied. 
For improved visualization, the imaginary values in panel (c) have been shifted within a small window. 
The curves are results of perturbation theory, following Eq.~(\ref{eq:D_l_main})~\cite{SM}.
(d) The average spectral transition point $\bar{\gamma}_{\star}^S L$ as a function of system size $L$ obtained in perturbation theory.
(e) The average scar eigenvalue spacing $\bar{\delta}$, at zero dephasing, plotted as a function of the corresponding sector dimension $D$. 
Data in panels (a)-(c) is obtained by exact diagonalization of the PXP model for a system size of $L\!=\!10$ with periodic boundary conditions.
}
\label{fig:pxp}
\end{figure}

\emph{PXP spectral transition---}Figures~\ref{fig:pxp}(a)-(b) illustrate the Liouvillian spectra for the PXP model for two dephasing rates. These results are qualitatively similar to the CL model in Fig.~\ref{fig:schematic}:  for weak dephasing $\gamma\!=\!0.0004$, the QMBS eigenvalues cluster around $\mathrm{Re}\big(\lambda^\prime\big)\!=\!0$, while at larger dephasing, $\gamma\!=\!0.04$, these eigenvalues pair-wisely shift along the real axis. The main difference with Fig.~\ref{fig:schematic} is that the PXP model does not possess the $\mathcal{T}_{-}$ superoperator obeying $\mathcal{T}_{-} \mathcal{L} \mathcal{T}_{-}^{-1}=-\mathcal{L}$~\cite{PhysRevX.13.031019}.
Hence, its Liouvillian spectrum, even at small $\gamma\!=\!0.0004$, is not confined to the real- and imaginary-axes and LPTS is already broken at that point, even for QMBS eigenvalues. However, the scar eigenvalues cluster around the axes at small $\gamma$ and exhibit similar movement under dephasing  as in the CL model. We next show that scar eigenvalues of the PXP Liouvillian indeed undergo sharp jumps upon sweeping the dephasing rate, similar to the CL model.

Figure~\ref{fig:pxp}(c) illustrates the imaginary part of eigenvalues and real part of normalized eigenvalues $\lambda^\prime/\gamma$ as a function of $\gamma$. Several eigenvalues with momentum  $q=0$ remain stationary at weak dephasing, while some pairs of eigenvalues abruptly bifurcate into two branches at specific dephasing rates $\gamma_\star^{(1,2\ldots)}$. 
In contrast, no such bifurcations occur in the momentum sector $q=\pi$, mirroring the behavior of the CL model with respect to inversion symmetry in Fig.~\ref{fig:CL_lpts}.  

Upon closer inspection of Fig.~\ref{fig:pxp}(c), we see that the process of eigenvalue bifurcation is smoother compared to the CL model in Fig.~\ref{fig:CL_lpts}(a). This can be understood in the framework of perturbation theory $
\mathcal{L}_\mathcal{S} \approx \mathcal{S}_0 + \mathcal{E}+ \gamma_{\mathrm{eff}}\mathcal{D}^\prime
$, suitably extended to the PXP model where the QMBS Hamiltonian $\hat S_0$ and the self-energy $\hat{\Sigma}$ are expressed in a dimerized spin-1 basis~\cite{Omiya2022,Omiya2023b}. The dissipator $\mathcal{D}^\prime=\sum_j \hat{\sigma}^z_{j}\otimes\hat{\sigma}^z_j$ splits into a linear and quadratic term upon mapping to the dimerized spin-1 basis -- see SM~\cite{SM}. When only the linear term is considered, the spectrum exhibits $\mathbb{P}\mathbb{T}$ symmetry and spectral transition. 
However, when both linear and quadratic term are included, the $\mathcal{T}_{-}$ symmetry is no longer present but we capture the rounded profile of the spectral evolution curves. The perturbative estimate, including both linear and quadratic terms, is plotted in Fig.~\ref{fig:pxp}(c) and shows good agreement with the numerics.

The typical Liouvillian spectral
transition point $\bar{\gamma}_\star^\mathrm{S}$ for the PXP scar eigenmodes can be estimated from the spacings $\bar{\delta}$ of nondegenerate scar eigenvalues at $\gamma\!=\!0$ according to the same expression, $\bar{\gamma}_\star^\mathrm{S} \!\sim\! \bar{\delta}\;\Vert\mathcal{D}^\prime\Vert^{-1}$, used previously for the CL model in the main text. As confirmed in Fig.~\ref{fig:pxp}(e),  $\bar{\delta}\!\sim\! LD^{-2}$ for thermal eigenvalues, while the scar $\bar{\delta}$ is nearly independent of  $D$. Thus, the typical critical point for PXP scar eigenmodes also scales as $\bar{\gamma}_\star^\mathrm{S} \!\sim\! L^{-1}$, consistent with the approximate scars in the CL model. 
This is also confirmed by the perturbation theory result in Fig.~\ref{fig:pxp}(d), where we keep only the linear term of the dissipator, allowing to unambiguously identify the spectral transition points and their mean value.

\begin{figure}[t]
\centering
\includegraphics[width=1\linewidth]{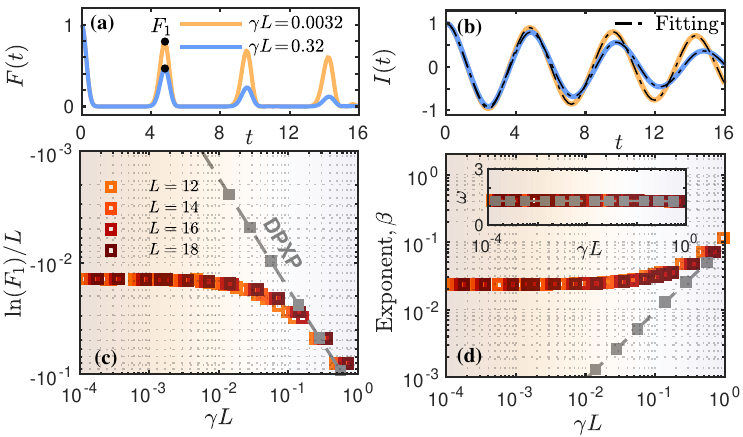}
\caption{The dynamics of the global fidelity $F(t)$ (a) and imbalance $I(t)$ (b) in the PXP model for two dephasing rates at system size $L=16$. The black circles in (a) denote the first fidelity revival peak, $F_1$, while the black dash-dotted lines in (b) are fits of the imbalance dynamics to $\bar{I}(t)$.
(c) The global fidelity density at the first revival, $\ln(F_1)/L$  and (d) the decay coefficient $\beta$ of density imbalance  for the PXP model as a function of dephasing $\gamma$. 
Parameters $(\beta,\omega)$ are obtained by fitting the imbalance dynamics over the time period $[0,5]$. These results are obtained by exact diagonalization.
}
\label{fig:pxp_dyn}
\end{figure}

Finally, similar to the CL model in the main text, we check for the robustness of scar signatures in the dynamics by performing time evolution generated by the Lindblad equation for the PXP model with pure dephasing. Here we take as initial state $|\rho_{\mathrm{Z}_2}\rangle\!\rangle \!=\! |\mathbb{Z}_2\rangle \!\otimes\! |\mathbb{Z}_2\rangle$ and the resulting fidelity density at the first revival, $\ln F_1/L$, as well as the imbalance decay $\beta$ and frequency $\omega$, are shown in Fig.~\ref{fig:pxp_dyn}. These results mirror those for the CL model, in particular $\beta$ has a similarly weak dependence on $\gamma L$ as long as  $\gamma\!<\!\gamma_\star^\mathrm{S}$. 

To confirm that these results are due to approximate scars, we consider a deformation of the PXP model~\cite{Choi2019,Khemani2019,Bull2020}  for which QMBSs assume a nearly exact form:
\begin{equation}\label{eq:DPXP}
\hat{H}_{\mathrm{DPXP}}=\hat{H}_{\mathrm{PXP}} + \sum^L_j\sum^{L/2}_{d=2}J_\mathrm{d} \hat{P}_{j-1}\hat{\sigma}^x_{j}\hat{P}_{j+1}\big(\hat{\sigma}^z_{j-d}+\hat{\sigma}^z_{j+d}\big).    
\end{equation}
 At the optimal coupling strength $J_\mathrm{d}\!=\!0.051(\phi^{d-1}\!+\!\phi^{1-d})^{-2}$, with $\phi=(1+\sqrt{5})/2$, the scar subspace contains $L+1$ eigenstates that are almost perfectly decoupled from the thermal bulk of the spectrum~\cite{Choi2019}.
 This DPXP model indeed behaves similarly to the exact scars in the CL model with $J_{\mathrm{x},j}\!=\!0$, e.g., the imbalance decay coefficient $\beta$ exhibits a similar power-law dependence on $\gamma$ in both cases. The similarity extends to the Liouvillian spectra, as the DPXP model also exhibits a spectral transition for infinitesimal $\gamma$~\cite{SM}.

\newpage 
\cleardoublepage 

\beginsupplement

\newpage

\pagenumbering{arabic} 

\onecolumngrid
\begin{center}
{\Large {\bf Supplementary Materials for}} \\ 
\vspace{0.25cm}
{\large {\bf ``Liouvillian Spectral Transition in Noisy Quantum Many-Body Scars''}}\\
\vspace{+0.5cm}
Jin-Lou Ma, Zexian Guo, Yu Gao, Zlatko Papi\'c, and Lei Ying
\end{center}
\vspace{+0.5cm}

\section{A. Liouvillian perturbation theory}\label{SSec:theory}

We focus on a class of quantum many-body scar (QMBS) systems whose Hamiltonian can be written as 
\begin{eqnarray}\label{eq:scarham}
 \hat{H} = \hat{S}_0 \oplus \hat{T} + \hat{V},   
\end{eqnarray}
where $\hat{S}_0$ and $\hat{T}$ denote the Hamiltonian of scar and thermal subsystems, respectively, while $\hat{V}$ represents the coupling between them. This general Hamiltonian includes a large class of QMBS models~\cite{serbyn2021quantum}. 

The effect of the thermal subspace $\hat T$ can be incorporated by defining the effective Hamiltonian for the scar subspace as $\hat{S} = \hat{S}_0+\hat{\Sigma}$, where $\hat{\Sigma}$ is the self-energy from the thermal subspace. If we neglect the coupling between the bath and the thermal subsystem, the resulting Liouvillian can be written as
\begin{equation}\label{eq:LeffSM}
\mathcal{S} \approx \mathcal{S}_{0} + \mathcal{E}+\gamma_{\mathrm{eff}}\mathcal{D}.
\end{equation}
Here, the Liouvillian of the exact scar subspace is denoted by $\mathcal{S}_{0}$, while the self-energy and the dissipator are
\begin{equation}\label{eq:epsilon_D}
    \mathcal{E}={-i\big(\hat{\Sigma} \otimes \hat{I}-\hat{I} \otimes \hat{\Sigma}\big)},
    \quad\quad
    \mathcal{D}=\mathcal{D}^\prime-\sum_j\hat{I}_{j}\otimes\hat{I}_j,
\quad\quad
\mathcal{D}^\prime=\sum_j\hat{\sigma}_j^z\otimes\hat{\sigma}_j^z.
\end{equation}
The self-energy can be  obtained via the non-equilibrium Green's function as shown below. When  $\mathcal{E}$ is zero, the problem reduces to the exact scar case. The effective dephasing strength, $\gamma_\mathrm{eff}$, in Eq.~(\ref{eq:LeffSM}) is model-dependent.

 The dephasing-free Liouvillian  of the  exact scar subspace is given by
\begin{equation}\label{seq:s0}
\mathcal{S}_{0}={-iJ\Big( \hat S_0 \otimes \hat{I}-\hat{I} \otimes \hat S_0 \Big)},
\end{equation}
with the corresponding eigenmodes and eigenvalues
\begin{equation}\label{seq:0th_eigenvalue}
    |\lambda_{(l,s)}^{(0)}\rangle\!\rangle = |E_{l+s}^\mathrm{S}\rangle \otimes |E_s^\mathrm{S}\rangle, 
    \quad\quad\quad\quad
    \lambda_{l}^{(0)}=i(E_{l+s}^\mathrm{S}-E_{s}^\mathrm{S})=2ilJ.
\end{equation}
The eigenmodes are labeled using indices $l=-N,-N+1,\cdots,N$ and $s = 0, 1,2,\cdots, N-|l|$.  $E_s^{\mathrm{S}}$ and $|E_s^{\mathrm{S}}\rangle$ is the scar eigenvalue and eigensate for the Hamitonian of $\hat{S}_0$, respectively.

\begin{figure}[bth]
\includegraphics[width=0.6\linewidth]{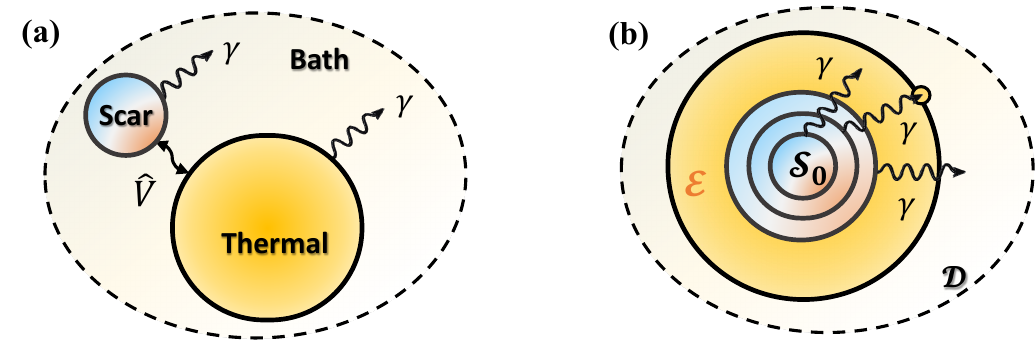}
\caption{
(a) Schematic of a general open quantum many-body system with approximate scars. The scar and thermal subsystems lose their quantum coherence to the bath with the dephasing rate $\gamma$, while interacting via the coupling strength $V$. 
(b) The Liouvillian spectral transition originates from the competition between self-energy, $\mathcal{E}$, and dissipation $\mathcal{D}$. A set of eigenmodes in Liouville space, denoted by blue concentric circles, are surrounded by the self-energy shell $\mathcal{E}$, due to the  thermal subspace. The latter suppresses those dephasing processes (wavy arrows) that do not exceed the boundary of the self-energy shell (yellow circle). 
}
\label{fig:selfenergy}
\end{figure}

\subsection{A1. The cases with $\mathcal{T}_{-}$ symmetry: CL model}

For concreteness, here we focus on the Creutz-ladder (CL) model introduced in the main text, where $\hat S_0$ takes the form of a free paramagnet Hamiltonian, to be specified below. We define the basis states  $|\!\!\Uparrow\rangle \equiv \big|\substack {\uparrow\\ \downarrow} \big\rangle, |\!\!\Downarrow\rangle \equiv \big|\substack {\downarrow\\ \uparrow} \big\rangle$~\cite{zhang2023many,Dong2023}, for which we can define the standard Pauli matrices $\hat{\mathbb{S}}^{x,y,z}$, e.g., $\hat{\mathbb{S}}^z \equiv |{\Uparrow}\rangle\langle\Downarrow\!\!| - |\!\!\Downarrow\rangle\langle{\Uparrow}|$ and similarly for $x$, $y$. With the help of these two basis vectors and Pauli matrices, we can define the QMBS Hamiltonian using a spin-1/2 operator $\hat S_0$ whose eigenstates  are the Dicke states of a free collective spin: 
\begin{equation}\label{seq:s0_E}
     \hat{S}_0\!=\!\sum_{j=1}^N(|\!\!\Uparrow \rangle\langle\Downarrow\!\!|_j+|\!\!\Downarrow\rangle\langle\Uparrow \!\!|_j)\equiv\!\sum_{j=1}^N\hat{\mathbb{S}}^x_j, 
     \quad\quad\quad
    |E_{s}^\mathrm{S}\rangle=\binom{N}{s}^{-\frac{1}{2}} \sum_{i_1 \neq \cdots \neq i_{s}}  \bigotimes_{j=1}^N\left\{\begin{array}{ll}
|+\rangle_j & j \in\left\{i_1, \ldots, i_{s}\right\} \\
|-\rangle_j & \text { otherwise }
\end{array} ,\right.
\end{equation}
where $|\pm \rangle_j = \big(|{\Uparrow}\rangle_j \pm  |{\Downarrow}\rangle_j\big)/{\sqrt{2}}$ are the spin bases states pointing along the $x$-direction, and $n_S=0,1,\ldots,N$ is the number of $|+\rangle$ appearing in the tensor product. In Eq.~(\ref{seq:s0_E}),  we have written the eigenstates belonging to the maximal spin representation.

To explain the structure of Liouvillian QMBS spectra, we set up a degenerate perturbation theory framework.  We consider the Liouvillian of the exact scar subspace, Eqs.~(\ref{seq:s0}) and~(\ref{seq:s0_E}), with eigenvalues
\begin{equation}\label{seq:eigenvalue0}
\lambda_{l}^{(0)}=i(E_{l+s}^\mathrm{S}-E_{s}^\mathrm{S}) = 2ilJ,
\end{equation} 
where $l = -N,-N+1,\cdots,N-1,N$. 
The eigenvalues $\lambda_{l}$ are equally spaced along the imaginary axis, with the real part equal to zero.
 
In the regime $\Vert\gamma_{\mathrm{eff}}\mathcal{D}^\prime+\mathcal{E}\Vert\ll \Vert\mathcal{S}_0\Vert$, we can use the Liouvillian perturbation theory, in which the traceless Liouvillian eigenvalues for the $l$th layer are approximately given by $\lambda^\prime_{l}\approx \lambda^{(0)}_{l} + \lambda^{(1)}_{l}$.
Here, the zeroth-order eigenvalues are given by the eigenvalues of the unperturbed system in Eq.~(\ref{seq:eigenvalue0}), while the first-order corrections to the eigenvalues, $\lambda_{l}^{(1)}$, are obtained by diagonalizing the projection of $\gamma_{\mathrm{eff}} \mathcal{D}^\prime + \mathcal{E}$ to the degenerate subspace of the $l$th layer, labeled by states $s = 0,1,2,\cdots,N-|l|$.

First, we consider the dephasing term. Recall that $|\pm \rangle_j = \big(|0\rangle_j \pm  |1\rangle_j\big)/{\sqrt{2}}$ and hence we have $\hat{\mathbb{S}}_j^z|\pm\rangle_j = |\mp\rangle_j$. The projection of $\mathcal{D}^\prime$ to the $l$th layer, $\mathcal{D}_l^\prime$, couples only $|\lambda_{(l,s)}^{(0)}\rangle\!\rangle$ and $|\lambda_{(l,s\pm1)}^{(0)}\rangle\!\rangle$, hence it can be written as 
\begin{equation}\label{eq:D_l}
\begin{aligned}
   \mathcal{D}_l^\prime=  \sum_{s=0}^{N-|l|-1} d_{(l,s)} \Big[\ \big|\lambda_{(l,s)}^{(0)}\big\rangle\!\big\rangle  \big\langle\!\big\langle\lambda_{(l,s+1)}^{(0)}\big|+\mathrm{h.c.}  \Big],
\end{aligned}
\end{equation}
where $d_{(l,s)}$ represents the coupling coefficient between the $s$-th and $(s+1)$-th degenerate eigenstates in the $l$-momentum layer.  Using the translation symmetry of $\sum_j \hat{\mathbb{S}}^z_j$, we have 
\begin{equation}
\begin{aligned}
d_{(l,s)}&=\sum_{j=1}^N\Big[\langle E_{s}^\mathrm{S}|\hat{\mathbb{S}}^z_j|E_{s+1}^\mathrm{S}\rangle \langle E_{s+|l|}^\mathrm{S}|\hat{\mathbb{S}}^z_j|E_{s+|l|+1}^\mathrm{S}\rangle\Big]
=N  \Big[\langle E_{s}^\mathrm{S}|\hat{\mathbb{S}}^z_1|E_{s+1}^\mathrm{S}\rangle \langle E_{s+|l|}^\mathrm{S}|\hat{\mathbb{S}}^z_1|E_{s+|l|+1}^\mathrm{S}\rangle\Big].
\end{aligned}
\end{equation}
Furthermore, one can show that
\begin{equation}
    \langle E_{s}^\mathrm{S}|\hat{\mathbb{S}}^z_1|E_{s+1}^\mathrm{S}\rangle = \frac{\binom{N-1}{s}}{\sqrt{\binom{N}{s}\binom{N}{s+1}}} = \frac{1}{N}\sqrt{(N-s)(s+1)}.
\end{equation}
Therefore, we have
\begin{equation}
    d_{(l,s)} =\frac{1}{N}\sqrt{\left(s+1\right)\left(N-s\right)\left(|l|+s+1\right)\left(N-|l|-s\right)}.
\end{equation}

Next, we consider the effect of self-energy.  The Hamiltonian in Eq.~(\ref{eq:scarham}) is generally non-integrable and impossible to solve exactly. Hence  we employ the non-equilibrium Green’s function approach to model the effect of thermal bulk on the scar subspace, see Ref.~\cite{Guo2023Origin} for details. This will allow us to obtain analytical insights into the interplay of scarring with the dephasing dissipation. The Green’s function of the scar subspace is expressed as
\begin{equation}
\hat{G}_\mathrm{S}(E)=\left[\left(E+i 0^{+}\right) \hat{I}-\hat{S}_0-\hat{\Sigma}(E)\right]^{-1} .
\label{eq:Green_Func}
\end{equation}
The self-energy is obtained by solving the self-consistent Dyson's equation:
\begin{equation}
\hat{\Sigma}(E)=\hat{\Gamma}^{\dagger}\big[E \hat{I}-\hat{S}_0-\hat{\Sigma}(E)\big]^{-1} \hat{\Gamma},
\label{eq:self_energy}
\end{equation}
where $\hat{\Sigma}(E)=\sum_{p_{\mathrm{S}}} \hat{\sigma}_{p_{\mathrm{S}}}(E)$ represents the self-energy matrix, with $\hat{\sigma}_{p_{\mathrm{S}}}(E)=\hat{\Gamma}_{p_{\mathrm{S}}}^{\dagger} \hat{G}_{\mathrm{S}}(E) \hat{\Gamma}_{p_{\mathrm{S}}} \text { and } \hat{\Gamma}_{p_{\mathrm{S}}}=\sqrt{\gamma_{p_{\mathrm{S}}}}\left|p_{\mathrm{S}}\right\rangle\left\langle p_{\mathrm{S}}\right|$. Here, $|p_{\mathrm{S}}\rangle$ denote the product states in the scar subspace, and $\gamma_{p_{\mathrm{S}}} $ represents the interaction strength between the $|p_{\mathrm{S}}\rangle$ and the thermal subspace. For the QMBS model we consider here, Eq.~(\ref{eq:scarham}), $\hat S_0$ corresponds to a hypercube and $|p_\mathrm{S}\rangle$ represent some of the computational basis states, while the remaining basis states $|p_\mathrm{T}\rangle$ span the thermal subspace. For example, in the CL model, the vertices of the hypercube are product states of dimers, $\{|p_\mathrm{S}\rangle\} = \{ \otimes_{j=1}^{N=L/2}|\mathbb{D}_j\rangle \}$, where each $|\mathbb{D}_j\rangle$ can be either $|\substack {\downarrow\\ \uparrow} \rangle_j$ or $|\substack {\uparrow\\ \downarrow} \rangle_j$. The thermal states $|p_\mathrm{T}\rangle$ then involve the remaining basis states that contain $|\substack {\uparrow\\ \uparrow} \rangle_j$ and $|\substack {\downarrow\\ \downarrow} \rangle_j$ states of dimers. 

For the CL model, the interaction strength $\gamma_{p_{\mathrm{S}}} $ is proportional to the hopping matrix element  $h_{p_{\mathrm{S}}}=\sum_{p_{\mathrm{T}}}\left\langle p_{\mathrm{T}}\right| \hat{V}\left|p_{\mathrm{S}}\right\rangle$  between the scar $|p_{\mathrm{S}}\rangle$ and the thermal subspace $|p_{\mathrm{T}}\rangle$~\cite{Guo2023Origin}. The expression for $\hat \Gamma$ can be derived analytically: 
\begin{equation}\label{eq:gammaCL}
\begin{aligned}
\hat{\Gamma}_{\mathrm{}} =\sum_{p_{\mathrm{S}}}\sqrt{\gamma_{p_{\mathrm{S}}}}|p_{\mathrm{S}}\rangle\langle p_{\mathrm{S}}| =\sum_{p_{\mathrm{S}}}|p_{\mathrm{S}}\rangle\langle p_{\mathrm{S}}| \sqrt{\sum_{p_{\mathrm{T}}}\left\langle p_{\mathrm{T}}\right| \hat{V}\left|p_{\mathrm{S}}\right\rangle} =\sum_{p_{\mathrm{S}}}|p_{\mathrm{S}}\rangle\langle p_{\mathrm{S}}|\sqrt{\sum_{j=1}^{N-1}J_{\mathrm{x},j}\hat{P}^{j,j+1}_{p_{\mathrm{S}}}},
\end{aligned}
\end{equation}
where $\hat{P}_{p_\mathrm{S}}^{j,j+1} = \langle \mathbb{D}_j|\mathbb{D}_{j+1}\rangle$ represents the overlap between two neighboring dimers in $|p_{\mathrm{S}}\rangle$. In the following, we use this expression for $\hat\Gamma$ to evaluate the self-energy from the Dyson equation.

We assume that the thermal subspace is much larger than the scar subspace, and the energy (initially in scar subspace) that leaks into the thermal subspace cannot return. Thus, we have that the off-diagonal term, $\langle E_{s}^\mathrm{S}|\hat{\Sigma}|E_{s^\prime}^\mathrm{S}\rangle$ with $s\neq s^\prime$, is much smaller than the diagonal term, $\langle E_{s}^\mathrm{S}|\hat{\Sigma}|E_{s}^\mathrm{S}\rangle$. $\langle E_{s}^\mathrm{S}|\hat{\Sigma}|E_{s}^\mathrm{S}\rangle$ is calculated by Eq.~(\ref{eq:Green_Func})-(\ref{eq:self_energy}). Here, we use approximation as $\langle E_{s}^\mathrm{S}|\hat{\Sigma}|E_{s}^\mathrm{S}\rangle\approx\langle E_{s}^\mathrm{S}|\hat{\Sigma}(E=(-N+2s)J)|E_{s}^\mathrm{S}\rangle$ and obtained the results by solving the self-consistence equation.  The latter diagonal element can be written as 
\begin{equation}\label{eq:epsilon_l}
\begin{aligned}
\mathcal{E}_l &= i \times \mathrm{Diag}\Big[\Sigma_{(l,0)},\ \Sigma_{(l,1)},\ \cdots, \Sigma_{(l,N-|l|)}\Big], \quad \Sigma_{(l,s)}=\langle E_{s}^\mathrm{S}|\hat{\Sigma}|E_{s}^\mathrm{S}\rangle-\langle E_{s+l}^\mathrm{S}|\hat{\Sigma}|E_{s+l}^\mathrm{S}\rangle.
\end{aligned}
\end{equation}

Finally, we obtain the first-order correction to the eigenvalues  $\lambda_{l}^{(1)}$ by diagonalizing the matrix $\big(\gamma_{\mathrm{eff}}\mathcal{D}_l^\prime+\mathcal{E}_l\big)$ defined in Eq.~(\ref{eq:D_l}) and (\ref{eq:epsilon_l}). As an example, consider the CL model with $N = 5$ and the layers $l =-N,-N+2,\cdots,N$. For example, layer $l = 3$ holds triple degeneracy, while layer $l = 1$ holds five-fold degeneracy. The corresponding first-order eigenvalues are given by:
\begin{itemize}
  \item
For the exact scar case $\mathcal{E} = 0$, we have $\lambda^{(1)}_{l=3} = 0, \pm \frac{4\sqrt{5}}{5}\gamma_{\mathrm{eff}}$ and $\lambda^{(1)}_{l=1} = 0, \pm \frac{2\sqrt{10}}{5}\gamma_{\mathrm{eff}}, \pm \frac{2\sqrt{46}}{5}\gamma_{\mathrm{eff}}$. The Liouvillian eigenvalues shift along the real axis whenever the dissipative factor $\gamma_{\mathrm{eff}}$ is non-zero;

\item 
For the approximate scar case $\mathcal{E}\neq 0$, the superoperator of the perturbation is more complicated and in the layers $l=-N,-N+2,\cdots,N$ takes the form
\begin{equation}
\begin{aligned}
\begin{split}
\quad\quad\mathcal{E}_5+\gamma_{\mathrm{eff}}\mathcal{D}^\prime_5&=(i\Sigma_{(5,0)}),\quad\quad
\mathcal{E}_3+\gamma_{\mathrm{eff}}\mathcal{D}^\prime_3=\left(\begin{array}{ccc}
i\Sigma_{(3,0)}& \sqrt{8/5}\gamma_{\mathrm{eff}} & 0  \\
\sqrt{8/5}\gamma_{\mathrm{eff}} & i\Sigma_{(3,1)} & \sqrt{8/5}\gamma_{\mathrm{eff}}\\
0 & \sqrt{8/5}\gamma_{\mathrm{eff}} & i\Sigma_{(3,0)} \\
\end{array}\right),  \\ \\
\mathcal{E}_1+\gamma_{\mathrm{eff}}\mathcal{D}^\prime_1&=\left(\begin{array}{ccccc}
i\Sigma_{(1,0)} & \sqrt{8/5}\gamma_{\mathrm{eff}} & 0 & 0 & 0 \\
\sqrt{8/5}\gamma_{\mathrm{eff}} & i\Sigma_{(1,1)} & \sqrt{72/25}\gamma_{\mathrm{eff}} & 0 & 0\\
0 & \sqrt{72/25}\gamma_{\mathrm{eff}} & i\Sigma_{(1,2)} & \sqrt{72/25}\gamma_{\mathrm{eff}} & 0\\
0 & 0 & \sqrt{72/25}\gamma_{\mathrm{eff}} & i\Sigma_{(1,1)} & \sqrt{8/5}\gamma_{\mathrm{eff}}\\
0 & 0 & 0 & \sqrt{8/5}\gamma_{\mathrm{eff}} & i\Sigma_{(1,0)}
\end{array}\right).
\end{split}
\end{aligned}
\end{equation}
    
Here we use $\langle E_{s}^\mathrm{S}|\hat{\Sigma}|E_{s}^\mathrm{S}\rangle = -\langle E_{N-s}^\mathrm{S}|\hat{\Sigma}|E_{N-s}^\mathrm{S}\rangle$ because the CL model holds inversion symmetry. Analogous expressions for the layers $l=-N+1,-N+3,\cdots,N-1$ are:
\begin{equation}
\begin{split}
\mathcal{E}_4+\gamma_{\mathrm{eff}}\mathcal{D}^\prime_4=\left(\begin{array}{ccc}
i\Sigma_{(4,0)}& \gamma_{\mathrm{eff}}  \\
\gamma_{\mathrm{eff}} & i\Sigma_{(4,0)}\\
\end{array}\right),  \quad
\mathcal{E}_2+\gamma_{\mathrm{eff}}\mathcal{D}^\prime_2=\left(\begin{array}{ccccc}
i\Sigma_{(2,0)} & \sqrt{9/5}\gamma_{\mathrm{eff}} & 0 & 0 \\
\sqrt{9/5}\gamma_{\mathrm{eff}} & i\Sigma_{(2,1)} & 8/5\gamma_{\mathrm{eff}} & 0\\
0 & 8/5\gamma_{\mathrm{eff}} & i\Sigma_{(2,1)} & \sqrt{9/5}\gamma_{\mathrm{eff}}\\
0 & 0 & \sqrt{9/5}\gamma_{\mathrm{eff}} & i\Sigma_{(2,0)} 
\end{array}\right).
\end{split}
\end{equation}

Then, we have the first-order eigenvalues of the perturbation $\big(\gamma_\mathrm{eff}\mathcal{D}_l^\prime+\mathcal{E}_l\big)$ as
\begin{equation}
\begin{split}
   \lambda_{l=5}^{(1)} =&\ i\Sigma_{(5,0)}, \quad\quad\quad\quad 
\lambda_{l=3}^{(1)} = i\Sigma_{(3,0)},\ i\Delta_{3,+}\pm\sqrt{\left(i\Delta_{3,-}\right)^2+16\gamma^2/5},  \\
        \lambda_{l=1}^{(1)} =& \    i\Delta_{1,+}\pm\sqrt{\left(i\Delta_{1,-}\right)^2+32\gamma^2/5} , \ \Big\{\ \lambda\ \Big|\ \lambda^3-i\Theta_2\lambda^2-\Theta_1\lambda+i\Theta_0=0\Big\},
\end{split}
\end{equation}
and
\begin{equation}
\begin{split}
   \lambda_{l=4}^{(1)} = i\Sigma_{(4,0)}\pm2\gamma,
   \quad\quad  \lambda_{l=2}^{(1)} =&i\Delta_{2,+}+8\gamma/5\pm\sqrt{\left(i\Delta_{2,-}+8\gamma/5\right)^2+36\gamma^2/5},\\&i\Delta_{2,+}-8\gamma/5\pm\sqrt{\left(i\Delta_{2,-}-8\gamma/5\right)^2+36\gamma^2/5},
\end{split}
\end{equation}
where $\Delta_{l,\pm}=\left(\Sigma_{(l,1)} \pm \Sigma_{(l,0)}\right)/2$ and we used $\gamma_{\mathrm{eff}}=2\gamma$ due to two neighboring sites (dimer) are used to define a `spin-$1/2$' in the scar subspace (i.e., $N=L/2$). Besides, we set the coefficients $\Theta_2 = \sum_s\Sigma_{(1,s)}$, $\Theta_1 = \sum_{s>s^\prime}\left(\Sigma_{(1,s)}\Sigma_{(1,s^\prime)}\right)+736\gamma^2/25$, and $\Theta_0 = \Sigma_{(1,0)}\Sigma_{(1,1)}\Sigma_{(1,2)}+\left(576\Sigma_{(1,0)}+32\Sigma_{(1,2)}\right)\gamma^2/5$. 
\end{itemize}

Furthermore, for $J_{\mathrm{x},j} = 0.1$, the parameters $ \langle E_{0}^\mathrm{S}|\hat{\Sigma}|E_{0}^\mathrm{S}\rangle\approx 0.064$, $\langle E_{1}^\mathrm{S}|\hat{\Sigma}|E_{1}^\mathrm{S}\rangle \approx 0.028$, $\langle E_{2}^\mathrm{S}|\hat{\Sigma}|E_{2}^\mathrm{S}\rangle \approx 0.0061$ can be calculated by using the non-equilibrium Green's function method in Eq.~(\ref{eq:self_energy}).  The first-order eigenvalue as the dephasing rate increases under these parameters were shown in the main text, illustrating the spectral transition point approximately aligns with the numerical results.

For the case $|l|=N$, we simply use non-degenerate perturbation theory and obtain  $\lambda^\prime_{N}\approx \lambda^{(0)}_{N} + \lambda^{(1)}_{N}$, with the first-order eigenvalue given by
\begin{equation}
\begin{aligned}
    \lambda^{(1)}_{|l|=N} = \pm\Big\langle\!\!\Big\langle\lambda_{(N,0)}^{(0)}\Big|\left(\gamma_\mathrm{eff}\mathcal{D}^\prime+\mathcal{E}\right)\Big|\lambda_{(N,0)}^{(0)}\Big\rangle\!\!\Big\rangle &= \pm\gamma_\mathrm{eff}\sum_{j}^N\big(\langle E_{N}^\mathrm{S}|\hat{\mathbb{S}}_j^z|E_{N}^\mathrm{S}\rangle\times
    \langle E_{0}^\mathrm{S}|\hat{\mathbb{S}}_j^z|E_{0}^\mathrm{S}\rangle\big)\pm i\left(\langle E_{N}^\mathrm{S}|\hat{\Sigma}|E_{N}^\mathrm{S}\rangle-\langle E_{0}^\mathrm{S}|\hat{\Sigma}|E_{0}^\mathrm{S}\rangle\right)\\&= \pm\gamma_\mathrm{eff}\sum_{j}^N\langle +|^{\otimes N}\hat{\mathbb{S}}_j^z|+\rangle^{\otimes N}\langle -|^{\otimes N}\hat{\mathbb{S}}_j^z|-\rangle^{\otimes N}\pm i\left(\langle E_{N}^\mathrm{S}|\hat{\Sigma}|E_{N}^\mathrm{S}\rangle-\langle E_{0}^\mathrm{S}|\hat{\Sigma}|E_{0}^\mathrm{S}\rangle\right)\\&=\pm i\left(\langle E_{N}^\mathrm{S}|\hat{\Sigma}|E_{N}^\mathrm{S}\rangle-\langle E_{0}^\mathrm{S}|\hat{\Sigma}|E_{0}^\mathrm{S}\rangle\right).
    \end{aligned}
\end{equation}
These expressions show that the first-order eigenvalues are purely imaginary. This implies that eigenvalues in the momentum layer of $|l|=N$ move along the imaginary axis under the perturbation $\big( \gamma_\mathrm{eff}\mathcal{D}^\prime +\mathcal{E}\big)$. 


A few comments are in order. The layer indices can be split into two groups, where QMBS layers $l=-N,-N+2,\cdots,N$ represent layers undergoing spectral transition in Fig.~\ref{fig:schematic} of the main text, and $l=-N+1,-N+3,\cdots,N-1$ are the remaining layers without spectral transition. The layers $l=\pm N$ are special in that they contain a single level $s=0$. Consequently, as we show in above, these QMBS Liouvillian eigenvalues remain on the imaginary axis for arbitrary values of the dephasing, while the other $l$ sectors generally distribute across the complex plane. Second, we note that our labeling of QMBS layers is unrelated with inversion symmetry quantum number (see Sec.~B for a discussion of symmetries of the Liouvillian). The latter is given by $p=(-1)^l$, therefore the spectral transition is found in different inversion-symmetric subspaces depending on the parity of $N$. 

\subsection{A2. The cases without $\mathcal{T}_{-}$ symmetry: PXP model}

The scar subspace of the PXP model can also be approximately described using the Hamiltonian decomposition in Eq.~(\ref{eq:scarham}). However, the QMBS Hamiltonian $\hat S_0$ and self-energy $\hat{\Sigma}$ should be replaced by an effective spin-1 paramagnet in the dimerized picture~\cite{Omiya2022,Omiya2023b}. First, we use an artificial $S = 1$ “blockspin” by defining $|0\rangle_b:=|\downarrow\!\downarrow\rangle_{b_\alpha, b_\beta},|+\rangle_b:=|\downarrow\uparrow\rangle_{b_\alpha, b_\beta}, |-\rangle_b:=|\uparrow\downarrow\rangle_{b_\alpha, b_\beta}$, where $b_\alpha(b_\beta)$ represents the left(right) site in the $b$-th block. Letting $N=L/2$ be the number of the blocks, the QMBS Hamiltonian $\hat S_0$ and self-energy $\hat{\Sigma}$ can be written as
\begin{equation}
\hat{S}_0=\sqrt{2} \sum_{b =1}^{N} \hat{\tau}_b^x,\quad\quad \hat{\Sigma}=-\sum_{b=1}^{N}(|+, 0\rangle+|0,-\rangle)\left\langle+,-\right|_{b, b+1}+\mathrm{h.c.},
\end{equation}
where $\hat{\tau}_{b}$ is the standard spin-$1$ Pauli matrices for $b$-th block in the PXP model. In this picture, the initial state and the eigenstates can be expressed as 
\begin{equation}
\label{eq:pxpeigen}
\left|\mathbb{Z}_2\right\rangle=\bigotimes^{N}_{b=1}|\downarrow\uparrow\rangle_{b_\alpha,b_\beta}=\bigotimes^{N}_{b=1}|+\rangle_{b}, \quad|E_s^\mathrm{S}\rangle =  \sqrt{\frac{1}{\mathcal{N}_{s}}}(\hat{J}^{-})^{N-s}\bigotimes^{N}_{b=1}|\widetilde{+}\rangle_{b},
\end{equation}
where $\hat{J}^{\pm}=\mp i\sum_b\hat{\tau}^{\pm}_b$ is the raising (lowering) operator, with $\hat{\tau}^{\pm}_b = \hat{\tau}^{y}_b\pm i\hat{\tau}^{z}_b$. The normalization factor $\mathcal{N}_{s}$ reads
\begin{equation}
\mathcal{N}_{s}=\prod_{M=s-N+1}^{N}\left[(N\left(N+1\right)-M(M-1)\right],
\end{equation}
where the index is chosen from $s=0,1,2,\cdots,L$. In addition, $|\widetilde{\pm}\rangle_b$ and $|\widetilde{0}\rangle_b$ are the eigenvectors of the local operator $\hat\tau_b^x$ with positive (negative) and zero eigenvalue.

Considering the Rydberg projector $P_{\mathrm{Ryd}}=\prod_{b}(\hat{I}-|+,-\rangle\langle+,-\mid)_{b, b+1}$, the eigenstates of PXP model can be approximated by $|E^{\mathrm{S}\ast}_{s}\rangle = P_{\mathrm{Ryd}}|E_{s}^\mathrm{S}\rangle$. Based on first-order perturbation theory, the energy shift $\Delta E_{s}=E_{s}^{\mathrm{S}\ast}-E_{s}^\mathrm{S}$ is given as 
\begin{equation}
\Delta E_s={\left\langle E_{s}^\mathrm{S}\right| \hat\Sigma\left|E_{s}^\mathrm{S}\right\rangle}=-\frac{1}{\mathcal{N}_{s}} \sum_{b=1}^N\left\langle E_{s}^\mathrm{S}\right|(|+, 0\rangle+|0,-\rangle)\left\langle+,-\right|_{b, b+1} +\mathrm{h.c.}\mid E_{s}^\mathrm{S}\rangle.
\end{equation}
After some algebraic manipulations, the energy correction from all the sites is thus estimated as~\cite{Omiya2022}
\begin{equation}\label{eq:pxpshift}
    {\left\langle E_{s}^\mathrm{S}\right| \hat\Sigma\left|E_{s}^\mathrm{S}\right\rangle}=-\frac{\sqrt{2}}{8} N \frac{\left|c_+\right|^2-\left|c_{-}\right|^2}{\mathcal{N}_{s}},
\end{equation}
where the coefficient $c_\pm$ reads
\begin{equation}
\begin{aligned}
c_+ & =\prod_{M=s-N-1}^{N-2} \sqrt{\left(N-2\right)\left(N-1\right)-M(M-1)}, \\
c_{-} & =s(s-1)(s-2)(s-3) \prod_{M=s-N+3}^{N-2} \sqrt{\left(N-2\right)\left(N-1\right)-M(M-1)} .
\end{aligned}
\end{equation}
Then we put the results of Eq.~(\ref{eq:pxpshift}) into Eq.~(\ref{eq:epsilon_l}) and obtain the diagonal matrix $\mathcal{E}_l$.

The next step is to calculate the projection of $\mathcal{D}^\prime$ to the $l$th layer, $\mathcal{D}_l^\prime$. In the spin-$1$ picture, the dephasing term is expressed as
\begin{equation}
\label{eq:pxpdissipator}
\begin{aligned}
&\mathcal{D}^{\prime}=\sum_b \hat{D}_b^\alpha \otimes \hat{D}_b^\alpha+\sum_b \hat{D}_b^\beta \otimes \hat{D}_b^\beta,\\ \quad &\hat{D}_b^\alpha=-|\downarrow\!\uparrow\rangle\langle\downarrow\!\uparrow|_{b_\alpha, b_\beta}-|\downarrow\!\downarrow\rangle\langle\downarrow\!\downarrow|_{b_\alpha, b_\beta}+|\uparrow\!\downarrow\rangle\langle\uparrow\!\downarrow|_{b_\alpha, b_\beta}=-\hat{\tau}_b^z-|0\rangle\langle 0|_b,\\\quad &\hat{D}_b^\beta=|\downarrow\!\uparrow\rangle\langle\downarrow\!\uparrow|_{b_\alpha, b_\beta}-|\downarrow\!\downarrow\rangle\langle\downarrow\!\downarrow|_{b_\alpha, b_\beta}-|\uparrow\!\downarrow\rangle\langle\uparrow\!\downarrow|_{b_\alpha, b_\beta}=\hat{\tau}_b^z-| 0\rangle\left\langle0\right|_b.
\end{aligned}
\end{equation}

Because of the Rydberg projector, the dissipator term is not exactly the linear Zeeman term as in the CL model. If we only consider the linear Zeeman term $\hat{\tau}^z$, we obtain something similar to the CL model in that the dissipator matrix $\mathcal{D}_l^\prime$ couples only $|\lambda_{(l,s)}^{(0)}\rangle\!\rangle$ and $|\lambda_{(l,s\pm 1)}^{(0)}\rangle\!\rangle$,
\begin{equation}\label{eq:D_l_pxp1}
\begin{aligned}
   \mathcal{D}_l^\prime=  \sum_{s=0}^{L-|l|} d_{(l,s)}^{(0)} \ \big|\lambda_{(l,s)}^{(0)}\big\rangle\!\big\rangle  \big\langle\!\big\langle\lambda_{(l,s)}^{(0)}\big| +\sum_{k=1}^{L-|l|}\sum_{s=0}^{L-|l|-k} d_{(l,s)}^{(k)} \Big[\ \big|\lambda_{(l,s)}^{(0)}\big\rangle\!\big\rangle  \big\langle\!\big\langle\lambda_{(l,s+k)}^{(0)}\big|+\mathrm{h.c.}  \Big],
\end{aligned}
\end{equation}
where $d^{(k)}_{(l,s)}$ represents the coupling coefficient between the $s$-th and $(s+k)$-th degenerate eigenstates in the $l$-momentum layer. It follows that $d^{(k)}_{(l,s)} = 0$ when $ k\neq1$. For PXP model, the corresponding eigenmodes and eigenvalues are written as
\begin{equation}\label{seq:0th_eigenvalue}
    |\lambda_{(l,s)}^{(0)}\rangle\!\rangle = |E_{l+s}^\mathrm{S}\rangle \otimes |E_s^\mathrm{S}\rangle, 
    \quad\quad\quad\quad
    \lambda_{l}^{(0)}=i(E_{l+s}^\mathrm{S}-E_{s}^\mathrm{S})=2\sqrt{2}ilJ.
\end{equation}

Using the translation symmetry of $\sum_i \hat{\tau}^z_i$, we have 
\begin{equation}
\begin{aligned}
d_{(l,s)}^{(1)}&=2\sum_{b=1}^N\Big[\langle E_{s}^\mathrm{S}|\hat{\tau}^z_b|E_{s+1}^\mathrm{S}\rangle \langle E_{s+|l|}^\mathrm{S}|\hat{\tau}^z_b|E_{s+|l|+1}^\mathrm{S}\rangle\Big]
=\frac{2}{N}  \bigg[\Big\langle E_{s}^\mathrm{S}\Big|\sum_b\hat{\tau}^z_b\Big|E_{s+1}^\mathrm{S}\Big\rangle \Big\langle E_{s+|l|}^\mathrm{S}\Big|\sum_b\hat{\tau}^z_b\Big|E_{s+|l|+1}^\mathrm{S}\Big\rangle\bigg].
\end{aligned}
\end{equation}
Noting that $\sum_b\hat{\tau}^z_b=(1/2)(\hat{J^+}+\hat{J^-})$ and using the definition of $|E_{s}^\mathrm{S}\rangle$ in the Eq.~(\ref{eq:pxpeigen}), we obtain 
\begin{equation}
\begin{aligned}
    \Big\langle E_{s}^\mathrm{S}\Big|\sum_b\hat{\tau}^z_b\Big|E_{s+1}^\mathrm{S}\Big\rangle=&\frac{1}{2}\left\langle E_{s}^\mathrm{S}\left|\hat{J}^+\right|E_{s+1}^\mathrm{S}\right\rangle+\frac{1}{2}\left\langle E_{s}^\mathrm{S}\left|\hat{J}^-\right|E_{s+1}^\mathrm{S}\right\rangle\\
    =&\frac12\sqrt{\frac{\mathcal{N}_{s}}{\mathcal{N}_{s+1}}}\left(\bigotimes^{N}_{b=1}\langle\widetilde{+}|_{b}\right)\left(\hat{J}^{-}\right)^{s-(s+1)}\hat{J}^+\left(\bigotimes^{N}_{b=1}|\widetilde{+}\rangle_{b}\right)+\\&\frac12\sqrt{\frac{\mathcal{N}_{s}}{\mathcal{N}_{s+1}}}\left(\bigotimes^{N}_{b=1}\langle\widetilde{+}|_{b}\right)\left(\hat{J}^{-}\right)^{s-(s+1)}\hat{J}^-\left(\bigotimes^{N}_{b=1}|\widetilde{+}\rangle_{b}\right)\\=& 0+\frac12\sqrt{\frac{\mathcal{N}_{s}}{\mathcal{N}_{s+1}}}.
\end{aligned}
\end{equation}
Hence, the coefficient $d_{(l,s)}^{(1)}$ can be written in a compact form 
    \begin{equation}
\begin{aligned}
d_{(l,s)}^{(1)}=\frac1{2N}\sqrt{\frac{\mathcal{N}_{s}}{\mathcal{N}_{s+1}}}\sqrt{\frac{\mathcal{N}_{s+|l|}}{\mathcal{N}_{s+|l|+1}}}
=\frac{1}{2N}&\sqrt{N\left(N+1\right)-(N-s)(N-s-1)}
\\
\times & \sqrt{N\left(N+1\right)-(N-s-|l|)(N-s-|l|-1)}.
\end{aligned}
\end{equation}

In the following paragraph, we show the effect of the term $|0\rangle\langle0|_b$. Noticing that $|0\rangle\langle0|_b=\hat{I}-(\hat{\tau}^z_b)^2$, we define this term as quadratic Zeeman term.   Once we introduce the quadratic term, the dissipator matrix $\mathcal{D}_l^\prime$ couples not only the adjacent term; thus we should also consider $d^{(k)}_{(l,s)}$ with $k\neq1$ in Eq.~(\ref{eq:D_l_pxp1}). We first rewrite the quadratic term as$|0\rangle\langle0|_b=\hat{I}-\left(\hat{\tau}^z_b\right)^2=1/2\left(\hat{\tau}^x_b\right)^2+1/4\left(\left(\hat{\tau}^+_b\right)^2+\left(\hat{\tau}^-_b\right)^2\right)$. Then, after acting the quadratic term on $|E_{s}^\mathrm{S}\rangle$, it overlaps with $|E_{s}^\mathrm{S}\rangle$ and $|E_{s\pm2}^\mathrm{S}\rangle$. Unlike the linear term, the quadratic term contributes to $d^{(0)}_{(l,s)}$ and $d^{(2)}_{(l,s)}$ as follows

\begin{equation}
\label{eq:pxp00term}
\begin{aligned}
d_{(l,s)}^{(0)}&=\frac12\sum_{b=1}^N\Big[\langle E_{s}^\mathrm{S}|(\hat{\tau}^x_b)^2|E_{s}^\mathrm{S}\rangle \langle E_{s+|l|}^\mathrm{S}|(\hat{\tau}^x_b)^2|E_{s+|l|}^\mathrm{S}\rangle\Big]
=\frac N2\langle E_{s}^\mathrm{S}|(\hat{\tau}^x_1)^2|E_{s}^\mathrm{S}\rangle \langle E_{s+|l|}^\mathrm{S}|(\hat{\tau}^x_1)^2|E_{s+|l|}^\mathrm{S}\rangle.\\
d_{(l,s)}^{(2)}&=\frac18\sum_{b=1}^N\Big[\langle E_{s}^\mathrm{S}|(\hat{\tau}^-_b)^2|E_{s+2}^\mathrm{S}\rangle \langle E_{s+|l|}^\mathrm{S}|(\hat{\tau}^-_b)^2|E_{s+|l|+2}^\mathrm{S}\rangle\Big]
=\frac N8\langle E_{s}^\mathrm{S}|(\hat{\tau}^-_1)^2|E_{s+2}^\mathrm{S}\rangle \langle E_{s+|l|}^\mathrm{S}|(\hat{\tau}^-_1)^2|E_{s+|l|+2}^\mathrm{S}\rangle.
\end{aligned}
\end{equation}

We first define the state $|T_{N,M}\rangle\equiv|E_{s}^\mathrm{S}\rangle$ with total spin $N$ and total magnetic quantum number $M=s-N$, then decompose it via Clebsch-Gordan coefficients
\begin{equation}
      |T_{N,M}\rangle = \sum_{m_1} C_{1,m_1; N-1,M-m_1}^{N,M} |T_{1,m_1}\rangle \otimes |T_{N-1,M-m_1}\rangle,
\end{equation}
where $m_1$ is magnetic quantum number of the first block. The probability for the first block to have $m_1 = \pm 1$ is:
\begin{equation}
  P_{\pm} = \left| C_{1,\pm1; N-1,M\mp1}^{N,M} \right|^2 = \frac{(N \pm M)(N \pm M - 1)}{2N(2N-1)}.
\end{equation}
The fully symmetric state $|T_{N,M}\rangle$ ensures identical expectation values for any single-block operator. Thus, $\langle (\hat{\tau}^x_{b})^2 \rangle$ is block-independent. Since $(\hat{\tau}^x_{1})^2$ has eigenvalue 1 for $m_1 = \pm 1$ and 0 for $m_1 = 0$, the total probability gives:
  \begin{equation}
  \label{eq:pxpMeq0}
  \langle T_{N,M}| (\hat{\tau}^x_{1})^2|T_{N,M} \rangle = P_+ + P_- = \frac{(N+M)(N+M-1) + (N-M)(N-M-1)}{2N(2N-1)} = \frac{N^2 + M^2 - N}{N(2N-1)}.
  \end{equation}

In addition, $(\hat{\tau}^-_{1})^2$ reduces $m_1$ by 2 units, mapping $|T_{ N,M+2}\rangle$ to a state with total $\tau^x = M$. Only the $m_1 = +1$ component contributes:
  \begin{equation}
  (\hat{\tau}^-_{1})^2 |T_{N, M+2}\rangle = 2  C_{1,+1; N-1,M+1}^{N,M+2}  |T_{1,-1}\rangle \otimes |T_{N-1,M+1}\rangle.
  \end{equation}  
  where the coefficients for maximal spin coupling are:
  \begin{equation}
  C_{1,+1; N-1,M+1}^{N,M+2} = \sqrt{\frac{(N+M+2)(N+M+1)}{2N(2N-1)}}, \quad \quad
  C_{1,-1; N-1,M+1}^{N,M} = \sqrt{\frac{(N-M)(N-M-1)}{2N(2N-1)}}.
\end{equation}
Then the inner product gives:
\begin{equation}
\label{eq:pxpMeq2}
  \langle T_{N,M} | (\hat{\tau}^-_{1})^2 | T_{N,M+2} \rangle = 2  C_{1,+1; N-1,M+1}^{N,M+2}  C_{1,-1; N-1,M+1}^{N,M} = \frac{\sqrt{(N+M+2)(N+M+1)(N-M)(N-M-1)}}{N(2N-1)}.
    \end{equation}
Using the results in Eq.~(\ref{eq:pxp00term}), Eq.~(\ref{eq:pxpMeq0}) and Eq.~(\ref{eq:pxpMeq2}) and substituting $M$ with $s-N$, the coefficient $d^{(k)}_{(l,s)}$ of PXP model can be concluded as

\begin{equation}
\label{eq:coefficientpxp}
\begin{aligned}
    d^{(0)}_{(l,s)}&=\frac{1}{2N(2N-1)^2}\Big[N(N-1)+(N-s)^2\Big]\Big[N(N-1)+(N-s-|l|)^2\Big],  \\
    d^{(1)}_{(l,s)}&=\frac{1}{2N}\sqrt{N\left(N+1\right)-(N-s)(N-s-1)}\sqrt{N\left(N+1\right)-(N-s-|l|)(N-s-|l|-1)}, \\
    d^{(2)}_{(l,s)}&=\frac{1}{8N(2N-1)^2}\sqrt{(2N-s)(2N-s-1)(s+1)(s+2)}\sqrt{(2N-s-|l|)(2N-s-|l|-1)(s+|l|+1)(s+|l|+2)} ,
\end{aligned}
\end{equation}
while $d^{(k\ge3)}_{(l,s)} = 0$. Using Eq.~(\ref{eq:coefficientpxp}) and Eq.~(\ref{eq:pxpshift}), we can construct the complete projection of $\left(\mathcal{E} \!+\! \gamma_{\mathrm{eff}} \mathcal{D}^\prime\right)$ to the degenerate subspace of the $l$th layer.  

\begin{figure}[tb]
\centering
\includegraphics[width=0.7\linewidth]{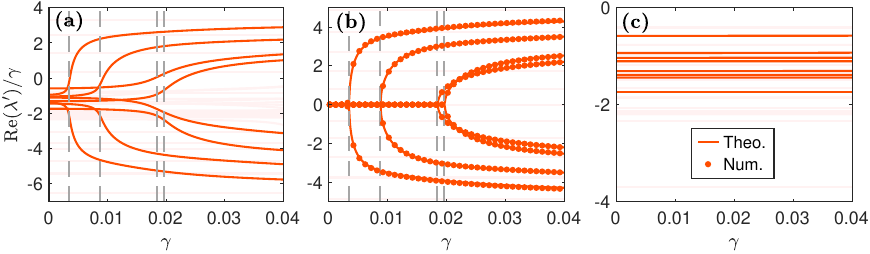}  
\caption{ The real part of PXP Liouvillian spectrum $\mathrm{Re}(\lambda^\prime/\gamma)$, as the dephasing rate $\gamma$ is varied. The results are calculated by (a) including the linear ($\hat{\tau}^z_b$) and the quadratic ($|0\rangle\langle0|_b=\hat{I}-(\hat{\tau}^z_b)^2$) terms, (b) only the linear term, (c) only the quadratic them. The vertical dashed lines indicate the spectral transition points. The system size is chosen by $L=10$. }\label{figSpxp}
\end{figure}

The results of perturbation theory of the PXP model are plotted in Fig.~(\ref{figSpxp}) against the numerical results for the same system size. While the full result is in good agreement with numerics, we see that dropping the quadratic term in Eq.~(\ref{eq:pxpdissipator}), shown in panel (b), already captures the spectral transitions. In fact, in this case we see the transitions become sharply defined, reminiscent of the CL model in the main text. This is not a coincidence since the perturbed PXP Liouvillian without the quadratic term obeys the $\mathbb{PT}$ symmetry, which is generated by the superoperator $\mathcal{T}_{-}$, 
\begin{equation}\label{eq:symmetry_operator}
   \mathcal{T}_- = \sqrt{\pm1}\Big[\Big(\prod_b \hat{\tau}_b^x\Big) \otimes \hat{I}\Big]\mathcal{K}, \quad \quad \mathcal{T}_{-} \mathcal{L} \mathcal{T}_{-}^{-1}=-\mathcal{L},
\end{equation}
where $\mathcal{K}$ is the complex-conjugation superoperator. Conversely, we do not observe any spectral transitions if we only consider the quadratic term and discard the linear term -- see panel (c) in Fig.~(\ref{figSpxp}). This suggests that for models like PXP, the spectral transition experienced by QMBS Liouvillian eigenvalues is not due to spontaneous breaking of an exact symmetry generated by $\mathcal{T}_{-}$.



\subsection{A3. Scaling behavior}

From the perspective of perturbation theory, the spectral transition phenomenon originates from the competition between dissipation $\gamma_{\mathrm{eff}}\mathcal{D}^\prime$ and the self-energy $\mathcal{E}$, as summarized in Fig.~\ref{fig:selfenergy}.  We can estimate the typical Liouvillian spectral transition point:
\begin{eqnarray}
 \bar{\gamma}_{\star}^\mathrm{S}\times\Vert\mathcal{D}^\prime\Vert \sim \Vert\mathcal{E}\Vert,   
\end{eqnarray}
where $\Vert\hat{A}\Vert = \sqrt{\lambda_\mathrm{max}(\hat{A}^\dagger\hat{A})}$ refers to the matrix norm of $\hat{A}$. The dissipation superoperator scales with $N$, $\Vert\mathcal{D}^\prime\Vert = \sum_{j=1}^N \Vert\hat{\sigma}_j^z\otimes\hat{\sigma}_j^z\Vert\sim N$. To estimate the scaling of self-energy for large $N$, recall the self-consistency condition, Eq.~(\ref{eq:self_energy}), and focus on the CL model, with $\hat{S}_0$ in Eq.~(\ref{seq:s0_E}). The maximal energy is $NJ$, so we have $\Vert E\hat{I}-\hat{S}_0\Vert$ scaling as $N$. According to Eq.~(\ref{eq:gammaCL}), 
\begin{equation}
\big\Vert\hat{\Gamma}\big\Vert = \sqrt{\max{\bigg(\sum_{j=1}^{N-1}J_{\mathrm{x},j}\hat{P}^{j,j+1}_{p_{\mathrm{S}}}\bigg)}}.
\end{equation}
The possible value of $\hat{P}^{j,j+1}_{p_{\mathrm{S}}}$ is $0$ and $1$, so $\Vert\hat{\Gamma}\Vert=\sqrt{\sum_{j=1}^{N-1}J_{\mathrm{x},j}}\sim\sqrt{N}$, where we assume that all the couplings $J_{\mathrm{x},j}$ are at the same order of magnitude. Approximately, from the self-consistence equation, we assume that $\Vert\hat{\Sigma}\Vert$ increase slower than $\Vert E\hat{I}-\hat{S}_0\Vert$, and we have
\begin{equation}
\big\Vert\hat{\Sigma}\big\Vert\times\big\Vert E\hat{I}-\hat{S}_0\big\Vert \sim \big\Vert\hat{\Gamma}\big\Vert^2.
\end{equation}
Then we obtain that $\Vert\hat{\Sigma}\Vert\sim \mathrm{const.}$, which agrees with the assumption we made that $\Vert\hat{\Sigma}\Vert$ increase slower than $\Vert E\hat{I}-\hat{S}_0\Vert$. Finally, we use Eq.~(\ref{eq:epsilon_D}) and have $\Vert\mathcal{E}\Vert\sim \mathrm{const.}$ in large-$N$ limit, so that $\bar{\gamma}_{\star}^\mathrm{S}\sim N^{-1}\sim L^{-1}$. As discussed in the main text, this scaling is in stark contrast with $\bar{\gamma}_\star \sim D^{-2}$ for thermal eigenvalues, where $D$ is the Hilbert space dimension.

\subsection{A4. Difference between scar subspaces with and without a spectral transition}

Here, in the framework of the perturbation theory, we will explain the difference between the two scar subspaces, $(\mathrm{S},{+}1)$ and $(\mathrm{S},{-}1)$, from the point of view of spectral transition in Fig.~\ref{fig:schematic}. We will work at small dephasing rates, such that $\vert\vert\mathcal{S}_0\vert\vert\gg\vert\vert\mathcal{E}\vert\vert\gg\vert\vert\gamma_{\text{eff}} \mathcal{D}^\prime\vert\vert$. Without dephasing and assuming a weak coupling between the scar subspace and thermal bulk, the dephasing-free eigenvalues read $\lambda_{(l,s)}^{\mathrm{DF}} = i(2lJ+\Sigma_{(l,s)})$. Previously, we showed that $\Sigma_{(l,s)} = \Sigma_{(l,N-|l|-s)}$, which cause $\left|\lambda_{(l,s)}^{\mathrm{DF}}\right\rangle$ to be degenerate with $\left|\lambda_{(l,N-|l|-s)}^{\mathrm{DF}}\right\rangle$. Once the dissipation rate $\gamma$ is nonzero, this degeneracy will be broken. A special case arises for the non-degenerate state present when $N-|l|$ is even, so that we can find an $s$ to satisfy $s=N-|l|-s$.

For the case $s\neq N-|l|-s$, degenerate perturbation theory is applicable when dissipation is very small. The general expression for the eigenvalue $\lambda_{(l,s)}$ at arbitrary order is given by sum-over-paths expansion:
\begin{equation}
\lambda_{(l,s)} = i(2lJ+\Sigma_{(l,s)}) + \sum_{n=1}^{\infty} \gamma_{\text{eff}}^{n} \sum_{\text{path}} \frac{\langle \lambda_{(l,s)}^{\mathrm{DF}} | \mathcal{D}_l^\prime | \lambda_{(l,p_1)}^{\mathrm{DF}} \rangle \langle \lambda_{(l,p_1)}^{\mathrm{DF}} | \mathcal{D}_l^\prime | \lambda_{(l,p_2)}^{\mathrm{DF}} \rangle \cdots \langle \lambda_{(l,p_{n-1})}^{\mathrm{DF}} | \mathcal{D}_l^\prime | \lambda_{(l,N-|l|-s)}^{\mathrm{DF}} \rangle}{\prod_{j=1}^{n-1} (i\Sigma_{(l,s)} - i\Sigma_{(l,p_j)})},
\end{equation}
where $\left\vert p_j\right\rangle$ are the intermediate states between $\left|\lambda_{(l,s)}^{\mathrm{DF}}\right\rangle$ and $\left|\lambda_{(l,N-|l|-s)}^{\mathrm{DF}}\right\rangle$.

The nature of the correction term depends critically on the order $n$. The shortest path between $\left|\lambda_{(l,s)}^{\mathrm{DF}}\right\rangle$ and $\left|\lambda_{(l,N-|l|-s)}^{\mathrm{DF}}\right\rangle$ costs $|N-|l|-2s|$ steps, which means only odd (even) order terms make a contribution when $N-|l|$ is odd (even). Consequently, the eigenvalues are all purely imaginary when $N-|l|$ is even, while they all acquire nonzero real part when $N-|l|$ is odd. For example, in the case $N=5, l = +3, s=0$, only the second-order term should be considered:
\begin{equation}
\gamma_{\text{eff}}^{2} \frac{\langle \lambda_{(3,0)}^{\mathrm{DF}} | \mathcal{D}_l^\prime | \lambda_{(3,1)}^{\mathrm{DF}} \rangle \langle \lambda_{(3,1)}^{\mathrm{DF}} | \mathcal{D}_l^\prime | \lambda_{(3,2)}^{\mathrm{DF}} \rangle}{(i\Sigma_{(3,0)} - i\Sigma_{(3,1)})}\in i \mathbb{R},
\end{equation}
which is purely imaginary. While in the case $N=5, l = +2, s=0$, the third-order term dominates:
\begin{equation}\gamma_{\text{eff}}^{3} \frac{\langle \lambda_{(2,0)}^{\mathrm{DF}} | \mathcal{D}_l^\prime | \lambda_{(2,1)}^{\mathrm{DF}} \rangle \langle \lambda_{(2,1)}^{\mathrm{DF}} | \mathcal{D}_l^\prime | \lambda_{(2,2)}^{\mathrm{DF}} \rangle \langle \lambda_{(2,2)}^{\mathrm{DF}} | \mathcal{D}_l^\prime | \lambda_{(2,3)}^{\mathrm{DF}} \rangle}{(i\Sigma_{(2,0)} - i\Sigma_{(2,1)}) (i\Sigma_{(2,0)} - i\Sigma_{(2,2)})}\in  \mathbb{R},\end{equation}
which is real.

For the case $s= N-|l|-s$, we should consider the non-degenerate expansion. Here, any path connecting the state back to itself must involve an even number of steps under the dissipative operator $\mathcal{D}_l^\prime$, consequently forcing all odd-order corrections to vanish identically. Its eigenvalue correction is given by:
\begin{equation}  
\lambda_{(l,s)} = i\Sigma_{(l,s)} + \sum_{n=1}^{\infty} \gamma_{\text{eff}}^{2n} \sum_{\text{path}} \frac{\langle \lambda_{(l,s)}^{\mathrm{DF}} | \mathcal{D}_l^\prime | \lambda_{(l,p_1)}^{\mathrm{DF}} \rangle \langle \lambda_{(l,p_1)}^{\mathrm{DF}} | \mathcal{D}_l^\prime | \lambda_{(l,p_2)}^{\mathrm{DF}} \rangle \cdots \langle \lambda_{(l,p_{2n-1})}^{\mathrm{DF}} | \mathcal{D}_l^\prime | \lambda_{(l,s)}^{\mathrm{DF}} \rangle}{\prod_{j=1}^{2n-1} (i\Sigma_{(l,s)} - i\Sigma_{(l,p_j)})}.
\end{equation}
Noticing the factor $i^{2n-1}$ in the denominator, the eigenvalue should be purely imaginary when dissipation is small.

The key distinction between two subspaces emerges from the parity of coupling path length. Within the  `$\mathrm{S},{-}1$' subspace, coupling between degenerate states necessitates an even number of applications of the dissipative operator $\mathcal{D}_l^\prime$. Consequently, for small $\gamma$, only even-order corrections are non-zero, resulting in the eigenvalue shifting purely along the imaginary axis. Conversely, in the `$\mathrm{S},{+}1$' subspace, the coupling mechanism allows odd-order corrections, leading to a shift along the real axis even for small dissipation. This distinction leads to the fact that the eigenvalues in `$\mathrm{S},{-}1$' subspace undergoes a spectral transition while the eigenvalues in `$\mathrm{S},{+}1$' do not. 

To conclude, both the thermal subspace and the bath contributes to the eigenvalue correction in `$\mathrm{S},{\pm}1$' subspace. However, only the eigenvalues in `$\mathrm{S},{-}1$' subspace undergo spectral transition due to the parity of coupling path length. Additionally, in the PXP model, the same analysis can be used to prove that the spectral transition only occurs exclusively in `$\mathrm{S},0$' subspace.

\section{B. Symmetries of the Liouvillian}

First, we briefly analyze the symmetries of the CL model and the PXP model in an open quantum setting. We denote the basis states for the Liouvillian $\mathcal{L}$ of the CL model by 
\begin{equation}
 |B\rangle\!\rangle= \left|\substack {u_{1}\\ d_{1} }...\substack {u_{i}\\ d_{i} }...\substack {u_{N}\\ d_{N} }\right\rangle\!\otimes\!|\substack {u^{\prime}_{1}\\ d^{\prime}_{1} }...\substack {u^{\prime}_{i}\\ d^{\prime}_{i} }...\substack {u^{\prime}_{N}\\ d^{\prime}_{N} }\rangle, \quad    u, d, u^\prime, d^\prime \in\{ \uparrow,\downarrow\}.
\end{equation}
The inversion symmetry superoperator $\mathcal{P}$ has the following action on basis states:
\begin{equation}
 \mathcal{P}|B\rangle\!\rangle=|\bar{B}\rangle\!\rangle= \left|\substack {\bar{u}_{1}\\ \bar{d}_{1} }...\substack {\bar{u}_{i}\\ \bar{d}_{i} }...\substack {\bar{u}_{N}\\ \bar{d}_{N} }\right\rangle\!\otimes\!|\substack {\bar{u}^{\prime}_{1}\\ \bar{d}^{\prime}_{1} }...\substack {\bar{u}^{\prime}_{i}\\ \bar{d}^{\prime}_{i} }...\substack {\bar{u}^{\prime}_{N}\\ \bar{d}^{\prime}_{N}}\rangle,   
\end{equation}
where $\bar{u}_i$ ($\bar{d}_i$) denotes the $i$th spin state opposite to $u_i$ ($d_i$). We denote the eigenvalues of $\mathcal{P}$ by $p=\pm 1$.

In addition, we should consider the reflection superoperator $\mathcal{R}$, 
\begin{equation} \mathcal{R}|B\rangle\!\rangle=|B^{\prime}\rangle\!\rangle= \left|\substack {u_{N}\\ d_{N} }...\substack {u_{i}\\ d_{i} }...\substack {u_{1}\\ d_{1} }\right\rangle\!\otimes\!|\substack {u^{\prime}_{N}\\ d^{\prime}_{N} }...\substack {u^{\prime}_{i}\\ d^{\prime}_{i} }...\substack {u^{\prime}_{1}\\ d^{\prime}_{1}}\rangle.
\end{equation}
The CL Hamiltonian of the two horizontal legs $H_{\rm{h},\alpha}$ remains unchanged under the action of inversion symmetry operator $\mathcal{P}$, but it maps to $-H_{\rm{h},\alpha}$ under the action of reflection symmetry operator $\mathcal{R}$ when the number of spins $N$ is odd. Hence, we will restrict to the cases when $N$ is even, when the eigenmodes can be classified according to the eigenvalues $r=\pm 1$ of $\mathcal{R}$.

On the other hands, for the PXP model, we denote the basis states for the Liouvillian $\mathcal{L}$ by $|B\rangle\!\rangle= \left|u_{1} ...u_{i}...u_{N}\right\rangle\!\otimes\!|u^{\prime}_{1}...u^{\prime}_{i}...u^{\prime}_{N}\rangle$. The translation symmetry superoperator $\mathcal{Q}$ is defined by 
\begin{equation}
 \mathcal{Q}|B\rangle\!\rangle=|B^{\prime}\rangle\!\rangle= \left|u_{N} ...u_{i+1}...u_{N-1}\right\rangle\!\otimes\!|u^{\prime}_{N}...u^{\prime}_{i+1}...u^{\prime}_{N-1}\rangle,   
\end{equation}
which we use to characterize the eigenmodes in terms of conserved momentum $q$. 

Next, we study the symmetry properties of the Liouvillian eigenmodes responsible for the spectral transition discussed in the main text. Recall that the unperturbed $\mathcal{S} _0$ has eigenmodes $|\lambda_{(l,s)}^{(0)}\rangle\!\rangle = |E_{l+s}^\mathrm{S}\rangle \otimes |E_{s}^\mathrm{S}\rangle$. Take $N = 3$ case as an example. There are $N+1=4$ eigenstates of $\hat S_0$ in the maximal spin representation:
\begin{equation}
    \begin{aligned}
   &|E_0^\mathrm{S}\rangle = |---\rangle ,  \quad \quad
   |E_1^\mathrm{S}\rangle = \frac{1}{\sqrt{3}}\Big(|+--\rangle + |-+-\rangle+ |--+\rangle\Big),\\& |E_2^\mathrm{S}\rangle=\frac{1}{\sqrt{3}}\Big(|++-\rangle + |+-+\rangle+ |-++\rangle\Big), \quad\quad |E_3^\mathrm{S}\rangle = |+++\rangle,
\end{aligned}
\end{equation}\label{seq:egN2}
with energies $-3J$, $-J$, $J$ and $3J$, respectively. The Liouvillian spectrum of $\mathcal{S}_0$ has a seven-layer structure with (imaginary) eigenvalues $0$, $\pm 2J$, $\pm 4J$, $\pm 6J$.

The Liouvillian spectrum of $(\mathcal{S}_0+\gamma\mathcal{D}^\prime)$ maintains the seven-layer structure, but now includes eigenvalues with non-zero real parts.
In the layer of $l=1$, there are two types of Liouvillian eigenmodes. Both of them are the linear superpositions of $|E_0^\mathrm{S}\rangle\otimes|E_1^\mathrm{S}\rangle$, $|E_1^\mathrm{S}\rangle\otimes|E_2^\mathrm{S}\rangle$, and $|E_2^\mathrm{S}\rangle\otimes|E_3^\mathrm{S}\rangle$, but with different  coefficients. One of them is purely imaginary, 
\begin{equation}
\begin{aligned}
&\frac{1}{\sqrt{2}}|E_0^\mathrm{S}\rangle\otimes|E_1^\mathrm{S}\rangle-\frac{1}{\sqrt{2}}|E_2^\mathrm{S}\rangle\otimes|E_3^\mathrm{S}\rangle,
\end{aligned}
\end{equation}
while the other one is more complex: 
\begin{equation}
    \begin{aligned}
        &\frac{1}{2}|E_0^\mathrm{S}\rangle\otimes|E_1^\mathrm{S}\rangle+\frac{1}{2}|E_2^\mathrm{S}\rangle\otimes|E_3^\mathrm{S}\rangle\pm \frac{1}{\sqrt{2}}|E_1^\mathrm{S}\rangle\otimes|E_2^\mathrm{S}\rangle .
    \end{aligned}
\end{equation}
Note that the former one does not include components of $|E_1^\mathrm{S}\rangle\otimes|E_2^\mathrm{S}\rangle = |\lambda^{(0)}_{(1,1)}\rangle\!\rangle$. More generally, we find that the pure imaginary Liouvillian eigenmodes in layer $l$ have zero overlap with those $|\lambda^{(0)}_{(l,s)}\rangle\!\rangle$ where $s$ is odd.

Furthermore, we find that the spectral transition phenomenon exists only in layers where $(N-|l|)$ is even, i.e., where $l =-N, -N+2,\ldots,N$.
Thus, studying the Liouvillian eigenmodes $\Big\{|\lambda^{(0)}_{(l,s)}\rangle\!\rangle\Big\}$ in such layers can help us to understand the nature of Liouvillian spectral transition.
We define a special Liouvillian vector
\begin{equation}   \big|\zeta\big\rangle\!\big\rangle= \sum_{(N-|l|)\in \mathrm{even}} \frac{1}{\sqrt{N-1}}\Big|\lambda^{(0)}_{(l,1)}\Big\rangle\!\!\Big\rangle,
\end{equation}
where the $N-1$ normalizing factor counts the total number of layers with spectral transition. We confirmed, for the CL model and  $\gamma< \gamma_{\star}^\mathrm{S}$, the overlap between $\big|\zeta\big\rangle\!\big\rangle$ and the eigenmodes in the layer $l^{\prime}$ is nearly zero. Once  $\gamma>\gamma_{\star}^\mathrm{S}$, the overlap rises  sharply to a finite number. This behavior mimics that of  the real part of Liouvillian eigenvalues.

According to the general symmetry classification of Lindbladians~\cite{PhysRevX.13.031019}, the operator responsible for $\mathbb{PT}$ symmetry is the $\mathcal{T}_-$ superoperator that obeys:
\begin{equation}\label{eq:tplusminus}
\mathcal{T}_{-} \mathcal{L} \mathcal{T}_{-}^{-1}=-\mathcal{L}, \; \mathcal{T}_{-}^2= \pm 1.
\end{equation}
The explicit form of $\mathcal{T}_{-}$ for the CL model is 
\begin{equation}
    \mathcal{T}_- = \sqrt{p_x}\Big[\Big(\prod_i \hat{\sigma}_i^x\Big) \otimes \hat{I}\Big]\mathcal{K},
\end{equation}
where $\mathcal{K}$ is the Liouvillian complex-conjugation operator, $\mathcal{K} \mathcal{L}\mathcal{K}^{-1} = 
\mathcal{L}^*$, and $\hat{I}$ is the identity matrix. The factor $p_x = \pm1$, which makes $\mathcal{T}_-^2 = \pm1$, respectively.

When $\gamma = 0$, $|-(\lambda^{\prime})^\ast\rangle\!\rangle$ and $|\lambda^{\prime}\rangle\!\rangle$ represent the same state which lies on the imaginary axis.  As the dephasing rate $\gamma$ increases until it exceeds the critical point $\gamma_\star^\mathrm{S}$, the symmetry of $\lambda^\prime=-(\lambda^\prime)^\ast$ is broken, while the eigenvectors continue to be related via  $\mathbb{P}\mathbb{T} |\lambda^{\prime} \rangle\! \rangle = |-(\lambda^{\prime})^\star\rangle\!\rangle$ remains. 
This indicates that, after the critical point, a few pairs of degenerate eigenvalues move symmetrically away from the imaginary axis along the real axis. 
In addition, a few eigenvalues stay at the imaginary axis regardless of the magnitude of $\gamma$. These eigenmodes can be found using the condition of 
\begin{equation}
\mathbb{P}\mathbb{T} |\lambda^{\prime} \rangle\! \rangle = |\lambda^{\prime}\rangle\!\rangle, \quad\quad\lambda^\prime = -(\lambda^{\prime})^\ast.
\end{equation}

We note that, for the PXP model, we cannot find a $\mathcal{T}_-$ operator that satisfies Eq.~(\ref{eq:tplusminus}). Without $\mathcal{T}_-$ symmetry, the Liouvillian spectrum is intrinsically asymmetric about the imaginary axis~\cite{PhysRevX.13.031019}, as confirmed by the numerical simulations. 
\begin{figure}[tb]
\centering
\includegraphics[width=0.6\linewidth]{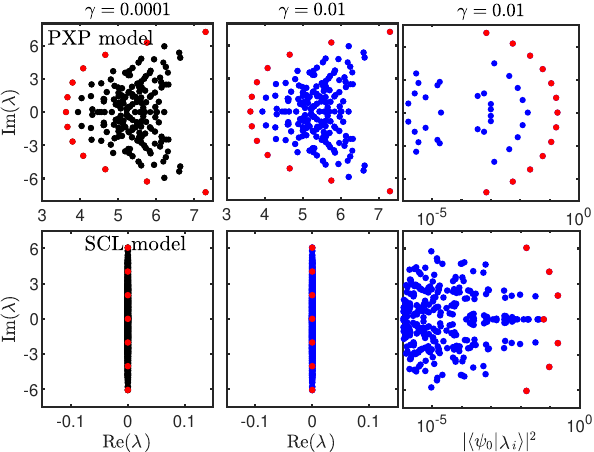}  
\caption{ The Liouvillian spectra of the effective non-Hermitian Hamiltonians for the PXP model and CL model ($J_{\textrm{x},j}=0.1$) at two dephasing rates $\gamma=0.0001,0.01$. The black and blue dots indicate the spectra of thermal states, while the red dots denote the approximate scars. The system size is $L=12$. Other parameters are the same as the main text}\label{figS2}
\end{figure}
\begin{figure*}[tb]
\centering
\includegraphics[width=1\linewidth]{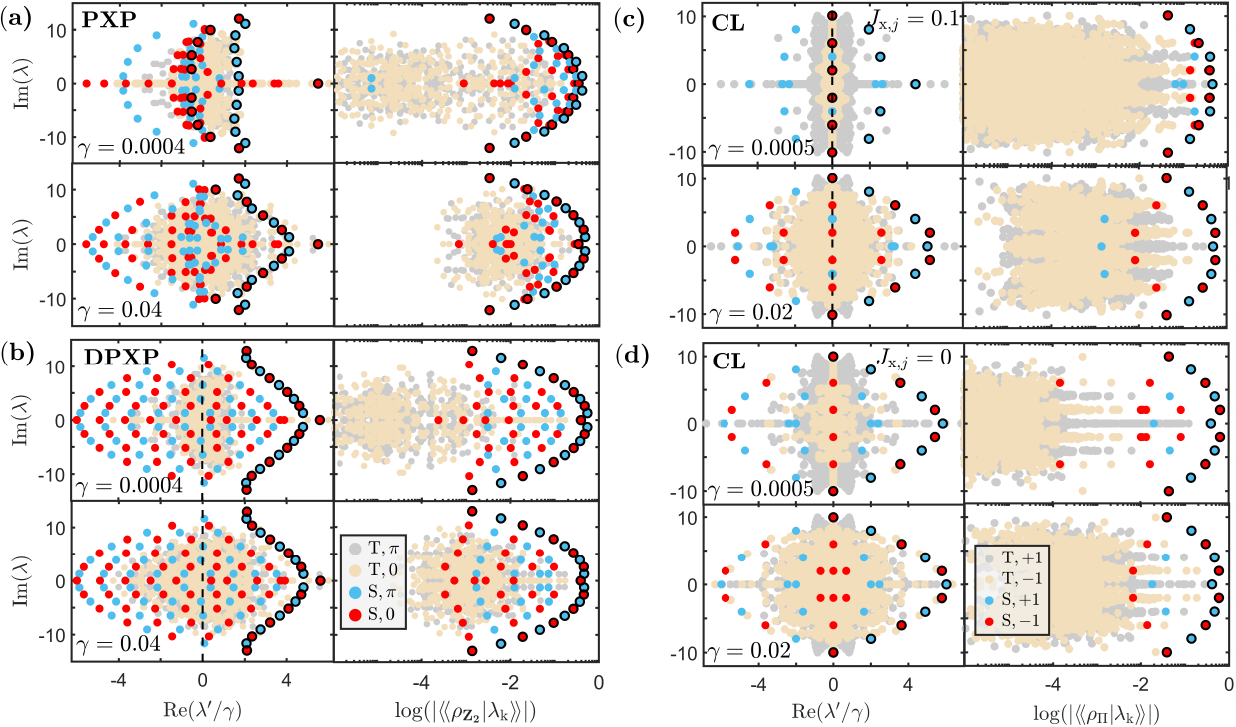} 
\caption{
(a)-(b) The Liouvillian spectra (left panel) and the overlap with $|\mathbb{Z}_2 \rangle \otimes |\mathbb{Z}_2\rangle$ (right panel) corresponding to  two dephasing rates, $\gamma=0.0004, 0.04$, of the PXP model and DPXP model. The scar (S) and thermal (T) states are labeled according to $q=0,\pi$ momenta under translation, as indicated in the legend.  (c)-(d) The case of the CL model with $J_{\mathrm{x},j}=0,0.1$. The states are labeled by $p=\pm1$, indicating the inversion symmetry is even (odd). The overlap is with the state $|\Pi \rangle \otimes |\Pi\rangle$, defined in the main text. In all plots, the black boundary dots indicate scar states with the local maximum overlap with the chosen initial state. The system size is $L=10$ in all plots. 
}\label{figS3}
\end{figure*}
\section{C. The origin of Liouvillian spectrum transition}\label{ssec:mechanism}

While a broader understanding of general types of dissipation and their effects on QMBSs is beyond the scope of current work, one immediate question is whether our results in the main text can be explained more simply by imposing certain approximations, such as considering the effective non-Hermitian description (as opposed to the full Lindblad dynamics). 

In Fig.~\ref{figS2} we numerically argue that spectrum transition due to dephasing does not exist in the non-Hermitian approximation which adds the term $i\gamma\sum^{L}_{j=1} \hat{\sigma}^{z}_{j}$ to the Hamiltonian. From the form of this term, it is clear that for the CL model in the sector with conserved magnetization $\sum^{L}_{j=1}\hat{\sigma}^{z}_{j}=0$, the non-Hermitian operator in the spin $z$-direction has no effect on the Liouville spectrum.
This is confirmed in Fig.~\ref{figS2}, which also shows that the non-Hermitian PXP model immediately spectrum transition by a very small $\gamma=10^{-4}$, in stark contrast with the results for its Liouvillian spectrum in the main text.

\section{D. Exact vs. approximate scars in CL and PXP models}

For the CL model, in the main text we emphasized a significant difference in the Liouvillian spectra of the two inversion symmetry  subspaces $p=\pm1$, and also between the cases of approximate ($J_{\mathrm{x},j}\neq 0$) vs. exact  ($J_{\mathrm{x},j}=0$) QMBSs. We highlight these differences in more detail in Fig.~\ref{figS3}(c)-(d). In the odd subspace $p=-1$, most approximate scars exhibit spectrum transition under weak dissipation, while in the even subspace $p=+1$ they have immediately spectrum transition. Under strong dissipation, the behavior of the two sectors tends to be consistent. In addition, we also observe that the maximum overlaps of scars are uniformly distributed in the Liouvillian spectral range across different subspaces. In the case of weak dissipation, more scars in $p=-1$ odd subspace  approach the zero point, $\mathrm{Re}(\lambda^{\prime}/\gamma)=0$. However, as the dephasing rate increases, they gradually move away from zero. 

For the PXP model, also shown in Fig.~\ref{figS3}(a)-(b), we find qualitatively similar behavior to the CL model, with a few minor differences. First, the spectrum transition occurs in different momentum $q$ sectors under translation, rather than inversion symmetry $p$ in the CL model. Second, there is no way to directly tune between approximate and exact scar limits. Instead, as a proxy for the latter, we consider a deformed PXP model (DPXP), introduced in the main text. With these provisos, the behavior of the PXP model largely mirrors that of CL model, with the main difference that $\mathbb{PT}$ symmetry is only approximately obeyed, e.g., see the scar states with momentum $q=0$ in Fig.~\ref{figS3}(a). As we explained above, this is attributed to a lack of $\mathcal{T}_{-}$ symmetry of the PXP Liouvillian. Nevertheles, these scarred eigenvalues still exhibit clustering around the imaginary axis at weak dephasing rates, giving rise to a certain level of robustness of QMBS signatures in dissipative dynamics, just like for the CL model with   $J_{\mathrm{x},j}\neq 0$.

\section{E. Nsumerical results on Liouvillian spectra of different models }\label{ssec.more_numerics}

\begin{figure}[htb]
\centering
\includegraphics[width=0.75\linewidth]{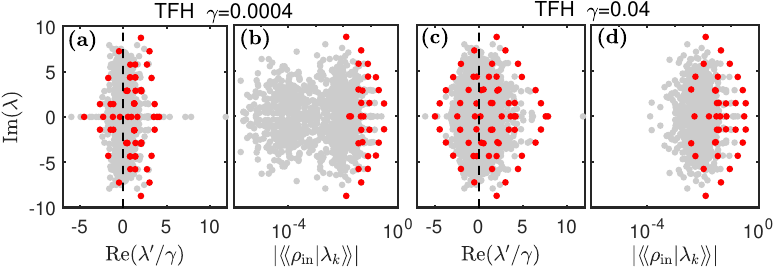}  
\caption{1D tilted Fermi-Hubbard (TFH) model, Eq.~(\ref{eq:TFH}), for two different dephasing rates. The gray dots indicate the Liouvillian spectra of thermal states, while the red dots indicate the spectra of scar states. The system size is $L=6$. }\label{figS4}
\end{figure}

Liouvillian spectrum transition can also be observed in the approximate scar states of a 1D tilted Fermi-Hubbard (TFH) model, Fig.~\ref{figS4}, whose Hamiltonian (after the  Jordan-Wigner transformation) is~\cite{Desaules2021Tilted}
\begin{equation}\label{eq:TFH}
\begin{aligned}
   \hat{H}_\mathrm{TFH}=J\sum_j^{L} \Big[\hat{\sigma}^{+}_{j,\uparrow}\hat{\sigma}^{-}_{j,\uparrow}\hat{P}_{j,\downarrow}(1-\hat{P}_{j,\downarrow}) 
    -\hat{\sigma}^{+}_{j,\downarrow}\hat{\sigma}^{-}_{j,\downarrow}\hat{P}_{j,\uparrow}(1-\hat{P}_{j,\uparrow})\Big],
\end{aligned}
\end{equation}
where the action of $\hat{\sigma}^+_{j,\uparrow(\downarrow)}$ is to create an
excitation on site $j$ with Jordan-Wigner spin up or down. $\hat{P}_{j,\uparrow(\downarrow)} =\big(\hat{I}+\hat{\sigma}^{z}_{j,\uparrow(\downarrow)}\big)/2$ the projector on the excited state of the Jordan-Wigner spin.

\begin{figure}[htb]
\centering
\includegraphics[width=0.9\linewidth]{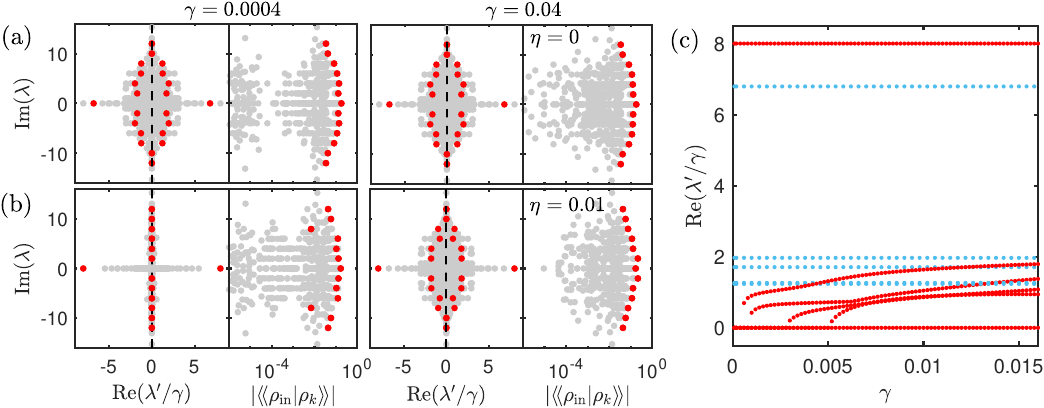}  
\caption{ Hilbert space fragmentation in the frustrated Heisenberg ladder model (\ref{eq:HeisLadder}). (a) and (b) correspond to the case of $\eta=0$ and $\eta=0.01$, respectively. The gray dots indicate the Liouville spectra of all states, while the red dots denote the spectra of a few typical local HSF states. (c) The variation of the Liouvillian spectrum $\mathrm{Re}(\lambda^{\prime}/\gamma)$ of a few typical local HSF states with the dephasing rate $\gamma$. The red (bule) dots are $\eta=0.01$ ($0$). The system size is $L=8$ and the parameter $J=1$. }\label{figS5}
\end{figure}

Beyond QMBSs, our conclusions are also applicable to Hilbert space fragmentation (HSF). As an example of HSF, we consider the spin-1/2 frustrated Heisenberg ladder~\cite{10.21468/SciPostPhys.11.4.074}:
\begin{equation}\label{eq:HeisLadder}
\begin{split}
\hat{H}_{\mathrm{FLH}} =& \hat{H}_{+}+\hat{H}_{-}+\hat{H}_{\mathrm{v}}+\hat{H}_{\mathrm{x}}, \quad \quad\quad\quad\mathrm{with}\quad\quad
\hat{H}_{\pm}=J(1-\eta)\sum^N_{j=1}\left[\left(\hat{\sigma}^{+}_{j,{\pm}}\hat{\sigma}^{-}_{j+1,\pm}+\mathrm{h.c.}\right)+\hat{\sigma}^{z}_{j,{\pm}}\hat{\sigma}^{z}_{j+1,\pm}\right],
\\
\hat{H}_{\mathrm{v}}=&\frac{1}{2}J \sum^N_{j=1}\left[\left(\hat{\sigma}^{+}_{j,-}\hat{\sigma}^{-}_{j,+ }+\mathrm{h.c.}\right)+\hat{\sigma}^{z}_{j,-}\hat{\sigma}^{z}_{j,+}\right],\quad
\hat{H}_{\mathrm{x}}=J(1+\eta)\sum^N_{j=1}\left[\left(\hat{\sigma}^{+}_{j,\pm}\hat{\sigma}^{-}_{j+1,\mp}+\mathrm{h.c.}\right)+\hat{\sigma}^{z}_{j,\pm}\hat{\sigma}^{z}_{j+1,\mp}\right].
\end{split}
\end{equation}
When $\eta$ equals zero, we have the complete frustration structure, resulting in a disconnected Hilbert space fragments. Small but non-zero $\eta$ corresponds to a weakly-connected HSF.

In Fig.~\ref{figS5}, we see that when the local HSF is not coupled with the thermalization part, the dephasing system remains in a state of Liouvillian spectrum transition, while conversely, a clear process of Liouvillian spectrum transition occurs. It should be noted that the red dots in Fig.~\ref{figS5}(a)-(b) represent a few typical local HSF states and do not include all local HSF states.

\section{F. Details of fitting the imbalance dynamics}\label{ssec:dyamics}

\begin{figure}[htb]
\centering
\includegraphics[width=0.7\linewidth]{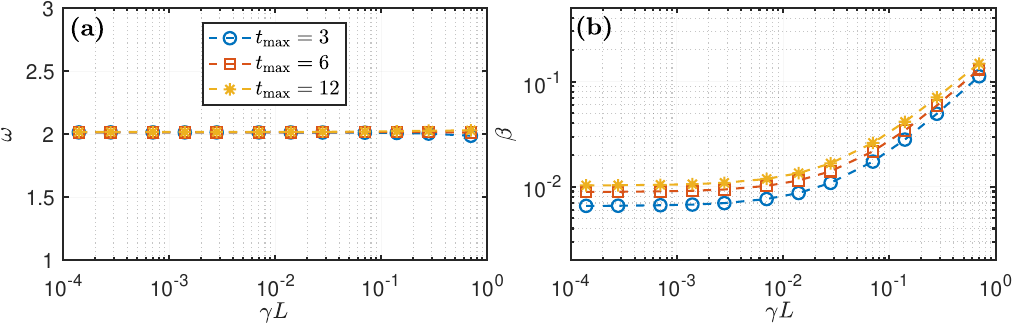}  
\caption{The coefficients $(\beta,\omega)$ of the imbalance fitting function $\bar{I}(t)$ as a function of dephasing $\gamma L$ for the CL model for different fitting intervals $t\in[0, t_{\rm{max}}]$ with $t_{\rm{max}}=3,6,12$. All data is for $L=14$, $J_\mathrm{h}\!=\!0.66$, $J_{\mathrm{x},j}\!\in\![0,0.2]$ drawn from a uniform distribution.}\label{figS6}
\end{figure}

Liouvillian spectrum transition can also be manifested through the dynamical behavior of the density imbalance 
$I(t)$ for the special initial state $|\rho(0)\rangle\!\rangle$, i.e.,  $|\Pi\rangle\otimes|\Pi\rangle$ state in the CL model.  These QMBS oscillations can be fitted by using $\bar{I}(t)= \exp(-\beta t)\cos(\omega t)$ over one or several time periods, and Fig.~\ref{figS6} demonstrates the robustness of such fits. Through our numerical analysis, we found that compared to fitting several revivals, fitting one period results in a slight shift for the decay coefficient $\beta$, but the Liouvillian spectrum transition point remains unchanged. Since it is numerically costly to obtain data over many periods in large system sizes, in the main text we have restricted our analysis to the first period. 

\begin{thebibliography}{97}%
\makeatletter
\providecommand \@ifxundefined [1]{%
 \@ifx{#1\undefined}
}%
\providecommand \@ifnum [1]{%
 \ifnum #1\expandafter \@firstoftwo
 \else \expandafter \@secondoftwo
 \fi
}%
\providecommand \@ifx [1]{%
 \ifx #1\expandafter \@firstoftwo
 \else \expandafter \@secondoftwo
 \fi
}%
\providecommand \natexlab [1]{#1}%
\providecommand \enquote  [1]{``#1''}%
\providecommand \bibnamefont  [1]{#1}%
\providecommand \bibfnamefont [1]{#1}%
\providecommand \citenamefont [1]{#1}%
\providecommand \href@noop [0]{\@secondoftwo}%
\providecommand \href [0]{\begingroup \@sanitize@url \@href}%
\providecommand \@href[1]{\@@startlink{#1}\@@href}%
\providecommand \@@href[1]{\endgroup#1\@@endlink}%
\providecommand \@sanitize@url [0]{\catcode `\\12\catcode `\$12\catcode `\&12\catcode `\#12\catcode `\^12\catcode `\_12\catcode `\%12\relax}%
\providecommand \@@startlink[1]{}%
\providecommand \@@endlink[0]{}%
\providecommand \url  [0]{\begingroup\@sanitize@url \@url }%
\providecommand \@url [1]{\endgroup\@href {#1}{\urlprefix }}%
\providecommand \urlprefix  [0]{URL }%
\providecommand \Eprint [0]{\href }%
\providecommand \doibase [0]{https://doi.org/}%
\providecommand \selectlanguage [0]{\@gobble}%
\providecommand \bibinfo  [0]{\@secondoftwo}%
\providecommand \bibfield  [0]{\@secondoftwo}%
\providecommand \translation [1]{[#1]}%
\providecommand \BibitemOpen [0]{}%
\providecommand \bibitemStop [0]{}%
\providecommand \bibitemNoStop [0]{.\EOS\space}%
\providecommand \EOS [0]{\spacefactor3000\relax}%
\providecommand \BibitemShut  [1]{\csname bibitem#1\endcsname}%
\let\auto@bib@innerbib\@empty
\bibitem [{\citenamefont {Bloch}\ \emph {et~al.}(2012)\citenamefont {Bloch}, \citenamefont {Dalibard},\ and\ \citenamefont {Nascimb{\`e}ne}}]{Bloch2012}%
  \BibitemOpen
  \bibfield  {author} {\bibinfo {author} {\bibfnamefont {I.}~\bibnamefont {Bloch}}, \bibinfo {author} {\bibfnamefont {J.}~\bibnamefont {Dalibard}},\ and\ \bibinfo {author} {\bibfnamefont {S.}~\bibnamefont {Nascimb{\`e}ne}},\ }\bibfield  {title} {\bibinfo {title} {Quantum simulations with ultracold quantum gases},\ }\href {https://doi.org/10.1038/nphys2259} {\bibfield  {journal} {\bibinfo  {journal} {Nat.~Phys.}\ }\textbf {\bibinfo {volume} {8}},\ \bibinfo {pages} {267} (\bibinfo {year} {2012})}\BibitemShut {NoStop}%
\bibitem [{\citenamefont {Georgescu}\ \emph {et~al.}(2014)\citenamefont {Georgescu}, \citenamefont {Ashhab},\ and\ \citenamefont {Nori}}]{Georgescu2014}%
  \BibitemOpen
  \bibfield  {author} {\bibinfo {author} {\bibfnamefont {I.~M.}\ \bibnamefont {Georgescu}}, \bibinfo {author} {\bibfnamefont {S.}~\bibnamefont {Ashhab}},\ and\ \bibinfo {author} {\bibfnamefont {F.}~\bibnamefont {Nori}},\ }\bibfield  {title} {\bibinfo {title} {Quantum simulation},\ }\href {https://doi.org/10.1103/RevModPhys.86.153} {\bibfield  {journal} {\bibinfo  {journal} {Rev. Mod. Phys.}\ }\textbf {\bibinfo {volume} {86}},\ \bibinfo {pages} {153} (\bibinfo {year} {2014})}\BibitemShut {NoStop}%
\bibitem [{\citenamefont {Kjaergaard}\ \emph {et~al.}(2020)\citenamefont {Kjaergaard}, \citenamefont {Schwartz}, \citenamefont {Braum\"{u}ller}, \citenamefont {Krantz}, \citenamefont {Wang}, \citenamefont {Gustavsson},\ and\ \citenamefont {Oliver}}]{Kjaergaard2020}%
  \BibitemOpen
  \bibfield  {author} {\bibinfo {author} {\bibfnamefont {M.}~\bibnamefont {Kjaergaard}}, \bibinfo {author} {\bibfnamefont {M.~E.}\ \bibnamefont {Schwartz}}, \bibinfo {author} {\bibfnamefont {J.}~\bibnamefont {Braum\"{u}ller}}, \bibinfo {author} {\bibfnamefont {P.}~\bibnamefont {Krantz}}, \bibinfo {author} {\bibfnamefont {J.~I.-J.}\ \bibnamefont {Wang}}, \bibinfo {author} {\bibfnamefont {S.}~\bibnamefont {Gustavsson}},\ and\ \bibinfo {author} {\bibfnamefont {W.~D.}\ \bibnamefont {Oliver}},\ }\bibfield  {title} {\bibinfo {title} {Superconducting qubits: Current state of play},\ }\href {https://doi.org/10.1146/annurev-conmatphys-031119-050605} {\bibfield  {journal} {\bibinfo  {journal} {Annual Review of Condensed Matter Physics}\ }\textbf {\bibinfo {volume} {11}},\ \bibinfo {pages} {369} (\bibinfo {year} {2020})}\BibitemShut {NoStop}%
\bibitem [{\citenamefont {Browaeys}\ and\ \citenamefont {Lahaye}(2020)}]{Browaeys2020}%
  \BibitemOpen
  \bibfield  {author} {\bibinfo {author} {\bibfnamefont {A.}~\bibnamefont {Browaeys}}\ and\ \bibinfo {author} {\bibfnamefont {T.}~\bibnamefont {Lahaye}},\ }\bibfield  {title} {\bibinfo {title} {Many-body physics with individually controlled {Rydberg} atoms},\ }\href {https://doi.org/10.1038/s41567-019-0733-z} {\bibfield  {journal} {\bibinfo  {journal} {Nat.~Phys.}\ }\textbf {\bibinfo {volume} {16}},\ \bibinfo {pages} {132} (\bibinfo {year} {2020})}\BibitemShut {NoStop}%
\bibitem [{\citenamefont {Monroe}\ \emph {et~al.}(2021)\citenamefont {Monroe}, \citenamefont {Campbell}, \citenamefont {Duan}, \citenamefont {Gong}, \citenamefont {Gorshkov}, \citenamefont {Hess}, \citenamefont {Islam}, \citenamefont {Kim}, \citenamefont {Linke}, \citenamefont {Pagano}, \citenamefont {Richerme}, \citenamefont {Senko},\ and\ \citenamefont {Yao}}]{MonroeRMP}%
  \BibitemOpen
  \bibfield  {author} {\bibinfo {author} {\bibfnamefont {C.}~\bibnamefont {Monroe}}, \bibinfo {author} {\bibfnamefont {W.~C.}\ \bibnamefont {Campbell}}, \bibinfo {author} {\bibfnamefont {L.-M.}\ \bibnamefont {Duan}}, \bibinfo {author} {\bibfnamefont {Z.-X.}\ \bibnamefont {Gong}}, \bibinfo {author} {\bibfnamefont {A.~V.}\ \bibnamefont {Gorshkov}}, \bibinfo {author} {\bibfnamefont {P.~W.}\ \bibnamefont {Hess}}, \bibinfo {author} {\bibfnamefont {R.}~\bibnamefont {Islam}}, \bibinfo {author} {\bibfnamefont {K.}~\bibnamefont {Kim}}, \bibinfo {author} {\bibfnamefont {N.~M.}\ \bibnamefont {Linke}}, \bibinfo {author} {\bibfnamefont {G.}~\bibnamefont {Pagano}}, \bibinfo {author} {\bibfnamefont {P.}~\bibnamefont {Richerme}}, \bibinfo {author} {\bibfnamefont {C.}~\bibnamefont {Senko}},\ and\ \bibinfo {author} {\bibfnamefont {N.~Y.}\ \bibnamefont {Yao}},\ }\bibfield  {title} {\bibinfo {title} {Programmable quantum simulations of spin systems with trapped ions},\ }\href {https://doi.org/10.1103/RevModPhys.93.025001}
  {\bibfield  {journal} {\bibinfo  {journal} {Rev. Mod. Phys.}\ }\textbf {\bibinfo {volume} {93}},\ \bibinfo {pages} {025001} (\bibinfo {year} {2021})}\BibitemShut {NoStop}%
\bibitem [{\citenamefont {Deutsch}(1991)}]{DeutschETH}%
  \BibitemOpen
  \bibfield  {author} {\bibinfo {author} {\bibfnamefont {J.~M.}\ \bibnamefont {Deutsch}},\ }\bibfield  {title} {\bibinfo {title} {Quantum statistical mechanics in a closed system},\ }\href {https://doi.org/10.1103/PhysRevA.43.2046} {\bibfield  {journal} {\bibinfo  {journal} {Phys. Rev. A}\ }\textbf {\bibinfo {volume} {43}},\ \bibinfo {pages} {2046} (\bibinfo {year} {1991})}\BibitemShut {NoStop}%
\bibitem [{\citenamefont {Srednicki}(1994)}]{SrednickiETH}%
  \BibitemOpen
  \bibfield  {author} {\bibinfo {author} {\bibfnamefont {M.}~\bibnamefont {Srednicki}},\ }\bibfield  {title} {\bibinfo {title} {Chaos and quantum thermalization},\ }\href {https://doi.org/10.1103/PhysRevE.50.888} {\bibfield  {journal} {\bibinfo  {journal} {Phys. Rev. E}\ }\textbf {\bibinfo {volume} {50}},\ \bibinfo {pages} {888} (\bibinfo {year} {1994})}\BibitemShut {NoStop}%
\bibitem [{\citenamefont {Rigol}\ \emph {et~al.}(2008)\citenamefont {Rigol}, \citenamefont {Dunjko},\ and\ \citenamefont {Olshanii}}]{RigolNature}%
  \BibitemOpen
  \bibfield  {author} {\bibinfo {author} {\bibfnamefont {M.}~\bibnamefont {Rigol}}, \bibinfo {author} {\bibfnamefont {V.}~\bibnamefont {Dunjko}},\ and\ \bibinfo {author} {\bibfnamefont {M.}~\bibnamefont {Olshanii}},\ }\bibfield  {title} {\bibinfo {title} {Thermalization and its mechanism for generic isolated quantum systems},\ }\href {https://doi.org/10.1038/nature06838} {\bibfield  {journal} {\bibinfo  {journal} {Nature}\ }\textbf {\bibinfo {volume} {452}},\ \bibinfo {pages} {854} (\bibinfo {year} {2008})}\BibitemShut {NoStop}%
\bibitem [{\citenamefont {D'Alessio}\ \emph {et~al.}(2016)\citenamefont {D'Alessio}, \citenamefont {Kafri}, \citenamefont {Polkovnikov},\ and\ \citenamefont {Rigol}}]{dAlessio2016}%
  \BibitemOpen
  \bibfield  {author} {\bibinfo {author} {\bibfnamefont {L.}~\bibnamefont {D'Alessio}}, \bibinfo {author} {\bibfnamefont {Y.}~\bibnamefont {Kafri}}, \bibinfo {author} {\bibfnamefont {A.}~\bibnamefont {Polkovnikov}},\ and\ \bibinfo {author} {\bibfnamefont {M.}~\bibnamefont {Rigol}},\ }\bibfield  {title} {\bibinfo {title} {From quantum chaos and eigenstate thermalization to statistical mechanics and thermodynamics},\ }\href {https://doi.org/10.1080/00018732.2016.1198134} {\bibfield  {journal} {\bibinfo  {journal} {Adv. Phys.}\ }\textbf {\bibinfo {volume} {65}},\ \bibinfo {pages} {239} (\bibinfo {year} {2016})}\BibitemShut {NoStop}%
\bibitem [{\citenamefont {Ueda}(2020)}]{Ueda2020}%
  \BibitemOpen
  \bibfield  {author} {\bibinfo {author} {\bibfnamefont {M.}~\bibnamefont {Ueda}},\ }\bibfield  {title} {\bibinfo {title} {Quantum equilibration, thermalization and prethermalization in ultracold atoms},\ }\href {https://doi.org/10.1038/s42254-020-0237-x} {\bibfield  {journal} {\bibinfo  {journal} {Nature Reviews Physics}\ }\textbf {\bibinfo {volume} {2}},\ \bibinfo {pages} {669} (\bibinfo {year} {2020})}\BibitemShut {NoStop}%
\bibitem [{\citenamefont {Serbyn}\ \emph {et~al.}(2021)\citenamefont {Serbyn}, \citenamefont {Abanin},\ and\ \citenamefont {Papi{\'c}}}]{serbyn2021quantum}%
  \BibitemOpen
  \bibfield  {author} {\bibinfo {author} {\bibfnamefont {M.}~\bibnamefont {Serbyn}}, \bibinfo {author} {\bibfnamefont {D.~A.}\ \bibnamefont {Abanin}},\ and\ \bibinfo {author} {\bibfnamefont {Z.}~\bibnamefont {Papi{\'c}}},\ }\bibfield  {title} {\bibinfo {title} {Quantum many-body scars and weak breaking of ergodicity},\ }\href {https://doi.org/https://doi.org/10.1038/s41567-021-01230-2} {\bibfield  {journal} {\bibinfo  {journal} {Nature Physics}\ }\textbf {\bibinfo {volume} {17}},\ \bibinfo {pages} {675} (\bibinfo {year} {2021})}\BibitemShut {NoStop}%
\bibitem [{\citenamefont {Moudgalya}\ \emph {et~al.}(2022)\citenamefont {Moudgalya}, \citenamefont {Bernevig},\ and\ \citenamefont {Regnault}}]{Moudgalya_2022}%
  \BibitemOpen
  \bibfield  {author} {\bibinfo {author} {\bibfnamefont {S.}~\bibnamefont {Moudgalya}}, \bibinfo {author} {\bibfnamefont {B.~A.}\ \bibnamefont {Bernevig}},\ and\ \bibinfo {author} {\bibfnamefont {N.}~\bibnamefont {Regnault}},\ }\bibfield  {title} {\bibinfo {title} {Quantum many-body scars and hilbert space fragmentation: a review of exact results},\ }\href {https://doi.org/10.1088/1361-6633/ac73a0} {\bibfield  {journal} {\bibinfo  {journal} {Reports on Progress in Physics}\ }\textbf {\bibinfo {volume} {85}},\ \bibinfo {pages} {086501} (\bibinfo {year} {2022})}\BibitemShut {NoStop}%
\bibitem [{\citenamefont {Chandran}\ \emph {et~al.}(2023)\citenamefont {Chandran}, \citenamefont {Iadecola}, \citenamefont {Khemani},\ and\ \citenamefont {Moessner}}]{ChadranReview}%
  \BibitemOpen
  \bibfield  {author} {\bibinfo {author} {\bibfnamefont {A.}~\bibnamefont {Chandran}}, \bibinfo {author} {\bibfnamefont {T.}~\bibnamefont {Iadecola}}, \bibinfo {author} {\bibfnamefont {V.}~\bibnamefont {Khemani}},\ and\ \bibinfo {author} {\bibfnamefont {R.}~\bibnamefont {Moessner}},\ }\bibfield  {title} {\bibinfo {title} {Quantum many-body scars: A quasiparticle perspective},\ }\href {https://doi.org/10.1146/annurev-conmatphys-031620-101617} {\bibfield  {journal} {\bibinfo  {journal} {Annual Review of Condensed Matter Physics}\ }\textbf {\bibinfo {volume} {14}},\ \bibinfo {pages} {443} (\bibinfo {year} {2023})}\BibitemShut {NoStop}%
\bibitem [{\citenamefont {Bernien}\ \emph {et~al.}(2017)\citenamefont {Bernien}, \citenamefont {Schwartz}, \citenamefont {Keesling}, \citenamefont {Levine}, \citenamefont {Omran}, \citenamefont {Pichler}, \citenamefont {Choi}, \citenamefont {Zibrov}, \citenamefont {Endres}, \citenamefont {Greiner} \emph {et~al.}}]{bernien2017probing}%
  \BibitemOpen
  \bibfield  {author} {\bibinfo {author} {\bibfnamefont {H.}~\bibnamefont {Bernien}}, \bibinfo {author} {\bibfnamefont {S.}~\bibnamefont {Schwartz}}, \bibinfo {author} {\bibfnamefont {A.}~\bibnamefont {Keesling}}, \bibinfo {author} {\bibfnamefont {H.}~\bibnamefont {Levine}}, \bibinfo {author} {\bibfnamefont {A.}~\bibnamefont {Omran}}, \bibinfo {author} {\bibfnamefont {H.}~\bibnamefont {Pichler}}, \bibinfo {author} {\bibfnamefont {S.}~\bibnamefont {Choi}}, \bibinfo {author} {\bibfnamefont {A.~S.}\ \bibnamefont {Zibrov}}, \bibinfo {author} {\bibfnamefont {M.}~\bibnamefont {Endres}}, \bibinfo {author} {\bibfnamefont {M.}~\bibnamefont {Greiner}}, \emph {et~al.},\ }\bibfield  {title} {\bibinfo {title} {Probing many-body dynamics on a 51-atom quantum simulator},\ }\href {https://doi.org/https://doi.org/10.1038/nature24622} {\bibfield  {journal} {\bibinfo  {journal} {Nature}\ }\textbf {\bibinfo {volume} {551}},\ \bibinfo {pages} {579} (\bibinfo {year} {2017})}\BibitemShut {NoStop}%
\bibitem [{\citenamefont {Turner}\ \emph {et~al.}(2018{\natexlab{a}})\citenamefont {Turner}, \citenamefont {Michailidis}, \citenamefont {Abanin}, \citenamefont {Serbyn},\ and\ \citenamefont {Papi{\'c}}}]{turner2018weak}%
  \BibitemOpen
  \bibfield  {author} {\bibinfo {author} {\bibfnamefont {C.~J.}\ \bibnamefont {Turner}}, \bibinfo {author} {\bibfnamefont {A.~A.}\ \bibnamefont {Michailidis}}, \bibinfo {author} {\bibfnamefont {D.~A.}\ \bibnamefont {Abanin}}, \bibinfo {author} {\bibfnamefont {M.}~\bibnamefont {Serbyn}},\ and\ \bibinfo {author} {\bibfnamefont {Z.}~\bibnamefont {Papi{\'c}}},\ }\bibfield  {title} {\bibinfo {title} {Weak ergodicity breaking from quantum many-body scars},\ }\href {https://doi.org/10.1038/s41567-018-0137-5} {\bibfield  {journal} {\bibinfo  {journal} {Nature Physics}\ }\textbf {\bibinfo {volume} {14}},\ \bibinfo {pages} {745} (\bibinfo {year} {2018}{\natexlab{a}})}\BibitemShut {NoStop}%
\bibitem [{\citenamefont {Ho}\ \emph {et~al.}(2019)\citenamefont {Ho}, \citenamefont {Choi}, \citenamefont {Pichler},\ and\ \citenamefont {Lukin}}]{Ho2019}%
  \BibitemOpen
  \bibfield  {author} {\bibinfo {author} {\bibfnamefont {W.~W.}\ \bibnamefont {Ho}}, \bibinfo {author} {\bibfnamefont {S.}~\bibnamefont {Choi}}, \bibinfo {author} {\bibfnamefont {H.}~\bibnamefont {Pichler}},\ and\ \bibinfo {author} {\bibfnamefont {M.~D.}\ \bibnamefont {Lukin}},\ }\bibfield  {title} {\bibinfo {title} {Periodic orbits, entanglement, and quantum many-body scars in constrained models: Matrix product state approach},\ }\href {https://doi.org/10.1103/PhysRevLett.122.040603} {\bibfield  {journal} {\bibinfo  {journal} {Phys. Rev. Lett.}\ }\textbf {\bibinfo {volume} {122}},\ \bibinfo {pages} {040603} (\bibinfo {year} {2019})}\BibitemShut {NoStop}%
\bibitem [{\citenamefont {Choi}\ \emph {et~al.}(2019)\citenamefont {Choi}, \citenamefont {Turner}, \citenamefont {Pichler}, \citenamefont {Ho}, \citenamefont {Michailidis}, \citenamefont {Papi\ifmmode~\acute{c}\else \'{c}\fi{}}, \citenamefont {Serbyn}, \citenamefont {Lukin},\ and\ \citenamefont {Abanin}}]{Choi2019}%
  \BibitemOpen
  \bibfield  {author} {\bibinfo {author} {\bibfnamefont {S.}~\bibnamefont {Choi}}, \bibinfo {author} {\bibfnamefont {C.~J.}\ \bibnamefont {Turner}}, \bibinfo {author} {\bibfnamefont {H.}~\bibnamefont {Pichler}}, \bibinfo {author} {\bibfnamefont {W.~W.}\ \bibnamefont {Ho}}, \bibinfo {author} {\bibfnamefont {A.~A.}\ \bibnamefont {Michailidis}}, \bibinfo {author} {\bibfnamefont {Z.}~\bibnamefont {Papi\ifmmode~\acute{c}\else \'{c}\fi{}}}, \bibinfo {author} {\bibfnamefont {M.}~\bibnamefont {Serbyn}}, \bibinfo {author} {\bibfnamefont {M.~D.}\ \bibnamefont {Lukin}},\ and\ \bibinfo {author} {\bibfnamefont {D.~A.}\ \bibnamefont {Abanin}},\ }\bibfield  {title} {\bibinfo {title} {Emergent su(2) dynamics and perfect quantum many-body scars},\ }\href {https://doi.org/10.1103/PhysRevLett.122.220603} {\bibfield  {journal} {\bibinfo  {journal} {Phys. Rev. Lett.}\ }\textbf {\bibinfo {volume} {122}},\ \bibinfo {pages} {220603} (\bibinfo {year} {2019})}\BibitemShut {NoStop}%
\bibitem [{\citenamefont {Surace}\ \emph {et~al.}(2020)\citenamefont {Surace}, \citenamefont {Mazza}, \citenamefont {Giudici}, \citenamefont {Lerose}, \citenamefont {Gambassi},\ and\ \citenamefont {Dalmonte}}]{Surace2020}%
  \BibitemOpen
  \bibfield  {author} {\bibinfo {author} {\bibfnamefont {F.~M.}\ \bibnamefont {Surace}}, \bibinfo {author} {\bibfnamefont {P.~P.}\ \bibnamefont {Mazza}}, \bibinfo {author} {\bibfnamefont {G.}~\bibnamefont {Giudici}}, \bibinfo {author} {\bibfnamefont {A.}~\bibnamefont {Lerose}}, \bibinfo {author} {\bibfnamefont {A.}~\bibnamefont {Gambassi}},\ and\ \bibinfo {author} {\bibfnamefont {M.}~\bibnamefont {Dalmonte}},\ }\bibfield  {title} {\bibinfo {title} {Lattice gauge theories and string dynamics in {Rydberg} atom quantum simulators},\ }\href {https://doi.org/10.1103/PhysRevX.10.021041} {\bibfield  {journal} {\bibinfo  {journal} {Phys. Rev. X}\ }\textbf {\bibinfo {volume} {10}},\ \bibinfo {pages} {021041} (\bibinfo {year} {2020})}\BibitemShut {NoStop}%
\bibitem [{\citenamefont {Khemani}\ \emph {et~al.}(2019)\citenamefont {Khemani}, \citenamefont {Laumann},\ and\ \citenamefont {Chandran}}]{Khemani2019}%
  \BibitemOpen
  \bibfield  {author} {\bibinfo {author} {\bibfnamefont {V.}~\bibnamefont {Khemani}}, \bibinfo {author} {\bibfnamefont {C.~R.}\ \bibnamefont {Laumann}},\ and\ \bibinfo {author} {\bibfnamefont {A.}~\bibnamefont {Chandran}},\ }\bibfield  {title} {\bibinfo {title} {Signatures of integrability in the dynamics of {Rydberg}-blockaded chains},\ }\href {https://doi.org/10.1103/PhysRevB.99.161101} {\bibfield  {journal} {\bibinfo  {journal} {Phys. Rev. B}\ }\textbf {\bibinfo {volume} {99}},\ \bibinfo {pages} {161101} (\bibinfo {year} {2019})}\BibitemShut {NoStop}%
\bibitem [{\citenamefont {Lin}\ and\ \citenamefont {Motrunich}(2019)}]{lin2019exact}%
  \BibitemOpen
  \bibfield  {author} {\bibinfo {author} {\bibfnamefont {C.-J.}\ \bibnamefont {Lin}}\ and\ \bibinfo {author} {\bibfnamefont {O.~I.}\ \bibnamefont {Motrunich}},\ }\bibfield  {title} {\bibinfo {title} {Exact quantum many-body scar states in the rydberg-blockaded atom chain},\ }\href {https://doi.org/10.1103/PhysRevLett.122.173401} {\bibfield  {journal} {\bibinfo  {journal} {Phys. Rev. Lett.}\ }\textbf {\bibinfo {volume} {122}},\ \bibinfo {pages} {173401} (\bibinfo {year} {2019})}\BibitemShut {NoStop}%
\bibitem [{\citenamefont {Lin}\ \emph {et~al.}(2020{\natexlab{a}})\citenamefont {Lin}, \citenamefont {Calvera},\ and\ \citenamefont {Hsieh}}]{lin2020quantum}%
  \BibitemOpen
  \bibfield  {author} {\bibinfo {author} {\bibfnamefont {C.-J.}\ \bibnamefont {Lin}}, \bibinfo {author} {\bibfnamefont {V.}~\bibnamefont {Calvera}},\ and\ \bibinfo {author} {\bibfnamefont {T.~H.}\ \bibnamefont {Hsieh}},\ }\bibfield  {title} {\bibinfo {title} {Quantum many-body scar states in two-dimensional rydberg atom arrays},\ }\href {https://doi.org/10.1103/PhysRevB.101.220304} {\bibfield  {journal} {\bibinfo  {journal} {Phys. Rev. B}\ }\textbf {\bibinfo {volume} {101}},\ \bibinfo {pages} {220304} (\bibinfo {year} {2020}{\natexlab{a}})}\BibitemShut {NoStop}%
\bibitem [{\citenamefont {Bluvstein}\ \emph {et~al.}(2021)\citenamefont {Bluvstein}, \citenamefont {Omran}, \citenamefont {Levine}, \citenamefont {Keesling}, \citenamefont {Semeghini}, \citenamefont {Ebadi}, \citenamefont {Wang}, \citenamefont {Michailidis}, \citenamefont {Maskara}, \citenamefont {Ho}, \citenamefont {Choi}, \citenamefont {Serbyn}, \citenamefont {Greiner}, \citenamefont {Vuletić},\ and\ \citenamefont {Lukin}}]{bluvstein2021controlling}%
  \BibitemOpen
  \bibfield  {author} {\bibinfo {author} {\bibfnamefont {D.}~\bibnamefont {Bluvstein}}, \bibinfo {author} {\bibfnamefont {A.}~\bibnamefont {Omran}}, \bibinfo {author} {\bibfnamefont {H.}~\bibnamefont {Levine}}, \bibinfo {author} {\bibfnamefont {A.}~\bibnamefont {Keesling}}, \bibinfo {author} {\bibfnamefont {G.}~\bibnamefont {Semeghini}}, \bibinfo {author} {\bibfnamefont {S.}~\bibnamefont {Ebadi}}, \bibinfo {author} {\bibfnamefont {T.~T.}\ \bibnamefont {Wang}}, \bibinfo {author} {\bibfnamefont {A.~A.}\ \bibnamefont {Michailidis}}, \bibinfo {author} {\bibfnamefont {N.}~\bibnamefont {Maskara}}, \bibinfo {author} {\bibfnamefont {W.~W.}\ \bibnamefont {Ho}}, \bibinfo {author} {\bibfnamefont {S.}~\bibnamefont {Choi}}, \bibinfo {author} {\bibfnamefont {M.}~\bibnamefont {Serbyn}}, \bibinfo {author} {\bibfnamefont {M.}~\bibnamefont {Greiner}}, \bibinfo {author} {\bibfnamefont {V.}~\bibnamefont {Vuletić}},\ and\ \bibinfo {author} {\bibfnamefont {M.~D.}\ \bibnamefont {Lukin}},\ }\bibfield  {title} {\bibinfo {title}
  {Controlling quantum many-body dynamics in driven rydberg atom arrays},\ }\href {https://doi.org/10.1126/science.abg2530} {\bibfield  {journal} {\bibinfo  {journal} {Science}\ }\textbf {\bibinfo {volume} {371}},\ \bibinfo {pages} {1355} (\bibinfo {year} {2021})}\BibitemShut {NoStop}%
\bibitem [{\citenamefont {Mondragon-Shem}\ \emph {et~al.}(2021)\citenamefont {Mondragon-Shem}, \citenamefont {Vavilov},\ and\ \citenamefont {Martin}}]{mondragon2021fate}%
  \BibitemOpen
  \bibfield  {author} {\bibinfo {author} {\bibfnamefont {I.}~\bibnamefont {Mondragon-Shem}}, \bibinfo {author} {\bibfnamefont {M.~G.}\ \bibnamefont {Vavilov}},\ and\ \bibinfo {author} {\bibfnamefont {I.}~\bibnamefont {Martin}},\ }\bibfield  {title} {\bibinfo {title} {Fate of quantum many-body scars in the presence of disorder},\ }\href {https://doi.org/10.1103/PRXQuantum.2.030349} {\bibfield  {journal} {\bibinfo  {journal} {PRX Quantum}\ }\textbf {\bibinfo {volume} {2}},\ \bibinfo {pages} {030349} (\bibinfo {year} {2021})}\BibitemShut {NoStop}%
\bibitem [{\citenamefont {Omiya}\ and\ \citenamefont {M\"uller}(2023{\natexlab{a}})}]{Omiya2022}%
  \BibitemOpen
  \bibfield  {author} {\bibinfo {author} {\bibfnamefont {K.}~\bibnamefont {Omiya}}\ and\ \bibinfo {author} {\bibfnamefont {M.}~\bibnamefont {M\"uller}},\ }\bibfield  {title} {\bibinfo {title} {Quantum many-body scars in bipartite rydberg arrays originating from hidden projector embedding},\ }\href {https://doi.org/10.1103/PhysRevA.107.023318} {\bibfield  {journal} {\bibinfo  {journal} {Phys. Rev. A}\ }\textbf {\bibinfo {volume} {107}},\ \bibinfo {pages} {023318} (\bibinfo {year} {2023}{\natexlab{a}})}\BibitemShut {NoStop}%
\bibitem [{\citenamefont {Zhao}\ \emph {et~al.}(2025)\citenamefont {Zhao}, \citenamefont {Datla}, \citenamefont {Tian}, \citenamefont {Aliyu},\ and\ \citenamefont {Loh}}]{zhao2024observationquantumthermalizationrestricted}%
  \BibitemOpen
  \bibfield  {author} {\bibinfo {author} {\bibfnamefont {L.}~\bibnamefont {Zhao}}, \bibinfo {author} {\bibfnamefont {P.~R.}\ \bibnamefont {Datla}}, \bibinfo {author} {\bibfnamefont {W.}~\bibnamefont {Tian}}, \bibinfo {author} {\bibfnamefont {M.~M.}\ \bibnamefont {Aliyu}},\ and\ \bibinfo {author} {\bibfnamefont {H.}~\bibnamefont {Loh}},\ }\bibfield  {title} {\bibinfo {title} {Observation of quantum thermalization restricted to hilbert space fragments and ${\mathbb{z}}_{2k}$ scars},\ }\href {https://doi.org/10.1103/PhysRevX.15.011035} {\bibfield  {journal} {\bibinfo  {journal} {Phys. Rev. X}\ }\textbf {\bibinfo {volume} {15}},\ \bibinfo {pages} {011035} (\bibinfo {year} {2025})}\BibitemShut {NoStop}%
\bibitem [{\citenamefont {Kerschbaumer}\ \emph {et~al.}(2024)\citenamefont {Kerschbaumer}, \citenamefont {Ljubotina}, \citenamefont {Serbyn},\ and\ \citenamefont {Desaules}}]{kerschbaumer2024quantummanybodyscarspxp}%
  \BibitemOpen
  \bibfield  {author} {\bibinfo {author} {\bibfnamefont {A.}~\bibnamefont {Kerschbaumer}}, \bibinfo {author} {\bibfnamefont {M.}~\bibnamefont {Ljubotina}}, \bibinfo {author} {\bibfnamefont {M.}~\bibnamefont {Serbyn}},\ and\ \bibinfo {author} {\bibfnamefont {J.-Y.}\ \bibnamefont {Desaules}},\ }\href {https://arxiv.org/abs/2410.18913} {\bibinfo {title} {Quantum many-body scars beyond the pxp model in rydberg simulators}} (\bibinfo {year} {2024}),\ \Eprint {https://arxiv.org/abs/2410.18913} {arXiv:2410.18913 [quant-ph]} \BibitemShut {NoStop}%
\bibitem [{\citenamefont {Ding}\ \emph {et~al.}(2024)\citenamefont {Ding}, \citenamefont {Bai}, \citenamefont {Liu}, \citenamefont {Shi}, \citenamefont {Guo}, \citenamefont {Li},\ and\ \citenamefont {Adams}}]{Ding2024RydbergClusters}%
  \BibitemOpen
  \bibfield  {author} {\bibinfo {author} {\bibfnamefont {D.}~\bibnamefont {Ding}}, \bibinfo {author} {\bibfnamefont {Z.}~\bibnamefont {Bai}}, \bibinfo {author} {\bibfnamefont {Z.}~\bibnamefont {Liu}}, \bibinfo {author} {\bibfnamefont {B.}~\bibnamefont {Shi}}, \bibinfo {author} {\bibfnamefont {G.}~\bibnamefont {Guo}}, \bibinfo {author} {\bibfnamefont {W.}~\bibnamefont {Li}},\ and\ \bibinfo {author} {\bibfnamefont {C.~S.}\ \bibnamefont {Adams}},\ }\bibfield  {title} {\bibinfo {title} {Ergodicity breaking from rydberg clusters in a driven-dissipative many-body system},\ }\href {https://doi.org/10.1126/sciadv.adl5893} {\bibfield  {journal} {\bibinfo  {journal} {Science Advances}\ }\textbf {\bibinfo {volume} {10}},\ \bibinfo {pages} {eadl5893} (\bibinfo {year} {2024})}\BibitemShut {NoStop}%
\bibitem [{\citenamefont {Ivanov}\ and\ \citenamefont {Motrunich}(2025)}]{ivanov2025exactarealawscareigenstates}%
  \BibitemOpen
  \bibfield  {author} {\bibinfo {author} {\bibfnamefont {A.~N.}\ \bibnamefont {Ivanov}}\ and\ \bibinfo {author} {\bibfnamefont {O.~I.}\ \bibnamefont {Motrunich}},\ }\href {https://arxiv.org/abs/2503.16327} {\bibinfo {title} {Many exact area-law scar eigenstates in the nonintegrable pxp and related models}} (\bibinfo {year} {2025}),\ \Eprint {https://arxiv.org/abs/2503.16327} {arXiv:2503.16327 [quant-ph]} \BibitemShut {NoStop}%
\bibitem [{\citenamefont {Moudgalya}\ \emph {et~al.}(2018)\citenamefont {Moudgalya}, \citenamefont {Regnault},\ and\ \citenamefont {Bernevig}}]{Moudgalya2018_2}%
  \BibitemOpen
  \bibfield  {author} {\bibinfo {author} {\bibfnamefont {S.}~\bibnamefont {Moudgalya}}, \bibinfo {author} {\bibfnamefont {N.}~\bibnamefont {Regnault}},\ and\ \bibinfo {author} {\bibfnamefont {B.~A.}\ \bibnamefont {Bernevig}},\ }\bibfield  {title} {\bibinfo {title} {Entanglement of exact excited states of affleck-kennedy-lieb-tasaki models: Exact results, many-body scars, and violation of the strong eigenstate thermalization hypothesis},\ }\href {https://doi.org/10.1103/PhysRevB.98.235156} {\bibfield  {journal} {\bibinfo  {journal} {Phys. Rev. B}\ }\textbf {\bibinfo {volume} {98}},\ \bibinfo {pages} {235156} (\bibinfo {year} {2018})}\BibitemShut {NoStop}%
\bibitem [{\citenamefont {Schecter}\ and\ \citenamefont {Iadecola}(2019)}]{Schecter2019}%
  \BibitemOpen
  \bibfield  {author} {\bibinfo {author} {\bibfnamefont {M.}~\bibnamefont {Schecter}}\ and\ \bibinfo {author} {\bibfnamefont {T.}~\bibnamefont {Iadecola}},\ }\bibfield  {title} {\bibinfo {title} {Weak ergodicity breaking and quantum many-body scars in spin-1 $xy$ magnets},\ }\href {https://doi.org/10.1103/PhysRevLett.123.147201} {\bibfield  {journal} {\bibinfo  {journal} {Phys. Rev. Lett.}\ }\textbf {\bibinfo {volume} {123}},\ \bibinfo {pages} {147201} (\bibinfo {year} {2019})}\BibitemShut {NoStop}%
\bibitem [{\citenamefont {Yang}\ \emph {et~al.}(2024)\citenamefont {Yang}, \citenamefont {Zhang}, \citenamefont {Li}, \citenamefont {Lin}, \citenamefont {Gopalakrishnan}, \citenamefont {Rigol},\ and\ \citenamefont {Lev}}]{yang2024phantom}%
  \BibitemOpen
  \bibfield  {author} {\bibinfo {author} {\bibfnamefont {K.}~\bibnamefont {Yang}}, \bibinfo {author} {\bibfnamefont {Y.}~\bibnamefont {Zhang}}, \bibinfo {author} {\bibfnamefont {K.-Y.}\ \bibnamefont {Li}}, \bibinfo {author} {\bibfnamefont {K.-Y.}\ \bibnamefont {Lin}}, \bibinfo {author} {\bibfnamefont {S.}~\bibnamefont {Gopalakrishnan}}, \bibinfo {author} {\bibfnamefont {M.}~\bibnamefont {Rigol}},\ and\ \bibinfo {author} {\bibfnamefont {B.~L.}\ \bibnamefont {Lev}},\ }\bibfield  {title} {\bibinfo {title} {Phantom energy in the nonlinear response of a quantum many-body scar state},\ }\href {https://doi.org/10.1126/science.adk8978} {\bibfield  {journal} {\bibinfo  {journal} {Science}\ }\textbf {\bibinfo {volume} {385}},\ \bibinfo {pages} {1063} (\bibinfo {year} {2024})}\BibitemShut {NoStop}%
\bibitem [{\citenamefont {Guo}\ \emph {et~al.}(2023)\citenamefont {Guo}, \citenamefont {Liu}, \citenamefont {Gao}, \citenamefont {Yang}, \citenamefont {Wang}, \citenamefont {Ma},\ and\ \citenamefont {Ying}}]{Guo2023Origin}%
  \BibitemOpen
  \bibfield  {author} {\bibinfo {author} {\bibfnamefont {Z.}~\bibnamefont {Guo}}, \bibinfo {author} {\bibfnamefont {B.}~\bibnamefont {Liu}}, \bibinfo {author} {\bibfnamefont {Y.}~\bibnamefont {Gao}}, \bibinfo {author} {\bibfnamefont {A.}~\bibnamefont {Yang}}, \bibinfo {author} {\bibfnamefont {J.}~\bibnamefont {Wang}}, \bibinfo {author} {\bibfnamefont {J.}~\bibnamefont {Ma}},\ and\ \bibinfo {author} {\bibfnamefont {L.}~\bibnamefont {Ying}},\ }\bibfield  {title} {\bibinfo {title} {Origin of hilbert-space quantum scars in unconstrained models},\ }\href {https://doi.org/10.1103/PhysRevB.108.075124} {\bibfield  {journal} {\bibinfo  {journal} {Phys. Rev. B}\ }\textbf {\bibinfo {volume} {108}},\ \bibinfo {pages} {075124} (\bibinfo {year} {2023})}\BibitemShut {NoStop}%
\bibitem [{\citenamefont {Hudomal}\ \emph {et~al.}(2020)\citenamefont {Hudomal}, \citenamefont {Vasi{\'{c}}}, \citenamefont {Regnault},\ and\ \citenamefont {Papi{\'{c}}}}]{Hudomal2020bosons}%
  \BibitemOpen
  \bibfield  {author} {\bibinfo {author} {\bibfnamefont {A.}~\bibnamefont {Hudomal}}, \bibinfo {author} {\bibfnamefont {I.}~\bibnamefont {Vasi{\'{c}}}}, \bibinfo {author} {\bibfnamefont {N.}~\bibnamefont {Regnault}},\ and\ \bibinfo {author} {\bibfnamefont {Z.}~\bibnamefont {Papi{\'{c}}}},\ }\bibfield  {title} {\bibinfo {title} {Quantum scars of bosons with correlated hopping},\ }\href {https://doi.org/10.1038/s42005-020-0364-9} {\bibfield  {journal} {\bibinfo  {journal} {Communications Physics}\ }\textbf {\bibinfo {volume} {3}},\ \bibinfo {pages} {99} (\bibinfo {year} {2020})}\BibitemShut {NoStop}%
\bibitem [{\citenamefont {Zhao}\ \emph {et~al.}(2020)\citenamefont {Zhao}, \citenamefont {Vovrosh}, \citenamefont {Mintert},\ and\ \citenamefont {Knolle}}]{Zhao2020Optical}%
  \BibitemOpen
  \bibfield  {author} {\bibinfo {author} {\bibfnamefont {H.}~\bibnamefont {Zhao}}, \bibinfo {author} {\bibfnamefont {J.}~\bibnamefont {Vovrosh}}, \bibinfo {author} {\bibfnamefont {F.}~\bibnamefont {Mintert}},\ and\ \bibinfo {author} {\bibfnamefont {J.}~\bibnamefont {Knolle}},\ }\bibfield  {title} {\bibinfo {title} {Quantum many-body scars in optical lattices},\ }\href {https://doi.org/10.1103/PhysRevLett.124.160604} {\bibfield  {journal} {\bibinfo  {journal} {Phys. Rev. Lett.}\ }\textbf {\bibinfo {volume} {124}},\ \bibinfo {pages} {160604} (\bibinfo {year} {2020})}\BibitemShut {NoStop}%
\bibitem [{\citenamefont {Desaules}\ \emph {et~al.}(2021)\citenamefont {Desaules}, \citenamefont {Hudomal}, \citenamefont {Turner},\ and\ \citenamefont {Papi\ifmmode~\acute{c}\else \'{c}\fi{}}}]{Desaules2021Tilted}%
  \BibitemOpen
  \bibfield  {author} {\bibinfo {author} {\bibfnamefont {J.-Y.}\ \bibnamefont {Desaules}}, \bibinfo {author} {\bibfnamefont {A.}~\bibnamefont {Hudomal}}, \bibinfo {author} {\bibfnamefont {C.~J.}\ \bibnamefont {Turner}},\ and\ \bibinfo {author} {\bibfnamefont {Z.}~\bibnamefont {Papi\ifmmode~\acute{c}\else \'{c}\fi{}}},\ }\bibfield  {title} {\bibinfo {title} {Proposal for realizing quantum scars in the tilted 1d fermi-hubbard model},\ }\href {https://doi.org/10.1103/PhysRevLett.126.210601} {\bibfield  {journal} {\bibinfo  {journal} {Phys. Rev. Lett.}\ }\textbf {\bibinfo {volume} {126}},\ \bibinfo {pages} {210601} (\bibinfo {year} {2021})}\BibitemShut {NoStop}%
\bibitem [{\citenamefont {Scherg}\ \emph {et~al.}(2021)\citenamefont {Scherg}, \citenamefont {Kohlert}, \citenamefont {Sala}, \citenamefont {Pollmann}, \citenamefont {Hebbe~Madhusudhana}, \citenamefont {Bloch},\ and\ \citenamefont {Aidelsburger}}]{scherg2021observing}%
  \BibitemOpen
  \bibfield  {author} {\bibinfo {author} {\bibfnamefont {S.}~\bibnamefont {Scherg}}, \bibinfo {author} {\bibfnamefont {T.}~\bibnamefont {Kohlert}}, \bibinfo {author} {\bibfnamefont {P.}~\bibnamefont {Sala}}, \bibinfo {author} {\bibfnamefont {F.}~\bibnamefont {Pollmann}}, \bibinfo {author} {\bibfnamefont {B.}~\bibnamefont {Hebbe~Madhusudhana}}, \bibinfo {author} {\bibfnamefont {I.}~\bibnamefont {Bloch}},\ and\ \bibinfo {author} {\bibfnamefont {M.}~\bibnamefont {Aidelsburger}},\ }\bibfield  {title} {\bibinfo {title} {Observing non-ergodicity due to kinetic constraints in tilted fermi-hubbard chains},\ }\href {https://doi.org/10.1038/s41467-021-24726-0} {\bibfield  {journal} {\bibinfo  {journal} {Nature Communications}\ }\textbf {\bibinfo {volume} {12}},\ \bibinfo {pages} {4490} (\bibinfo {year} {2021})}\BibitemShut {NoStop}%
\bibitem [{\citenamefont {Su}\ \emph {et~al.}(2023)\citenamefont {Su}, \citenamefont {Sun}, \citenamefont {Hudomal}, \citenamefont {Desaules}, \citenamefont {Zhou}, \citenamefont {Yang}, \citenamefont {Halimeh}, \citenamefont {Yuan}, \citenamefont {Papi\ifmmode~\acute{c}\else \'{c}\fi{}},\ and\ \citenamefont {Pan}}]{Su2023}%
  \BibitemOpen
  \bibfield  {author} {\bibinfo {author} {\bibfnamefont {G.-X.}\ \bibnamefont {Su}}, \bibinfo {author} {\bibfnamefont {H.}~\bibnamefont {Sun}}, \bibinfo {author} {\bibfnamefont {A.}~\bibnamefont {Hudomal}}, \bibinfo {author} {\bibfnamefont {J.-Y.}\ \bibnamefont {Desaules}}, \bibinfo {author} {\bibfnamefont {Z.-Y.}\ \bibnamefont {Zhou}}, \bibinfo {author} {\bibfnamefont {B.}~\bibnamefont {Yang}}, \bibinfo {author} {\bibfnamefont {J.~C.}\ \bibnamefont {Halimeh}}, \bibinfo {author} {\bibfnamefont {Z.-S.}\ \bibnamefont {Yuan}}, \bibinfo {author} {\bibfnamefont {Z.}~\bibnamefont {Papi\ifmmode~\acute{c}\else \'{c}\fi{}}},\ and\ \bibinfo {author} {\bibfnamefont {J.-W.}\ \bibnamefont {Pan}},\ }\bibfield  {title} {\bibinfo {title} {Observation of many-body scarring in a {Bose-Hubbard} quantum simulator},\ }\href {https://doi.org/10.1103/PhysRevResearch.5.023010} {\bibfield  {journal} {\bibinfo  {journal} {Phys. Rev. Res.}\ }\textbf {\bibinfo {volume} {5}},\ \bibinfo {pages} {023010} (\bibinfo {year}
  {2023})}\BibitemShut {NoStop}%
\bibitem [{\citenamefont {Adler}\ \emph {et~al.}(2024)\citenamefont {Adler}, \citenamefont {Wei}, \citenamefont {Will}, \citenamefont {Srakaew}, \citenamefont {Agrawal}, \citenamefont {Weckesser}, \citenamefont {Moessner}, \citenamefont {Pollmann}, \citenamefont {Bloch},\ and\ \citenamefont {Zeiher}}]{Adler2024}%
  \BibitemOpen
  \bibfield  {author} {\bibinfo {author} {\bibfnamefont {D.}~\bibnamefont {Adler}}, \bibinfo {author} {\bibfnamefont {D.}~\bibnamefont {Wei}}, \bibinfo {author} {\bibfnamefont {M.}~\bibnamefont {Will}}, \bibinfo {author} {\bibfnamefont {K.}~\bibnamefont {Srakaew}}, \bibinfo {author} {\bibfnamefont {S.}~\bibnamefont {Agrawal}}, \bibinfo {author} {\bibfnamefont {P.}~\bibnamefont {Weckesser}}, \bibinfo {author} {\bibfnamefont {R.}~\bibnamefont {Moessner}}, \bibinfo {author} {\bibfnamefont {F.}~\bibnamefont {Pollmann}}, \bibinfo {author} {\bibfnamefont {I.}~\bibnamefont {Bloch}},\ and\ \bibinfo {author} {\bibfnamefont {J.}~\bibnamefont {Zeiher}},\ }\bibfield  {title} {\bibinfo {title} {Observation of hilbert space fragmentation and fractonic excitations in 2d},\ }\href {https://doi.org/10.1038/s41586-024-08188-0} {\bibfield  {journal} {\bibinfo  {journal} {Nature}\ }\textbf {\bibinfo {volume} {636}},\ \bibinfo {pages} {80} (\bibinfo {year} {2024})}\BibitemShut {NoStop}%
\bibitem [{\citenamefont {Zhang}\ \emph {et~al.}(2023)\citenamefont {Zhang}, \citenamefont {Dong}, \citenamefont {Gao}, \citenamefont {Zhao}, \citenamefont {Hao}, \citenamefont {Desaules}, \citenamefont {Guo}, \citenamefont {Chen}, \citenamefont {Deng}, \citenamefont {Liu} \emph {et~al.}}]{zhang2023many}%
  \BibitemOpen
  \bibfield  {author} {\bibinfo {author} {\bibfnamefont {P.}~\bibnamefont {Zhang}}, \bibinfo {author} {\bibfnamefont {H.}~\bibnamefont {Dong}}, \bibinfo {author} {\bibfnamefont {Y.}~\bibnamefont {Gao}}, \bibinfo {author} {\bibfnamefont {L.}~\bibnamefont {Zhao}}, \bibinfo {author} {\bibfnamefont {J.}~\bibnamefont {Hao}}, \bibinfo {author} {\bibfnamefont {J.-Y.}\ \bibnamefont {Desaules}}, \bibinfo {author} {\bibfnamefont {Q.}~\bibnamefont {Guo}}, \bibinfo {author} {\bibfnamefont {J.}~\bibnamefont {Chen}}, \bibinfo {author} {\bibfnamefont {J.}~\bibnamefont {Deng}}, \bibinfo {author} {\bibfnamefont {B.}~\bibnamefont {Liu}}, \emph {et~al.},\ }\bibfield  {title} {\bibinfo {title} {Many-body hilbert space scarring on a superconducting processor},\ }\href {https://doi.org/10.1038/s41567-022-01784-9} {\bibfield  {journal} {\bibinfo  {journal} {Nature Physics}\ }\textbf {\bibinfo {volume} {19}},\ \bibinfo {pages} {120} (\bibinfo {year} {2023})}\BibitemShut {NoStop}%
\bibitem [{\citenamefont {Larsen}\ \emph {et~al.}(2024)\citenamefont {Larsen}, \citenamefont {Nielsen}, \citenamefont {Eckardt},\ and\ \citenamefont {Petiziol}}]{larsen2024experimentalprotocolobservingsingle}%
  \BibitemOpen
  \bibfield  {author} {\bibinfo {author} {\bibfnamefont {P.~G.}\ \bibnamefont {Larsen}}, \bibinfo {author} {\bibfnamefont {A.~E.~B.}\ \bibnamefont {Nielsen}}, \bibinfo {author} {\bibfnamefont {A.}~\bibnamefont {Eckardt}},\ and\ \bibinfo {author} {\bibfnamefont {F.}~\bibnamefont {Petiziol}},\ }\href {https://arxiv.org/abs/2410.14613} {\bibinfo {title} {Experimental protocol for observing single quantum many-body scars with transmon qubits}} (\bibinfo {year} {2024}),\ \Eprint {https://arxiv.org/abs/2410.14613} {arXiv:2410.14613 [quant-ph]} \BibitemShut {NoStop}%
\bibitem [{\citenamefont {Shiraishi}\ and\ \citenamefont {Mori}(2017)}]{ShiraishiMori}%
  \BibitemOpen
  \bibfield  {author} {\bibinfo {author} {\bibfnamefont {N.}~\bibnamefont {Shiraishi}}\ and\ \bibinfo {author} {\bibfnamefont {T.}~\bibnamefont {Mori}},\ }\bibfield  {title} {\bibinfo {title} {Systematic construction of counterexamples to the eigenstate thermalization hypothesis},\ }\href {https://doi.org/10.1103/PhysRevLett.119.030601} {\bibfield  {journal} {\bibinfo  {journal} {Phys. Rev. Lett.}\ }\textbf {\bibinfo {volume} {119}},\ \bibinfo {pages} {030601} (\bibinfo {year} {2017})}\BibitemShut {NoStop}%
\bibitem [{\citenamefont {Ok}\ \emph {et~al.}(2019)\citenamefont {Ok}, \citenamefont {Choo}, \citenamefont {Mudry}, \citenamefont {Castelnovo}, \citenamefont {Chamon},\ and\ \citenamefont {Neupert}}]{NeupertScars}%
  \BibitemOpen
  \bibfield  {author} {\bibinfo {author} {\bibfnamefont {S.}~\bibnamefont {Ok}}, \bibinfo {author} {\bibfnamefont {K.}~\bibnamefont {Choo}}, \bibinfo {author} {\bibfnamefont {C.}~\bibnamefont {Mudry}}, \bibinfo {author} {\bibfnamefont {C.}~\bibnamefont {Castelnovo}}, \bibinfo {author} {\bibfnamefont {C.}~\bibnamefont {Chamon}},\ and\ \bibinfo {author} {\bibfnamefont {T.}~\bibnamefont {Neupert}},\ }\bibfield  {title} {\bibinfo {title} {Topological many-body scar states in dimensions one, two, and three},\ }\href {https://doi.org/10.1103/PhysRevResearch.1.033144} {\bibfield  {journal} {\bibinfo  {journal} {Phys. Rev. Research}\ }\textbf {\bibinfo {volume} {1}},\ \bibinfo {pages} {033144} (\bibinfo {year} {2019})}\BibitemShut {NoStop}%
\bibitem [{\citenamefont {Shibata}\ \emph {et~al.}(2020)\citenamefont {Shibata}, \citenamefont {Yoshioka},\ and\ \citenamefont {Katsura}}]{Shibata202Onsager}%
  \BibitemOpen
  \bibfield  {author} {\bibinfo {author} {\bibfnamefont {N.}~\bibnamefont {Shibata}}, \bibinfo {author} {\bibfnamefont {N.}~\bibnamefont {Yoshioka}},\ and\ \bibinfo {author} {\bibfnamefont {H.}~\bibnamefont {Katsura}},\ }\bibfield  {title} {\bibinfo {title} {Onsager's scars in disordered spin chains},\ }\href {https://doi.org/10.1103/PhysRevLett.124.180604} {\bibfield  {journal} {\bibinfo  {journal} {Phys. Rev. Lett.}\ }\textbf {\bibinfo {volume} {124}},\ \bibinfo {pages} {180604} (\bibinfo {year} {2020})}\BibitemShut {NoStop}%
\bibitem [{\citenamefont {McClarty}\ \emph {et~al.}(2020)\citenamefont {McClarty}, \citenamefont {Haque}, \citenamefont {Sen},\ and\ \citenamefont {Richter}}]{McClarty2020}%
  \BibitemOpen
  \bibfield  {author} {\bibinfo {author} {\bibfnamefont {P.~A.}\ \bibnamefont {McClarty}}, \bibinfo {author} {\bibfnamefont {M.}~\bibnamefont {Haque}}, \bibinfo {author} {\bibfnamefont {A.}~\bibnamefont {Sen}},\ and\ \bibinfo {author} {\bibfnamefont {J.}~\bibnamefont {Richter}},\ }\bibfield  {title} {\bibinfo {title} {Disorder-free localization and many-body quantum scars from magnetic frustration},\ }\href {https://doi.org/10.1103/PhysRevB.102.224303} {\bibfield  {journal} {\bibinfo  {journal} {Phys. Rev. B}\ }\textbf {\bibinfo {volume} {102}},\ \bibinfo {pages} {224303} (\bibinfo {year} {2020})}\BibitemShut {NoStop}%
\bibitem [{\citenamefont {Lee}\ \emph {et~al.}(2020)\citenamefont {Lee}, \citenamefont {Melendrez}, \citenamefont {Pal},\ and\ \citenamefont {Changlani}}]{Lee2020Colored}%
  \BibitemOpen
  \bibfield  {author} {\bibinfo {author} {\bibfnamefont {K.}~\bibnamefont {Lee}}, \bibinfo {author} {\bibfnamefont {R.}~\bibnamefont {Melendrez}}, \bibinfo {author} {\bibfnamefont {A.}~\bibnamefont {Pal}},\ and\ \bibinfo {author} {\bibfnamefont {H.~J.}\ \bibnamefont {Changlani}},\ }\bibfield  {title} {\bibinfo {title} {Exact three-colored quantum scars from geometric frustration},\ }\href {https://doi.org/10.1103/PhysRevB.101.241111} {\bibfield  {journal} {\bibinfo  {journal} {Phys. Rev. B}\ }\textbf {\bibinfo {volume} {101}},\ \bibinfo {pages} {241111} (\bibinfo {year} {2020})}\BibitemShut {NoStop}%
\bibitem [{\citenamefont {Pakrouski}\ \emph {et~al.}(2020)\citenamefont {Pakrouski}, \citenamefont {Pallegar}, \citenamefont {Popov},\ and\ \citenamefont {Klebanov}}]{pakrouski2020many}%
  \BibitemOpen
  \bibfield  {author} {\bibinfo {author} {\bibfnamefont {K.}~\bibnamefont {Pakrouski}}, \bibinfo {author} {\bibfnamefont {P.~N.}\ \bibnamefont {Pallegar}}, \bibinfo {author} {\bibfnamefont {F.~K.}\ \bibnamefont {Popov}},\ and\ \bibinfo {author} {\bibfnamefont {I.~R.}\ \bibnamefont {Klebanov}},\ }\bibfield  {title} {\bibinfo {title} {Many-body scars as a group invariant sector of hilbert space},\ }\href {https://doi.org/10.1103/PhysRevLett.125.230602} {\bibfield  {journal} {\bibinfo  {journal} {Phys. Rev. Lett.}\ }\textbf {\bibinfo {volume} {125}},\ \bibinfo {pages} {230602} (\bibinfo {year} {2020})}\BibitemShut {NoStop}%
\bibitem [{\citenamefont {Mark}\ and\ \citenamefont {Motrunich}(2020)}]{Mark2020eta}%
  \BibitemOpen
  \bibfield  {author} {\bibinfo {author} {\bibfnamefont {D.~K.}\ \bibnamefont {Mark}}\ and\ \bibinfo {author} {\bibfnamefont {O.~I.}\ \bibnamefont {Motrunich}},\ }\bibfield  {title} {\bibinfo {title} {$\ensuremath{\eta}$-pairing states as true scars in an extended hubbard model},\ }\href {https://doi.org/10.1103/PhysRevB.102.075132} {\bibfield  {journal} {\bibinfo  {journal} {Phys. Rev. B}\ }\textbf {\bibinfo {volume} {102}},\ \bibinfo {pages} {075132} (\bibinfo {year} {2020})}\BibitemShut {NoStop}%
\bibitem [{\citenamefont {Ren}\ \emph {et~al.}(2021)\citenamefont {Ren}, \citenamefont {Liang},\ and\ \citenamefont {Fang}}]{ren2021quasisymmetry}%
  \BibitemOpen
  \bibfield  {author} {\bibinfo {author} {\bibfnamefont {J.}~\bibnamefont {Ren}}, \bibinfo {author} {\bibfnamefont {C.}~\bibnamefont {Liang}},\ and\ \bibinfo {author} {\bibfnamefont {C.}~\bibnamefont {Fang}},\ }\bibfield  {title} {\bibinfo {title} {Quasisymmetry groups and many-body scar dynamics},\ }\href {https://doi.org/10.1103/PhysRevLett.126.120604} {\bibfield  {journal} {\bibinfo  {journal} {Phys. Rev. Lett.}\ }\textbf {\bibinfo {volume} {126}},\ \bibinfo {pages} {120604} (\bibinfo {year} {2021})}\BibitemShut {NoStop}%
\bibitem [{\citenamefont {Mohapatra}\ and\ \citenamefont {Balram}(2023)}]{Mohapatra2023}%
  \BibitemOpen
  \bibfield  {author} {\bibinfo {author} {\bibfnamefont {S.}~\bibnamefont {Mohapatra}}\ and\ \bibinfo {author} {\bibfnamefont {A.~C.}\ \bibnamefont {Balram}},\ }\bibfield  {title} {\bibinfo {title} {Pronounced quantum many-body scars in the one-dimensional spin-1 kitaev model},\ }\href {https://doi.org/10.1103/PhysRevB.107.235121} {\bibfield  {journal} {\bibinfo  {journal} {Phys. Rev. B}\ }\textbf {\bibinfo {volume} {107}},\ \bibinfo {pages} {235121} (\bibinfo {year} {2023})}\BibitemShut {NoStop}%
\bibitem [{\citenamefont {Kolb}\ and\ \citenamefont {Pakrouski}(2023)}]{kolb2023stability}%
  \BibitemOpen
  \bibfield  {author} {\bibinfo {author} {\bibfnamefont {P.}~\bibnamefont {Kolb}}\ and\ \bibinfo {author} {\bibfnamefont {K.}~\bibnamefont {Pakrouski}},\ }\bibfield  {title} {\bibinfo {title} {Stability of the many-body scars in fermionic spin-1/2 models},\ }\href {https://doi.org/10.1103/PRXQuantum.4.040348} {\bibfield  {journal} {\bibinfo  {journal} {PRX Quantum}\ }\textbf {\bibinfo {volume} {4}},\ \bibinfo {pages} {040348} (\bibinfo {year} {2023})}\BibitemShut {NoStop}%
\bibitem [{\citenamefont {Srivatsa}\ \emph {et~al.}(2023)\citenamefont {Srivatsa}, \citenamefont {Yarloo}, \citenamefont {Moessner},\ and\ \citenamefont {Nielsen}}]{Srivatsa2023}%
  \BibitemOpen
  \bibfield  {author} {\bibinfo {author} {\bibfnamefont {N.~S.}\ \bibnamefont {Srivatsa}}, \bibinfo {author} {\bibfnamefont {H.}~\bibnamefont {Yarloo}}, \bibinfo {author} {\bibfnamefont {R.}~\bibnamefont {Moessner}},\ and\ \bibinfo {author} {\bibfnamefont {A.~E.~B.}\ \bibnamefont {Nielsen}},\ }\bibfield  {title} {\bibinfo {title} {Mobility edges through inverted quantum many-body scarring},\ }\href {https://doi.org/10.1103/PhysRevB.108.L100202} {\bibfield  {journal} {\bibinfo  {journal} {Phys. Rev. B}\ }\textbf {\bibinfo {volume} {108}},\ \bibinfo {pages} {L100202} (\bibinfo {year} {2023})}\BibitemShut {NoStop}%
\bibitem [{\citenamefont {Desaules}\ \emph {et~al.}(2023)\citenamefont {Desaules}, \citenamefont {Banerjee}, \citenamefont {Hudomal}, \citenamefont {Papi\ifmmode~\acute{c}\else \'{c}\fi{}}, \citenamefont {Sen},\ and\ \citenamefont {Halimeh}}]{Desaules2023Schwinger}%
  \BibitemOpen
  \bibfield  {author} {\bibinfo {author} {\bibfnamefont {J.-Y.}\ \bibnamefont {Desaules}}, \bibinfo {author} {\bibfnamefont {D.}~\bibnamefont {Banerjee}}, \bibinfo {author} {\bibfnamefont {A.}~\bibnamefont {Hudomal}}, \bibinfo {author} {\bibfnamefont {Z.}~\bibnamefont {Papi\ifmmode~\acute{c}\else \'{c}\fi{}}}, \bibinfo {author} {\bibfnamefont {A.}~\bibnamefont {Sen}},\ and\ \bibinfo {author} {\bibfnamefont {J.~C.}\ \bibnamefont {Halimeh}},\ }\bibfield  {title} {\bibinfo {title} {Weak ergodicity breaking in the schwinger model},\ }\href {https://doi.org/10.1103/PhysRevB.107.L201105} {\bibfield  {journal} {\bibinfo  {journal} {Phys. Rev. B}\ }\textbf {\bibinfo {volume} {107}},\ \bibinfo {pages} {L201105} (\bibinfo {year} {2023})}\BibitemShut {NoStop}%
\bibitem [{\citenamefont {Halimeh}\ \emph {et~al.}(2023)\citenamefont {Halimeh}, \citenamefont {Barbiero}, \citenamefont {Hauke}, \citenamefont {Grusdt},\ and\ \citenamefont {Bohrdt}}]{Halimeh2023robustquantummany}%
  \BibitemOpen
  \bibfield  {author} {\bibinfo {author} {\bibfnamefont {J.~C.}\ \bibnamefont {Halimeh}}, \bibinfo {author} {\bibfnamefont {L.}~\bibnamefont {Barbiero}}, \bibinfo {author} {\bibfnamefont {P.}~\bibnamefont {Hauke}}, \bibinfo {author} {\bibfnamefont {F.}~\bibnamefont {Grusdt}},\ and\ \bibinfo {author} {\bibfnamefont {A.}~\bibnamefont {Bohrdt}},\ }\bibfield  {title} {\bibinfo {title} {Robust quantum many-body scars in lattice gauge theories},\ }\href {https://doi.org/10.22331/q-2023-05-15-1004} {\bibfield  {journal} {\bibinfo  {journal} {{Quantum}}\ }\textbf {\bibinfo {volume} {7}},\ \bibinfo {pages} {1004} (\bibinfo {year} {2023})}\BibitemShut {NoStop}%
\bibitem [{\citenamefont {Bu\ifmmode~\check{c}\else \v{c}\fi{}a}(2023)}]{Buca2023}%
  \BibitemOpen
  \bibfield  {author} {\bibinfo {author} {\bibfnamefont {B.}~\bibnamefont {Bu\ifmmode~\check{c}\else \v{c}\fi{}a}},\ }\bibfield  {title} {\bibinfo {title} {Unified theory of local quantum many-body dynamics: Eigenoperator thermalization theorems},\ }\href {https://doi.org/10.1103/PhysRevX.13.031013} {\bibfield  {journal} {\bibinfo  {journal} {Phys. Rev. X}\ }\textbf {\bibinfo {volume} {13}},\ \bibinfo {pages} {031013} (\bibinfo {year} {2023})}\BibitemShut {NoStop}%
\bibitem [{\citenamefont {Gotta}\ \emph {et~al.}(2023)\citenamefont {Gotta}, \citenamefont {Moudgalya},\ and\ \citenamefont {Mazza}}]{Gotta2023}%
  \BibitemOpen
  \bibfield  {author} {\bibinfo {author} {\bibfnamefont {L.}~\bibnamefont {Gotta}}, \bibinfo {author} {\bibfnamefont {S.}~\bibnamefont {Moudgalya}},\ and\ \bibinfo {author} {\bibfnamefont {L.}~\bibnamefont {Mazza}},\ }\bibfield  {title} {\bibinfo {title} {Asymptotic quantum many-body scars},\ }\href {https://doi.org/10.1103/PhysRevLett.131.190401} {\bibfield  {journal} {\bibinfo  {journal} {Phys. Rev. Lett.}\ }\textbf {\bibinfo {volume} {131}},\ \bibinfo {pages} {190401} (\bibinfo {year} {2023})}\BibitemShut {NoStop}%
\bibitem [{\citenamefont {Moudgalya}\ and\ \citenamefont {Motrunich}(2024)}]{Moudgalya2024}%
  \BibitemOpen
  \bibfield  {author} {\bibinfo {author} {\bibfnamefont {S.}~\bibnamefont {Moudgalya}}\ and\ \bibinfo {author} {\bibfnamefont {O.~I.}\ \bibnamefont {Motrunich}},\ }\bibfield  {title} {\bibinfo {title} {Exhaustive characterization of quantum many-body scars using commutant algebras},\ }\href {https://doi.org/10.1103/PhysRevX.14.041069} {\bibfield  {journal} {\bibinfo  {journal} {Phys. Rev. X}\ }\textbf {\bibinfo {volume} {14}},\ \bibinfo {pages} {041069} (\bibinfo {year} {2024})}\BibitemShut {NoStop}%
\bibitem [{\citenamefont {Evrard}\ \emph {et~al.}(2024)\citenamefont {Evrard}, \citenamefont {Pizzi}, \citenamefont {Mistakidis},\ and\ \citenamefont {Dag}}]{Evrard2024}%
  \BibitemOpen
  \bibfield  {author} {\bibinfo {author} {\bibfnamefont {B.}~\bibnamefont {Evrard}}, \bibinfo {author} {\bibfnamefont {A.}~\bibnamefont {Pizzi}}, \bibinfo {author} {\bibfnamefont {S.~I.}\ \bibnamefont {Mistakidis}},\ and\ \bibinfo {author} {\bibfnamefont {C.~B.}\ \bibnamefont {Dag}},\ }\bibfield  {title} {\bibinfo {title} {Quantum scars and regular eigenstates in a chaotic spinor condensate},\ }\href {https://doi.org/10.1103/PhysRevLett.132.020401} {\bibfield  {journal} {\bibinfo  {journal} {Phys. Rev. Lett.}\ }\textbf {\bibinfo {volume} {132}},\ \bibinfo {pages} {020401} (\bibinfo {year} {2024})}\BibitemShut {NoStop}%
\bibitem [{\citenamefont {Logari\ifmmode~\acute{c}\else \'{c}\fi{}}\ \emph {et~al.}(2024)\citenamefont {Logari\ifmmode~\acute{c}\else \'{c}\fi{}}, \citenamefont {Dooley}, \citenamefont {Pappalardi},\ and\ \citenamefont {Goold}}]{Dooley2024Dual}%
  \BibitemOpen
  \bibfield  {author} {\bibinfo {author} {\bibfnamefont {L.}~\bibnamefont {Logari\ifmmode~\acute{c}\else \'{c}\fi{}}}, \bibinfo {author} {\bibfnamefont {S.}~\bibnamefont {Dooley}}, \bibinfo {author} {\bibfnamefont {S.}~\bibnamefont {Pappalardi}},\ and\ \bibinfo {author} {\bibfnamefont {J.}~\bibnamefont {Goold}},\ }\bibfield  {title} {\bibinfo {title} {Quantum many-body scars in dual-unitary circuits},\ }\href {https://doi.org/10.1103/PhysRevLett.132.010401} {\bibfield  {journal} {\bibinfo  {journal} {Phys. Rev. Lett.}\ }\textbf {\bibinfo {volume} {132}},\ \bibinfo {pages} {010401} (\bibinfo {year} {2024})}\BibitemShut {NoStop}%
\bibitem [{\citenamefont {Pizzi}\ \emph {et~al.}(2024)\citenamefont {Pizzi}, \citenamefont {Evrard}, \citenamefont {Dag},\ and\ \citenamefont {Knolle}}]{pizzi2024quantumscarsmanybodysystems}%
  \BibitemOpen
  \bibfield  {author} {\bibinfo {author} {\bibfnamefont {A.}~\bibnamefont {Pizzi}}, \bibinfo {author} {\bibfnamefont {B.}~\bibnamefont {Evrard}}, \bibinfo {author} {\bibfnamefont {C.~B.}\ \bibnamefont {Dag}},\ and\ \bibinfo {author} {\bibfnamefont {J.}~\bibnamefont {Knolle}},\ }\href {https://arxiv.org/abs/2408.10301} {\bibinfo {title} {Quantum scars in many-body systems}} (\bibinfo {year} {2024}),\ \Eprint {https://arxiv.org/abs/2408.10301} {arXiv:2408.10301 [quant-ph]} \BibitemShut {NoStop}%
\bibitem [{\citenamefont {Dooley}(2021)}]{Dooley2021}%
  \BibitemOpen
  \bibfield  {author} {\bibinfo {author} {\bibfnamefont {S.}~\bibnamefont {Dooley}},\ }\bibfield  {title} {\bibinfo {title} {Robust quantum sensing in strongly interacting systems with many-body scars},\ }\href {https://doi.org/10.1103/PRXQuantum.2.020330} {\bibfield  {journal} {\bibinfo  {journal} {PRX Quantum}\ }\textbf {\bibinfo {volume} {2}},\ \bibinfo {pages} {020330} (\bibinfo {year} {2021})}\BibitemShut {NoStop}%
\bibitem [{\citenamefont {Desaules}\ \emph {et~al.}(2022)\citenamefont {Desaules}, \citenamefont {Pietracaprina}, \citenamefont {Papi\ifmmode~\acute{c}\else \'{c}\fi{}}, \citenamefont {Goold},\ and\ \citenamefont {Pappalardi}}]{Desaules2022QFI}%
  \BibitemOpen
  \bibfield  {author} {\bibinfo {author} {\bibfnamefont {J.-Y.}\ \bibnamefont {Desaules}}, \bibinfo {author} {\bibfnamefont {F.}~\bibnamefont {Pietracaprina}}, \bibinfo {author} {\bibfnamefont {Z.}~\bibnamefont {Papi\ifmmode~\acute{c}\else \'{c}\fi{}}}, \bibinfo {author} {\bibfnamefont {J.}~\bibnamefont {Goold}},\ and\ \bibinfo {author} {\bibfnamefont {S.}~\bibnamefont {Pappalardi}},\ }\bibfield  {title} {\bibinfo {title} {Extensive multipartite entanglement from su(2) quantum many-body scars},\ }\href {https://doi.org/10.1103/PhysRevLett.129.020601} {\bibfield  {journal} {\bibinfo  {journal} {Phys. Rev. Lett.}\ }\textbf {\bibinfo {volume} {129}},\ \bibinfo {pages} {020601} (\bibinfo {year} {2022})}\BibitemShut {NoStop}%
\bibitem [{\citenamefont {Dooley}\ \emph {et~al.}(2023)\citenamefont {Dooley}, \citenamefont {Pappalardi},\ and\ \citenamefont {Goold}}]{Dooley2023}%
  \BibitemOpen
  \bibfield  {author} {\bibinfo {author} {\bibfnamefont {S.}~\bibnamefont {Dooley}}, \bibinfo {author} {\bibfnamefont {S.}~\bibnamefont {Pappalardi}},\ and\ \bibinfo {author} {\bibfnamefont {J.}~\bibnamefont {Goold}},\ }\bibfield  {title} {\bibinfo {title} {Entanglement enhanced metrology with quantum many-body scars},\ }\href {https://doi.org/10.1103/PhysRevB.107.035123} {\bibfield  {journal} {\bibinfo  {journal} {Phys. Rev. B}\ }\textbf {\bibinfo {volume} {107}},\ \bibinfo {pages} {035123} (\bibinfo {year} {2023})}\BibitemShut {NoStop}%
\bibitem [{\citenamefont {Omran}\ \emph {et~al.}(2019)\citenamefont {Omran}, \citenamefont {Levine}, \citenamefont {Keesling}, \citenamefont {Semeghini}, \citenamefont {Wang}, \citenamefont {Ebadi}, \citenamefont {Bernien}, \citenamefont {Zibrov}, \citenamefont {Pichler}, \citenamefont {Choi}, \citenamefont {Cui}, \citenamefont {Rossignolo}, \citenamefont {Rembold}, \citenamefont {Montangero}, \citenamefont {Calarco}, \citenamefont {Endres}, \citenamefont {Greiner}, \citenamefont {Vuletić},\ and\ \citenamefont {Lukin}}]{omran2019GHZ}%
  \BibitemOpen
  \bibfield  {author} {\bibinfo {author} {\bibfnamefont {A.}~\bibnamefont {Omran}}, \bibinfo {author} {\bibfnamefont {H.}~\bibnamefont {Levine}}, \bibinfo {author} {\bibfnamefont {A.}~\bibnamefont {Keesling}}, \bibinfo {author} {\bibfnamefont {G.}~\bibnamefont {Semeghini}}, \bibinfo {author} {\bibfnamefont {T.~T.}\ \bibnamefont {Wang}}, \bibinfo {author} {\bibfnamefont {S.}~\bibnamefont {Ebadi}}, \bibinfo {author} {\bibfnamefont {H.}~\bibnamefont {Bernien}}, \bibinfo {author} {\bibfnamefont {A.~S.}\ \bibnamefont {Zibrov}}, \bibinfo {author} {\bibfnamefont {H.}~\bibnamefont {Pichler}}, \bibinfo {author} {\bibfnamefont {S.}~\bibnamefont {Choi}}, \bibinfo {author} {\bibfnamefont {J.}~\bibnamefont {Cui}}, \bibinfo {author} {\bibfnamefont {M.}~\bibnamefont {Rossignolo}}, \bibinfo {author} {\bibfnamefont {P.}~\bibnamefont {Rembold}}, \bibinfo {author} {\bibfnamefont {S.}~\bibnamefont {Montangero}}, \bibinfo {author} {\bibfnamefont {T.}~\bibnamefont {Calarco}}, \bibinfo {author} {\bibfnamefont {M.}~\bibnamefont
  {Endres}}, \bibinfo {author} {\bibfnamefont {M.}~\bibnamefont {Greiner}}, \bibinfo {author} {\bibfnamefont {V.}~\bibnamefont {Vuletić}},\ and\ \bibinfo {author} {\bibfnamefont {M.~D.}\ \bibnamefont {Lukin}},\ }\bibfield  {title} {\bibinfo {title} {Generation and manipulation of schrödinger cat states in rydberg atom arrays},\ }\href {https://doi.org/10.1126/science.aax9743} {\bibfield  {journal} {\bibinfo  {journal} {Science}\ }\textbf {\bibinfo {volume} {365}},\ \bibinfo {pages} {570} (\bibinfo {year} {2019})}\BibitemShut {NoStop}%
\bibitem [{\citenamefont {Albash}\ and\ \citenamefont {Lidar}(2015)}]{albash2015decoherence}%
  \BibitemOpen
  \bibfield  {author} {\bibinfo {author} {\bibfnamefont {T.}~\bibnamefont {Albash}}\ and\ \bibinfo {author} {\bibfnamefont {D.~A.}\ \bibnamefont {Lidar}},\ }\bibfield  {title} {\bibinfo {title} {Decoherence in adiabatic quantum computation},\ }\href {https://doi.org/10.1103/PhysRevA.91.062320} {\bibfield  {journal} {\bibinfo  {journal} {Phys. Rev. A}\ }\textbf {\bibinfo {volume} {91}},\ \bibinfo {pages} {062320} (\bibinfo {year} {2015})}\BibitemShut {NoStop}%
\bibitem [{\citenamefont {Yao}\ \emph {et~al.}(2023)\citenamefont {Yao}, \citenamefont {Xiang}, \citenamefont {Guo}, \citenamefont {Bao}, \citenamefont {Yang}, \citenamefont {Song}, \citenamefont {Shi}, \citenamefont {Zhu}, \citenamefont {Jin}, \citenamefont {Chen}, \citenamefont {Xu}, \citenamefont {Zhu}, \citenamefont {Shen}, \citenamefont {Wang}, \citenamefont {Zhang}, \citenamefont {Wu}, \citenamefont {Zou}, \citenamefont {Zhang}, \citenamefont {Li}, \citenamefont {Wang}, \citenamefont {Song}, \citenamefont {Cheng}, \citenamefont {Mondaini}, \citenamefont {Wang}, \citenamefont {You}, \citenamefont {Zhu}, \citenamefont {Ying},\ and\ \citenamefont {Guo}}]{WOS:001022795400001}%
  \BibitemOpen
  \bibfield  {author} {\bibinfo {author} {\bibfnamefont {Y.}~\bibnamefont {Yao}}, \bibinfo {author} {\bibfnamefont {L.}~\bibnamefont {Xiang}}, \bibinfo {author} {\bibfnamefont {Z.}~\bibnamefont {Guo}}, \bibinfo {author} {\bibfnamefont {Z.}~\bibnamefont {Bao}}, \bibinfo {author} {\bibfnamefont {Y.-F.}\ \bibnamefont {Yang}}, \bibinfo {author} {\bibfnamefont {Z.}~\bibnamefont {Song}}, \bibinfo {author} {\bibfnamefont {H.}~\bibnamefont {Shi}}, \bibinfo {author} {\bibfnamefont {X.}~\bibnamefont {Zhu}}, \bibinfo {author} {\bibfnamefont {F.}~\bibnamefont {Jin}}, \bibinfo {author} {\bibfnamefont {J.}~\bibnamefont {Chen}}, \bibinfo {author} {\bibfnamefont {S.}~\bibnamefont {Xu}}, \bibinfo {author} {\bibfnamefont {Z.}~\bibnamefont {Zhu}}, \bibinfo {author} {\bibfnamefont {F.}~\bibnamefont {Shen}}, \bibinfo {author} {\bibfnamefont {N.}~\bibnamefont {Wang}}, \bibinfo {author} {\bibfnamefont {C.}~\bibnamefont {Zhang}}, \bibinfo {author} {\bibfnamefont {Y.}~\bibnamefont {Wu}}, \bibinfo {author} {\bibfnamefont
  {Y.}~\bibnamefont {Zou}}, \bibinfo {author} {\bibfnamefont {P.}~\bibnamefont {Zhang}}, \bibinfo {author} {\bibfnamefont {H.}~\bibnamefont {Li}}, \bibinfo {author} {\bibfnamefont {Z.}~\bibnamefont {Wang}}, \bibinfo {author} {\bibfnamefont {C.}~\bibnamefont {Song}}, \bibinfo {author} {\bibfnamefont {C.}~\bibnamefont {Cheng}}, \bibinfo {author} {\bibfnamefont {R.}~\bibnamefont {Mondaini}}, \bibinfo {author} {\bibfnamefont {H.}~\bibnamefont {Wang}}, \bibinfo {author} {\bibfnamefont {J.~Q.}\ \bibnamefont {You}}, \bibinfo {author} {\bibfnamefont {S.-Y.}\ \bibnamefont {Zhu}}, \bibinfo {author} {\bibfnamefont {L.}~\bibnamefont {Ying}},\ and\ \bibinfo {author} {\bibfnamefont {Q.}~\bibnamefont {Guo}},\ }\bibfield  {title} {\bibinfo {title} {Observation of many-body fock space dynamics in two dimensions},\ }\href {https://doi.org/10.1038/s41567-023-02133-0} {\bibfield  {journal} {\bibinfo  {journal} {Nature Physics}\ }\textbf {\bibinfo {volume} {19}},\ \bibinfo {pages} {1459} (\bibinfo {year} {2023})}\BibitemShut
  {NoStop}%
\bibitem [{\citenamefont {Olmos}\ \emph {et~al.}(2012)\citenamefont {Olmos}, \citenamefont {Lesanovsky},\ and\ \citenamefont {Garrahan}}]{Olmos2012}%
  \BibitemOpen
  \bibfield  {author} {\bibinfo {author} {\bibfnamefont {B.}~\bibnamefont {Olmos}}, \bibinfo {author} {\bibfnamefont {I.}~\bibnamefont {Lesanovsky}},\ and\ \bibinfo {author} {\bibfnamefont {J.~P.}\ \bibnamefont {Garrahan}},\ }\bibfield  {title} {\bibinfo {title} {Facilitated spin models of dissipative quantum glasses},\ }\href {https://doi.org/10.1103/PhysRevLett.109.020403} {\bibfield  {journal} {\bibinfo  {journal} {Phys. Rev. Lett.}\ }\textbf {\bibinfo {volume} {109}},\ \bibinfo {pages} {020403} (\bibinfo {year} {2012})}\BibitemShut {NoStop}%
\bibitem [{\citenamefont {Macieszczak}\ \emph {et~al.}(2016)\citenamefont {Macieszczak}, \citenamefont {Gu{\c{t}}{\u{a}}}, \citenamefont {Lesanovsky},\ and\ \citenamefont {Garrahan}}]{Macieszczak2016}%
  \BibitemOpen
  \bibfield  {author} {\bibinfo {author} {\bibfnamefont {K.}~\bibnamefont {Macieszczak}}, \bibinfo {author} {\bibfnamefont {M.}~\bibnamefont {Gu{\c{t}}{\u{a}}}}, \bibinfo {author} {\bibfnamefont {I.}~\bibnamefont {Lesanovsky}},\ and\ \bibinfo {author} {\bibfnamefont {J.~P.}\ \bibnamefont {Garrahan}},\ }\bibfield  {title} {\bibinfo {title} {Towards a theory of metastability in open quantum dynamics},\ }\href {https://doi.org/10.1103/PhysRevLett.116.240404} {\bibfield  {journal} {\bibinfo  {journal} {Physical review letters}\ }\textbf {\bibinfo {volume} {116}},\ \bibinfo {pages} {240404} (\bibinfo {year} {2016})}\BibitemShut {NoStop}%
\bibitem [{\citenamefont {Bu{\v{c}}a}\ \emph {et~al.}(2019)\citenamefont {Bu{\v{c}}a}, \citenamefont {Tindall},\ and\ \citenamefont {Jaksch}}]{Buca2019}%
  \BibitemOpen
  \bibfield  {author} {\bibinfo {author} {\bibfnamefont {B.}~\bibnamefont {Bu{\v{c}}a}}, \bibinfo {author} {\bibfnamefont {J.}~\bibnamefont {Tindall}},\ and\ \bibinfo {author} {\bibfnamefont {D.}~\bibnamefont {Jaksch}},\ }\bibfield  {title} {\bibinfo {title} {Non-stationary coherent quantum many-body dynamics through dissipation},\ }\href {https://doi.org/10.1038/s41467-019-09757-y} {\bibfield  {journal} {\bibinfo  {journal} {Nature Communications}\ }\textbf {\bibinfo {volume} {10}},\ \bibinfo {pages} {1730} (\bibinfo {year} {2019})}\BibitemShut {NoStop}%
\bibitem [{\citenamefont {Booker}\ \emph {et~al.}(2020)\citenamefont {Booker}, \citenamefont {Buča},\ and\ \citenamefont {Jaksch}}]{Booker2020}%
  \BibitemOpen
  \bibfield  {author} {\bibinfo {author} {\bibfnamefont {C.}~\bibnamefont {Booker}}, \bibinfo {author} {\bibfnamefont {B.}~\bibnamefont {Buča}},\ and\ \bibinfo {author} {\bibfnamefont {D.}~\bibnamefont {Jaksch}},\ }\bibfield  {title} {\bibinfo {title} {Non-stationarity and dissipative time crystals: spectral properties and finite-size effects},\ }\href {https://doi.org/10.1088/1367-2630/ababc4} {\bibfield  {journal} {\bibinfo  {journal} {New Journal of Physics}\ }\textbf {\bibinfo {volume} {22}},\ \bibinfo {pages} {085007} (\bibinfo {year} {2020})}\BibitemShut {NoStop}%
\bibitem [{\citenamefont {Gambetta}\ \emph {et~al.}(2019)\citenamefont {Gambetta}, \citenamefont {Carollo}, \citenamefont {Marcuzzi}, \citenamefont {Garrahan},\ and\ \citenamefont {Lesanovsky}}]{Gambetta2019}%
  \BibitemOpen
  \bibfield  {author} {\bibinfo {author} {\bibfnamefont {F.~M.}\ \bibnamefont {Gambetta}}, \bibinfo {author} {\bibfnamefont {F.}~\bibnamefont {Carollo}}, \bibinfo {author} {\bibfnamefont {M.}~\bibnamefont {Marcuzzi}}, \bibinfo {author} {\bibfnamefont {J.~P.}\ \bibnamefont {Garrahan}},\ and\ \bibinfo {author} {\bibfnamefont {I.}~\bibnamefont {Lesanovsky}},\ }\bibfield  {title} {\bibinfo {title} {Discrete time crystals in the absence of manifest symmetries or disorder in open quantum systems},\ }\href {https://doi.org/10.1103/PhysRevLett.122.015701} {\bibfield  {journal} {\bibinfo  {journal} {Phys. Rev. Lett.}\ }\textbf {\bibinfo {volume} {122}},\ \bibinfo {pages} {015701} (\bibinfo {year} {2019})}\BibitemShut {NoStop}%
\bibitem [{\citenamefont {Ke\ss{}ler}\ \emph {et~al.}(2021)\citenamefont {Ke\ss{}ler}, \citenamefont {Kongkhambut}, \citenamefont {Georges}, \citenamefont {Mathey}, \citenamefont {Cosme},\ and\ \citenamefont {Hemmerich}}]{Kessler2021}%
  \BibitemOpen
  \bibfield  {author} {\bibinfo {author} {\bibfnamefont {H.}~\bibnamefont {Ke\ss{}ler}}, \bibinfo {author} {\bibfnamefont {P.}~\bibnamefont {Kongkhambut}}, \bibinfo {author} {\bibfnamefont {C.}~\bibnamefont {Georges}}, \bibinfo {author} {\bibfnamefont {L.}~\bibnamefont {Mathey}}, \bibinfo {author} {\bibfnamefont {J.~G.}\ \bibnamefont {Cosme}},\ and\ \bibinfo {author} {\bibfnamefont {A.}~\bibnamefont {Hemmerich}},\ }\bibfield  {title} {\bibinfo {title} {Observation of a dissipative time crystal},\ }\href {https://doi.org/10.1103/PhysRevLett.127.043602} {\bibfield  {journal} {\bibinfo  {journal} {Phys. Rev. Lett.}\ }\textbf {\bibinfo {volume} {127}},\ \bibinfo {pages} {043602} (\bibinfo {year} {2021})}\BibitemShut {NoStop}%
\bibitem [{\citenamefont {Wang}\ \emph {et~al.}(2024)\citenamefont {Wang}, \citenamefont {Yuan}, \citenamefont {Zhang}, \citenamefont {Wang}, \citenamefont {Deng},\ and\ \citenamefont {Duan}}]{Wang2024DecoherenceFree}%
  \BibitemOpen
  \bibfield  {author} {\bibinfo {author} {\bibfnamefont {H.-R.}\ \bibnamefont {Wang}}, \bibinfo {author} {\bibfnamefont {D.}~\bibnamefont {Yuan}}, \bibinfo {author} {\bibfnamefont {S.-Y.}\ \bibnamefont {Zhang}}, \bibinfo {author} {\bibfnamefont {Z.}~\bibnamefont {Wang}}, \bibinfo {author} {\bibfnamefont {D.-L.}\ \bibnamefont {Deng}},\ and\ \bibinfo {author} {\bibfnamefont {L.-M.}\ \bibnamefont {Duan}},\ }\bibfield  {title} {\bibinfo {title} {Embedding quantum many-body scars into decoherence-free subspaces},\ }\href {https://doi.org/10.1103/PhysRevLett.132.150401} {\bibfield  {journal} {\bibinfo  {journal} {Phys. Rev. Lett.}\ }\textbf {\bibinfo {volume} {132}},\ \bibinfo {pages} {150401} (\bibinfo {year} {2024})}\BibitemShut {NoStop}%
\bibitem [{\citenamefont {Jiang}\ \emph {et~al.}(2025)\citenamefont {Jiang}, \citenamefont {Xu}, \citenamefont {Yang}, \citenamefont {Hou}, \citenamefont {Wang},\ and\ \citenamefont {Pan}}]{jiang2025robustnessquantummanybodyscars}%
  \BibitemOpen
  \bibfield  {author} {\bibinfo {author} {\bibfnamefont {X.-P.}\ \bibnamefont {Jiang}}, \bibinfo {author} {\bibfnamefont {M.}~\bibnamefont {Xu}}, \bibinfo {author} {\bibfnamefont {X.}~\bibnamefont {Yang}}, \bibinfo {author} {\bibfnamefont {H.}~\bibnamefont {Hou}}, \bibinfo {author} {\bibfnamefont {Y.}~\bibnamefont {Wang}},\ and\ \bibinfo {author} {\bibfnamefont {L.}~\bibnamefont {Pan}},\ }\href {https://arxiv.org/abs/2501.00886} {\bibinfo {title} {Robustness of quantum many-body scars in the presence of markovian bath}} (\bibinfo {year} {2025}),\ \Eprint {https://arxiv.org/abs/2501.00886} {arXiv:2501.00886 [cond-mat.dis-nn]} \BibitemShut {NoStop}%
\bibitem [{\citenamefont {García-García}\ \emph {et~al.}(2025)\citenamefont {García-García}, \citenamefont {Lu}, \citenamefont {Sá},\ and\ \citenamefont {Verbaarschot}}]{garcíagarcía2025lindbladmanybodyscars}%
  \BibitemOpen
  \bibfield  {author} {\bibinfo {author} {\bibfnamefont {A.~M.}\ \bibnamefont {García-García}}, \bibinfo {author} {\bibfnamefont {Z.}~\bibnamefont {Lu}}, \bibinfo {author} {\bibfnamefont {L.}~\bibnamefont {Sá}},\ and\ \bibinfo {author} {\bibfnamefont {J.~J.~M.}\ \bibnamefont {Verbaarschot}},\ }\href {https://arxiv.org/abs/2503.06665} {\bibinfo {title} {Lindblad many-body scars}} (\bibinfo {year} {2025}),\ \Eprint {https://arxiv.org/abs/2503.06665} {arXiv:2503.06665 [quant-ph]} \BibitemShut {NoStop}%
\bibitem [{\citenamefont {Lin}\ \emph {et~al.}(2020{\natexlab{b}})\citenamefont {Lin}, \citenamefont {Chandran},\ and\ \citenamefont {Motrunich}}]{Lin2020Pert}%
  \BibitemOpen
  \bibfield  {author} {\bibinfo {author} {\bibfnamefont {C.-J.}\ \bibnamefont {Lin}}, \bibinfo {author} {\bibfnamefont {A.}~\bibnamefont {Chandran}},\ and\ \bibinfo {author} {\bibfnamefont {O.~I.}\ \bibnamefont {Motrunich}},\ }\bibfield  {title} {\bibinfo {title} {Slow thermalization of exact quantum many-body scar states under perturbations},\ }\href {https://doi.org/10.1103/PhysRevResearch.2.033044} {\bibfield  {journal} {\bibinfo  {journal} {Phys. Rev. Res.}\ }\textbf {\bibinfo {volume} {2}},\ \bibinfo {pages} {033044} (\bibinfo {year} {2020}{\natexlab{b}})}\BibitemShut {NoStop}%
\bibitem [{\citenamefont {Surace}\ \emph {et~al.}(2021)\citenamefont {Surace}, \citenamefont {Votto}, \citenamefont {Lazo}, \citenamefont {Silva}, \citenamefont {Dalmonte},\ and\ \citenamefont {Giudici}}]{Surace2021Pert}%
  \BibitemOpen
  \bibfield  {author} {\bibinfo {author} {\bibfnamefont {F.~M.}\ \bibnamefont {Surace}}, \bibinfo {author} {\bibfnamefont {M.}~\bibnamefont {Votto}}, \bibinfo {author} {\bibfnamefont {E.~G.}\ \bibnamefont {Lazo}}, \bibinfo {author} {\bibfnamefont {A.}~\bibnamefont {Silva}}, \bibinfo {author} {\bibfnamefont {M.}~\bibnamefont {Dalmonte}},\ and\ \bibinfo {author} {\bibfnamefont {G.}~\bibnamefont {Giudici}},\ }\bibfield  {title} {\bibinfo {title} {Exact many-body scars and their stability in constrained quantum chains},\ }\href {https://doi.org/10.1103/PhysRevB.103.104302} {\bibfield  {journal} {\bibinfo  {journal} {Phys. Rev. B}\ }\textbf {\bibinfo {volume} {103}},\ \bibinfo {pages} {104302} (\bibinfo {year} {2021})}\BibitemShut {NoStop}%
\bibitem [{\citenamefont {Moudgalya}\ \emph {et~al.}(2020)\citenamefont {Moudgalya}, \citenamefont {Regnault},\ and\ \citenamefont {Bernevig}}]{moudgalya2020eta}%
  \BibitemOpen
  \bibfield  {author} {\bibinfo {author} {\bibfnamefont {S.}~\bibnamefont {Moudgalya}}, \bibinfo {author} {\bibfnamefont {N.}~\bibnamefont {Regnault}},\ and\ \bibinfo {author} {\bibfnamefont {B.~A.}\ \bibnamefont {Bernevig}},\ }\bibfield  {title} {\bibinfo {title} {$\ensuremath{\eta}$-pairing in hubbard models: From spectrum generating algebras to quantum many-body scars},\ }\href {https://doi.org/10.1103/PhysRevB.102.085140} {\bibfield  {journal} {\bibinfo  {journal} {Phys. Rev. B}\ }\textbf {\bibinfo {volume} {102}},\ \bibinfo {pages} {085140} (\bibinfo {year} {2020})}\BibitemShut {NoStop}%
\bibitem [{\citenamefont {Mark}\ \emph {et~al.}(2020)\citenamefont {Mark}, \citenamefont {Lin},\ and\ \citenamefont {Motrunich}}]{Mark2020}%
  \BibitemOpen
  \bibfield  {author} {\bibinfo {author} {\bibfnamefont {D.~K.}\ \bibnamefont {Mark}}, \bibinfo {author} {\bibfnamefont {C.-J.}\ \bibnamefont {Lin}},\ and\ \bibinfo {author} {\bibfnamefont {O.~I.}\ \bibnamefont {Motrunich}},\ }\bibfield  {title} {\bibinfo {title} {Unified structure for exact towers of scar states in the affleck-kennedy-lieb-tasaki and other models},\ }\href {https://doi.org/10.1103/PhysRevB.101.195131} {\bibfield  {journal} {\bibinfo  {journal} {Phys. Rev. B}\ }\textbf {\bibinfo {volume} {101}},\ \bibinfo {pages} {195131} (\bibinfo {year} {2020})}\BibitemShut {NoStop}%
\bibitem [{\citenamefont {O'Dea}\ \emph {et~al.}(2020)\citenamefont {O'Dea}, \citenamefont {Burnell}, \citenamefont {Chandran},\ and\ \citenamefont {Khemani}}]{ODea2020Tunnels}%
  \BibitemOpen
  \bibfield  {author} {\bibinfo {author} {\bibfnamefont {N.}~\bibnamefont {O'Dea}}, \bibinfo {author} {\bibfnamefont {F.}~\bibnamefont {Burnell}}, \bibinfo {author} {\bibfnamefont {A.}~\bibnamefont {Chandran}},\ and\ \bibinfo {author} {\bibfnamefont {V.}~\bibnamefont {Khemani}},\ }\bibfield  {title} {\bibinfo {title} {From tunnels to towers: Quantum scars from lie algebras and $q$-deformed lie algebras},\ }\href {https://doi.org/10.1103/PhysRevResearch.2.043305} {\bibfield  {journal} {\bibinfo  {journal} {Phys. Rev. Res.}\ }\textbf {\bibinfo {volume} {2}},\ \bibinfo {pages} {043305} (\bibinfo {year} {2020})}\BibitemShut {NoStop}%
\bibitem [{\citenamefont {Bull}\ \emph {et~al.}(2020)\citenamefont {Bull}, \citenamefont {Desaules},\ and\ \citenamefont {Papi\ifmmode~\acute{c}\else \'{c}\fi{}}}]{Bull2020}%
  \BibitemOpen
  \bibfield  {author} {\bibinfo {author} {\bibfnamefont {K.}~\bibnamefont {Bull}}, \bibinfo {author} {\bibfnamefont {J.-Y.}\ \bibnamefont {Desaules}},\ and\ \bibinfo {author} {\bibfnamefont {Z.}~\bibnamefont {Papi\ifmmode~\acute{c}\else \'{c}\fi{}}},\ }\bibfield  {title} {\bibinfo {title} {Quantum scars as embeddings of weakly broken {Lie} algebra representations},\ }\href {https://doi.org/10.1103/PhysRevB.101.165139} {\bibfield  {journal} {\bibinfo  {journal} {Phys. Rev. B}\ }\textbf {\bibinfo {volume} {101}},\ \bibinfo {pages} {165139} (\bibinfo {year} {2020})}\BibitemShut {NoStop}%
\bibitem [{\citenamefont {Dong}\ \emph {et~al.}(2023)\citenamefont {Dong}, \citenamefont {Desaules}, \citenamefont {Gao}, \citenamefont {Wang}, \citenamefont {Guo}, \citenamefont {Chen}, \citenamefont {Zou}, \citenamefont {Jin}, \citenamefont {Zhu}, \citenamefont {Zhang}, \citenamefont {Li}, \citenamefont {Wang}, \citenamefont {Guo}, \citenamefont {Zhang}, \citenamefont {Ying},\ and\ \citenamefont {Papić}}]{Dong2023}%
  \BibitemOpen
  \bibfield  {author} {\bibinfo {author} {\bibfnamefont {H.}~\bibnamefont {Dong}}, \bibinfo {author} {\bibfnamefont {J.-Y.}\ \bibnamefont {Desaules}}, \bibinfo {author} {\bibfnamefont {Y.}~\bibnamefont {Gao}}, \bibinfo {author} {\bibfnamefont {N.}~\bibnamefont {Wang}}, \bibinfo {author} {\bibfnamefont {Z.}~\bibnamefont {Guo}}, \bibinfo {author} {\bibfnamefont {J.}~\bibnamefont {Chen}}, \bibinfo {author} {\bibfnamefont {Y.}~\bibnamefont {Zou}}, \bibinfo {author} {\bibfnamefont {F.}~\bibnamefont {Jin}}, \bibinfo {author} {\bibfnamefont {X.}~\bibnamefont {Zhu}}, \bibinfo {author} {\bibfnamefont {P.}~\bibnamefont {Zhang}}, \bibinfo {author} {\bibfnamefont {H.}~\bibnamefont {Li}}, \bibinfo {author} {\bibfnamefont {Z.}~\bibnamefont {Wang}}, \bibinfo {author} {\bibfnamefont {Q.}~\bibnamefont {Guo}}, \bibinfo {author} {\bibfnamefont {J.}~\bibnamefont {Zhang}}, \bibinfo {author} {\bibfnamefont {L.}~\bibnamefont {Ying}},\ and\ \bibinfo {author} {\bibfnamefont {Z.}~\bibnamefont {Papić}},\ }\bibfield  {title} {\bibinfo
  {title} {Disorder-tunable entanglement at infinite temperature},\ }\href {https://doi.org/10.1126/sciadv.adj3822} {\bibfield  {journal} {\bibinfo  {journal} {Science Advances}\ }\textbf {\bibinfo {volume} {9}},\ \bibinfo {pages} {eadj3822} (\bibinfo {year} {2023})}\BibitemShut {NoStop}%
\bibitem [{\citenamefont {Prosen}(2012)}]{PhysRevLett.109.090404}%
  \BibitemOpen
  \bibfield  {author} {\bibinfo {author} {\bibfnamefont {T.}~\bibnamefont {Prosen}},\ }\bibfield  {title} {\bibinfo {title} {$\mathbb{P}\mathbb{T}$-symmetric quantum liouvillean dynamics},\ }\href {https://doi.org/10.1103/PhysRevLett.109.090404} {\bibfield  {journal} {\bibinfo  {journal} {Phys. Rev. Lett.}\ }\textbf {\bibinfo {volume} {109}},\ \bibinfo {pages} {090404} (\bibinfo {year} {2012})}\BibitemShut {NoStop}%
\bibitem [{\citenamefont {Chen}\ \emph {et~al.}(2023)\citenamefont {Chen}, \citenamefont {Chen},\ and\ \citenamefont {Zhu}}]{Qianqian2023nonHerm}%
  \BibitemOpen
  \bibfield  {author} {\bibinfo {author} {\bibfnamefont {Q.}~\bibnamefont {Chen}}, \bibinfo {author} {\bibfnamefont {S.~A.}\ \bibnamefont {Chen}},\ and\ \bibinfo {author} {\bibfnamefont {Z.}~\bibnamefont {Zhu}},\ }\bibfield  {title} {\bibinfo {title} {{Weak ergodicity breaking in non-Hermitian many-body systems}},\ }\href {https://doi.org/10.21468/SciPostPhys.15.2.052} {\bibfield  {journal} {\bibinfo  {journal} {SciPost Phys.}\ }\textbf {\bibinfo {volume} {15}},\ \bibinfo {pages} {052} (\bibinfo {year} {2023})}\BibitemShut {NoStop}%
\bibitem [{\citenamefont {Shen}\ \emph {et~al.}(2024)\citenamefont {Shen}, \citenamefont {Qin}, \citenamefont {Desaules}, \citenamefont {Papi\ifmmode~\acute{c}\else \'{c}\fi{}},\ and\ \citenamefont {Lee}}]{Shen2024}%
  \BibitemOpen
  \bibfield  {author} {\bibinfo {author} {\bibfnamefont {R.}~\bibnamefont {Shen}}, \bibinfo {author} {\bibfnamefont {F.}~\bibnamefont {Qin}}, \bibinfo {author} {\bibfnamefont {J.-Y.}\ \bibnamefont {Desaules}}, \bibinfo {author} {\bibfnamefont {Z.}~\bibnamefont {Papi\ifmmode~\acute{c}\else \'{c}\fi{}}},\ and\ \bibinfo {author} {\bibfnamefont {C.~H.}\ \bibnamefont {Lee}},\ }\bibfield  {title} {\bibinfo {title} {Enhanced many-body quantum scars from the non-hermitian fock skin effect},\ }\href {https://doi.org/10.1103/PhysRevLett.133.216601} {\bibfield  {journal} {\bibinfo  {journal} {Phys. Rev. Lett.}\ }\textbf {\bibinfo {volume} {133}},\ \bibinfo {pages} {216601} (\bibinfo {year} {2024})}\BibitemShut {NoStop}%
\bibitem [{\citenamefont {Iadecola}\ and\ \citenamefont {\ifmmode \check{Z}\else \v{Z}\fi{}nidari\ifmmode~\check{c}\else \v{c}\fi{}}(2019)}]{Iadecola2019Ladders}%
  \BibitemOpen
  \bibfield  {author} {\bibinfo {author} {\bibfnamefont {T.}~\bibnamefont {Iadecola}}\ and\ \bibinfo {author} {\bibfnamefont {M.}~\bibnamefont {\ifmmode \check{Z}\else \v{Z}\fi{}nidari\ifmmode~\check{c}\else \v{c}\fi{}}},\ }\bibfield  {title} {\bibinfo {title} {Exact localized and ballistic eigenstates in disordered chaotic spin ladders and the fermi-hubbard model},\ }\href {https://doi.org/10.1103/PhysRevLett.123.036403} {\bibfield  {journal} {\bibinfo  {journal} {Phys. Rev. Lett.}\ }\textbf {\bibinfo {volume} {123}},\ \bibinfo {pages} {036403} (\bibinfo {year} {2019})}\BibitemShut {NoStop}%
\bibitem [{SM()}]{SM}%
  \BibitemOpen
  \href@noop {} {}\bibinfo {note} {See Supplemental Material [url] for details of the derivations and results for other models.}\BibitemShut {Stop}%
\bibitem [{\citenamefont {Breuer}\ and\ \citenamefont {Petruccione}(2002)}]{breuer2002theory}%
  \BibitemOpen
  \bibfield  {author} {\bibinfo {author} {\bibfnamefont {H.-P.}\ \bibnamefont {Breuer}}\ and\ \bibinfo {author} {\bibfnamefont {F.}~\bibnamefont {Petruccione}},\ }\href@noop {} {\emph {\bibinfo {title} {The theory of open quantum systems}}}\ (\bibinfo  {publisher} {Oxford University Press, USA},\ \bibinfo {year} {2002})\BibitemShut {NoStop}%
\bibitem [{\citenamefont {Fishman}\ \emph {et~al.}(2022)\citenamefont {Fishman}, \citenamefont {White},\ and\ \citenamefont {Stoudenmire}}]{ITensor}%
  \BibitemOpen
  \bibfield  {author} {\bibinfo {author} {\bibfnamefont {M.}~\bibnamefont {Fishman}}, \bibinfo {author} {\bibfnamefont {S.~R.}\ \bibnamefont {White}},\ and\ \bibinfo {author} {\bibfnamefont {E.~M.}\ \bibnamefont {Stoudenmire}},\ }\bibfield  {title} {\bibinfo {title} {{The ITensor Software Library for Tensor Network Calculations}},\ }\href {https://doi.org/10.21468/SciPostPhysCodeb.4} {\bibfield  {journal} {\bibinfo  {journal} {SciPost Phys. Codebases}\ ,\ \bibinfo {pages} {4}} (\bibinfo {year} {2022})}\BibitemShut {NoStop}%
\bibitem [{\citenamefont {S\'a}\ \emph {et~al.}(2023)\citenamefont {S\'a}, \citenamefont {Ribeiro},\ and\ \citenamefont {Prosen}}]{PhysRevX.13.031019}%
  \BibitemOpen
  \bibfield  {author} {\bibinfo {author} {\bibfnamefont {L.}~\bibnamefont {S\'a}}, \bibinfo {author} {\bibfnamefont {P.}~\bibnamefont {Ribeiro}},\ and\ \bibinfo {author} {\bibfnamefont {T.~c.~v.}\ \bibnamefont {Prosen}},\ }\bibfield  {title} {\bibinfo {title} {Symmetry classification of many-body lindbladians: Tenfold way and beyond},\ }\href {https://doi.org/10.1103/PhysRevX.13.031019} {\bibfield  {journal} {\bibinfo  {journal} {Phys. Rev. X}\ }\textbf {\bibinfo {volume} {13}},\ \bibinfo {pages} {031019} (\bibinfo {year} {2023})}\BibitemShut {NoStop}%
\bibitem [{\citenamefont {Desaules}\ \emph {et~al.}(2024)\citenamefont {Desaules}, \citenamefont {Gustafson}, \citenamefont {Li}, \citenamefont {Papi\ifmmode~\acute{c}\else \'{c}\fi{}},\ and\ \citenamefont {Halimeh}}]{Desaules2024FiniteTemperature}%
  \BibitemOpen
  \bibfield  {author} {\bibinfo {author} {\bibfnamefont {J.-Y.}\ \bibnamefont {Desaules}}, \bibinfo {author} {\bibfnamefont {E.~J.}\ \bibnamefont {Gustafson}}, \bibinfo {author} {\bibfnamefont {A.~C.~Y.}\ \bibnamefont {Li}}, \bibinfo {author} {\bibfnamefont {Z.}~\bibnamefont {Papi\ifmmode~\acute{c}\else \'{c}\fi{}}},\ and\ \bibinfo {author} {\bibfnamefont {J.~C.}\ \bibnamefont {Halimeh}},\ }\bibfield  {title} {\bibinfo {title} {Robust finite-temperature many-body scarring on a quantum computer},\ }\href {https://doi.org/10.1103/PhysRevA.110.042606} {\bibfield  {journal} {\bibinfo  {journal} {Phys. Rev. A}\ }\textbf {\bibinfo {volume} {110}},\ \bibinfo {pages} {042606} (\bibinfo {year} {2024})}\BibitemShut {NoStop}%
\bibitem [{\citenamefont {Jurcevic}\ and\ \citenamefont {Govia}(2022)}]{jurcevic2022effective}%
  \BibitemOpen
  \bibfield  {author} {\bibinfo {author} {\bibfnamefont {P.}~\bibnamefont {Jurcevic}}\ and\ \bibinfo {author} {\bibfnamefont {L.~C.}\ \bibnamefont {Govia}},\ }\bibfield  {title} {\bibinfo {title} {Effective qubit dephasing induced by spectator-qubit relaxation},\ }\href {https://doi.org/10.1088/2058-9565/ac8cad} {\bibfield  {journal} {\bibinfo  {journal} {Quantum Science and Technology}\ }\textbf {\bibinfo {volume} {7}},\ \bibinfo {pages} {045033} (\bibinfo {year} {2022})}\BibitemShut {NoStop}%
\bibitem [{\citenamefont {Levine}\ \emph {et~al.}(2018)\citenamefont {Levine}, \citenamefont {Keesling}, \citenamefont {Omran}, \citenamefont {Bernien}, \citenamefont {Schwartz}, \citenamefont {Zibrov}, \citenamefont {Endres}, \citenamefont {Greiner}, \citenamefont {Vuleti\ifmmode~\acute{c}\else \'{c}\fi{}},\ and\ \citenamefont {Lukin}}]{PhysRevLett.121.123603}%
  \BibitemOpen
  \bibfield  {author} {\bibinfo {author} {\bibfnamefont {H.}~\bibnamefont {Levine}}, \bibinfo {author} {\bibfnamefont {A.}~\bibnamefont {Keesling}}, \bibinfo {author} {\bibfnamefont {A.}~\bibnamefont {Omran}}, \bibinfo {author} {\bibfnamefont {H.}~\bibnamefont {Bernien}}, \bibinfo {author} {\bibfnamefont {S.}~\bibnamefont {Schwartz}}, \bibinfo {author} {\bibfnamefont {A.~S.}\ \bibnamefont {Zibrov}}, \bibinfo {author} {\bibfnamefont {M.}~\bibnamefont {Endres}}, \bibinfo {author} {\bibfnamefont {M.}~\bibnamefont {Greiner}}, \bibinfo {author} {\bibfnamefont {V.}~\bibnamefont {Vuleti\ifmmode~\acute{c}\else \'{c}\fi{}}},\ and\ \bibinfo {author} {\bibfnamefont {M.~D.}\ \bibnamefont {Lukin}},\ }\bibfield  {title} {\bibinfo {title} {High-fidelity control and entanglement of rydberg-atom qubits},\ }\href {https://doi.org/10.1103/PhysRevLett.121.123603} {\bibfield  {journal} {\bibinfo  {journal} {Phys. Rev. Lett.}\ }\textbf {\bibinfo {volume} {121}},\ \bibinfo {pages} {123603} (\bibinfo {year} {2018})}\BibitemShut
  {NoStop}%
\bibitem [{\citenamefont {Hahn}\ \emph {et~al.}(2021)\citenamefont {Hahn}, \citenamefont {McClarty},\ and\ \citenamefont {Luitz}}]{10.21468/SciPostPhys.11.4.074}%
  \BibitemOpen
  \bibfield  {author} {\bibinfo {author} {\bibfnamefont {D.}~\bibnamefont {Hahn}}, \bibinfo {author} {\bibfnamefont {P.~A.}\ \bibnamefont {McClarty}},\ and\ \bibinfo {author} {\bibfnamefont {D.~J.}\ \bibnamefont {Luitz}},\ }\bibfield  {title} {\bibinfo {title} {{Information dynamics in a model with Hilbert space fragmentation}},\ }\href {https://doi.org/10.21468/SciPostPhys.11.4.074} {\bibfield  {journal} {\bibinfo  {journal} {SciPost Phys.}\ }\textbf {\bibinfo {volume} {11}},\ \bibinfo {pages} {074} (\bibinfo {year} {2021})}\BibitemShut {NoStop}%
\bibitem [{\citenamefont {Fendley}\ \emph {et~al.}(2004)\citenamefont {Fendley}, \citenamefont {Sengupta},\ and\ \citenamefont {Sachdev}}]{FendleySachdev}%
  \BibitemOpen
  \bibfield  {author} {\bibinfo {author} {\bibfnamefont {P.}~\bibnamefont {Fendley}}, \bibinfo {author} {\bibfnamefont {K.}~\bibnamefont {Sengupta}},\ and\ \bibinfo {author} {\bibfnamefont {S.}~\bibnamefont {Sachdev}},\ }\bibfield  {title} {\bibinfo {title} {Competing density-wave orders in a one-dimensional hard-boson model},\ }\href {https://doi.org/10.1103/PhysRevB.69.075106} {\bibfield  {journal} {\bibinfo  {journal} {Phys. Rev. B}\ }\textbf {\bibinfo {volume} {69}},\ \bibinfo {pages} {075106} (\bibinfo {year} {2004})}\BibitemShut {NoStop}%
\bibitem [{\citenamefont {Lesanovsky}\ and\ \citenamefont {Katsura}(2012)}]{Lesanovsky2012}%
  \BibitemOpen
  \bibfield  {author} {\bibinfo {author} {\bibfnamefont {I.}~\bibnamefont {Lesanovsky}}\ and\ \bibinfo {author} {\bibfnamefont {H.}~\bibnamefont {Katsura}},\ }\bibfield  {title} {\bibinfo {title} {Interacting {Fibonacci} anyons in a {Rydberg} gas},\ }\href {https://doi.org/10.1103/PhysRevA.86.041601} {\bibfield  {journal} {\bibinfo  {journal} {Phys. Rev. A}\ }\textbf {\bibinfo {volume} {86}},\ \bibinfo {pages} {041601(R)} (\bibinfo {year} {2012})}\BibitemShut {NoStop}%
\bibitem [{\citenamefont {Turner}\ \emph {et~al.}(2018{\natexlab{b}})\citenamefont {Turner}, \citenamefont {Michailidis}, \citenamefont {Abanin}, \citenamefont {Serbyn},\ and\ \citenamefont {Papi\ifmmode~\acute{c}\else \'{c}\fi{}}}]{Turner2018PRB}%
  \BibitemOpen
  \bibfield  {author} {\bibinfo {author} {\bibfnamefont {C.~J.}\ \bibnamefont {Turner}}, \bibinfo {author} {\bibfnamefont {A.~A.}\ \bibnamefont {Michailidis}}, \bibinfo {author} {\bibfnamefont {D.~A.}\ \bibnamefont {Abanin}}, \bibinfo {author} {\bibfnamefont {M.}~\bibnamefont {Serbyn}},\ and\ \bibinfo {author} {\bibfnamefont {Z.}~\bibnamefont {Papi\ifmmode~\acute{c}\else \'{c}\fi{}}},\ }\bibfield  {title} {\bibinfo {title} {Quantum scarred eigenstates in a {Rydberg} atom chain: Entanglement, breakdown of thermalization, and stability to perturbations},\ }\href {https://doi.org/10.1103/PhysRevB.98.155134} {\bibfield  {journal} {\bibinfo  {journal} {Phys. Rev. B}\ }\textbf {\bibinfo {volume} {98}},\ \bibinfo {pages} {155134} (\bibinfo {year} {2018}{\natexlab{b}})}\BibitemShut {NoStop}%
\bibitem [{\citenamefont {Omiya}\ and\ \citenamefont {M\"uller}(2023{\natexlab{b}})}]{Omiya2023b}%
  \BibitemOpen
  \bibfield  {author} {\bibinfo {author} {\bibfnamefont {K.}~\bibnamefont {Omiya}}\ and\ \bibinfo {author} {\bibfnamefont {M.}~\bibnamefont {M\"uller}},\ }\bibfield  {title} {\bibinfo {title} {Fractionalization paves the way to local projector embeddings of quantum many-body scars},\ }\href {https://doi.org/10.1103/PhysRevB.108.054412} {\bibfield  {journal} {\bibinfo  {journal} {Phys. Rev. B}\ }\textbf {\bibinfo {volume} {108}},\ \bibinfo {pages} {054412} (\bibinfo {year} {2023}{\natexlab{b}})}\BibitemShut {NoStop}%
\end{thebibliography}
\end{document}